\documentclass[reqno]{amsart}
\usepackage{amsfonts}
\usepackage[colorlinks=true]{hyperref}
\usepackage{paralist}

\newtheorem{theorem}{Theorem}[section]

\newtheorem{lemma}[theorem]{Lemma}
\newtheorem{proposition}[theorem]{Proposition}

\theoremstyle{definition}

\newtheorem{remark}{Remark}

\input{tcilatex}

\begin{document}
\title[Neoclassical theory of electric charges]{Some mathematical problems
in a neoclassical theory of electric charges}
\author{Anatoli Babin }
\address{Department of Mathematics \\
University of California at Irvine \\
Irvine, CA 92697-3875, U.S.A.}
\email{ababine@math.uci.edu}
\thanks{Supported by AFOSR grant number FA9550-04-1-0359.}
\author{Alexander Figotin}
\address{Department of Mathematics \\
University of California at Irvine \\
Irvine, CA 92697-3875, U.S.A.}
\email{afigotin@math.uci.edu}
\date{}
\subjclass{35Q61; 35Q55; 35Q60; 35Q70; 35P30}
\keywords{ Maxwell equations, nonlinear Schr\"{o}dinger equation, Newton's
law, Lorentz force, hydrogen atom, nonlinear eigenvalue problem}
\dedicatory{Dedicated to Roger Temam on the occasion of his 70th birthday. }

\begin{abstract}
We study here a number of mathematical problems related to our recently
introduced neoclassical theory for electromagnetic phenomena in which
charges are represented by complex valued wave functions as in the Schr\"{o}%
dinger wave mechanics. In the non-relativistic case the dynamics of
elementary charges is governed by a system of nonlinear Schr\"{o}dinger
equations coupled with the electromagnetic fields, and we prove that if the
wave functions of charges are well separated and localized their centers
converge to trajectories of the classical point charges governed by the
Newton's equations with the Lorentz forces. We also found exact solutions in
the form of localized accelerating solitons. Our studies of a class of time
multiharmonic solutions of the same field equations show that they satisfy
Planck-Einstein relation and that the energy levels of the nonlinear
eigenvalue problem for the hydrogen atom converge to the well-known energy
levels of the linear Schr\"{o}dinger operator when the free charge size is
much larger than the Bohr radius.
\end{abstract}

\maketitle

\section{Introduction}

It is well known that the concept of a point charge interacting with the
electromagnetic (EM) field has fundamental problems. Indeed, in the
classical electrodynamics the evolution of a point charge $q$ of a mass $m$
in an external electromagnetic (EM) field is governed by Newton's equation 
\begin{equation}
\frac{\mathrm{d}}{\mathrm{d}t}\left[ m\mathbf{v}\left( t\right) \right] =q%
\left[ \mathbf{E}\left( t,\mathbf{r}\left( t\right) \right) +\frac{1}{%
\mathrm{c}}\mathbf{v}\left( t\right) \times \mathbf{B}\left( t,\mathbf{r}%
\left( t\right) \right) \right]  \label{pchar1}
\end{equation}%
where $\mathbf{r}$ and $\mathbf{v}=\mathbf{\dot{r}=}\frac{\mathrm{d}\mathbf{r%
}}{\mathrm{d}t}$ are respectively the charge's position and velocity, $%
\mathbf{E}\left( t,\mathbf{r}\right) $ and $\mathbf{B}\left( t,\mathbf{r}%
\right) $ are the electric field and the magnetic induction, and the
right-hand side of the equation (\ref{pchar1}) is the Lorentz force. On
other hand, if the charge's time-dependent position and velocity are
respectively $\mathbf{r}$ and $\mathbf{v}$ then the corresponding EM field
is described by the Maxwell equations%
\begin{equation}
\frac{1}{\mathrm{c}}\frac{\partial \mathbf{B}}{\partial t}+\nabla \times 
\mathbf{E}=\boldsymbol{0},\ \nabla \cdot \mathbf{B}=0,  \label{psys1}
\end{equation}%
\begin{equation}
\frac{1}{\mathrm{c}}\frac{\partial \mathbf{E}}{\partial t}-\nabla \times 
\mathbf{B}=-\frac{4\pi }{\mathrm{c}}q\delta \left( \mathbf{x}-\mathbf{r}%
\left( t\right) \right) \mathbf{v}\left( t\right) ,\ \nabla \cdot \mathbf{E}%
=4\pi q\delta \left( \mathbf{x}-\mathbf{r}\left( t\right) \right) ,
\label{psys2}
\end{equation}%
where $\delta $ is the Dirac delta-function,$\ \mathbf{v}\left( t\right) =%
\mathbf{\dot{r}}\left( t\right) $, $\mathrm{c}$ is speed of light. If one
would like to consider equations (\ref{pchar1})-(\ref{psys2}) as a closed
system "charge-EM field" there is a problem. Its origin is a singularity of
the EM field exactly at the position of the point charge, as, for instance,
for the electrostatic field $\mathbf{E}$ with the Coulomb's potential $\frac{%
q}{\left\vert \mathbf{x}-\mathbf{r}\right\vert }$ with a singularity at $%
\mathbf{x}=\mathbf{r}$. If one wants to stay within the classical
electromagentic theory, a possible remedy is the introduction of an extended
charge which, though very small, is not a point. There are two most well
known models for such an extended charge: the semi-relativistic Abraham
rigid charge model (a rigid sphere with spherically symmetric charge
distribution), \cite[Sections 2.4, 4.1, 10.2, 13]{Spohn}, \cite[Sections 2.2]%
{Rohrlich}, and the Lorentz relativistically covariant model which was
studied and advanced in \cite{Appel Kiessling}, \cite[Sections 16]{Jackson}, 
\cite{Nodvik}, \cite{Pearle1}, \cite[Sections 2, 6]{Rohrlich}, \cite%
{Schwinger}, \cite[Sections 2.5, 4.2, 10.1]{Spohn}, \cite{Yaghjian}. Poincar%
\'{e} suggested in 1905-1906, \cite{Poincare} (see also \cite[Sections
16.4-16.6]{Jackson}, \cite[Sections 2.3, 6.1- 6.3]{Rohrlich}, \cite[Section
63]{Pauli RT}, \cite{Schwinger}, \cite[Section 4.2]{Yaghjian} and references
there in), to add to the Lorentz-Abraham model non-electromagnetic cohesive
forces which balance the charge internal repulsive electromagnetic forces
and remarkably restore also the covariance of the entire model.

Another well known problem of a physical nature with point charges and the
classical electrodynamics is related to the Rutherford planetary model for
the hydrogen atom. The planetary model of the hydrogen atom is inconsistent
with physically observed phenomena such as the stability of atoms and the
discreteness of the hydrogen energy levels. That inconsistency lead as we
well know to introduction of non-classical models, such as the Bohr model
and later on the Schr\"{o}dinger's model of the hydrogen atom.

To address the above mentioned problems with the classical electrodynamics
we introduced recently in \cite{BF4}, \cite{BF5} \emph{wave-corpuscle
mechanics} (WCM) for charges which is conceived as one mechanics for
macroscopic and atomic scales. This model is \emph{neoclassical} in the
sense that it is based on the classical concept of the electromagnetic field
but an elementary charge is not a point but a \emph{wave-corpuscle}
described by a complex-valued scalar wave function $\psi $ with the density $%
\left\vert \psi \left( t,\mathbf{x}\right) \right\vert ^{2}$\ which is not
given a probabilistic interpretation. The dynamics and shape of a wave
corpuscle is governed by a nonlinear Klein-Gordon or a nonlinear Schr\"{o}%
dinger equation in relativistic and nonrelativistic cases respectively, and
a wave-corpuscle is defined as a special type of solutions to these
equations. Consequently an elementary charge in the WCM does not have a
prescribed geometry which differs it from the classical Abraham and Lorentz
models and more recent developments (see \cite{BambusiG93}, \cite{Kiessling2}%
, \cite{Komech09}, \cite{ImaikinKV06}, \cite{Spohn} and references therein).
Another important difference is that Newton's equations are not postulated
as in (\ref{pchar1}) but rather are derived from the field equations in a
non-relativistic regime when charges are well separated and localized.
Physical aspects of the model are discussed in detail in \cite{BF4}, \cite%
{BF5}, where the Poincar\'{e} type cohesive forces are constructed somewhat
differently then we propose here. The primary focus of this paper is on
studies of mathematical properties of the field equations of the WCM.

The paper is organized as follows. In the first two subsections of this
Introduction we describe the system relativistic and non-relativistic
Lagrangians and the corresponding field equations, and in the third one we
consider the relevant nonlinearities and their examples. Importantly, the
nonlinearity depends on a size parameter $a>0$ associated with a free
solution (ground state). Section 2 is devoted to remote interaction regimes
for many charges. We prove there that in the case of the non-relativistic
field equations when $a\rightarrow 0$ (macroscopic limit) the centers of the
interacting charges defined by the formula 
\begin{equation*}
\mathbf{r}^{\ell }\left( t\right) =\int_{\mathbb{R}^{3}}\mathbf{x}\left\vert
\psi ^{\ell }\left( t,\mathbf{x}\right) \right\vert ^{2}\,\mathrm{d}\mathbf{%
x,}\ell =1,...,N,
\end{equation*}%
converge to the solutions of Newton's equations with the Lorentz forces if $%
\psi ^{\ell }$ remain localized. We also provide examples of exact solutions
of the field equations in the form of accelerating solitons for which the
localization assumption holds. In Section 3 we study systems of bound
charges and time multiharmonic solutions to the field equations. We consider
there, in particular, a connection, discovered in a different setting in 
\cite{Bialynicki}, between the Planck-Einstein energy-frequency relation and
the logarithmic nonlinearity as well as some dynamical issues related to the
logarithmic nonlinearity closely related to results in \cite{Cazenave83}, 
\cite{CazenaveHaraux80}. We continue then with a study of a system of two
charges with the logarithmic nonlinearity as a model for the hydrogen atom.
We prove that if $\kappa =a_{1}/a\rightarrow 0$, where $a_{1}$\ is the Bohr
radius, then the lower energy levels associated with this model converge to
the well known energy levels of the linear Schr\"{o}dinger operator for the
hydrogen atom. The proof is based on an approach developed in \cite%
{BerestyckiLions83I}, \cite{BerestyckiLions83II} with certain modifications.

\subsection{Relativistic Lagrangian and field equations\label{S:rel}}

\ Let us consider a system of $N$ charges interacting directly only with the
EM field described by its $4$-vector potential $\left( \varphi ,\mathbf{A}%
\right) $. The charges are described by their wave functions $\psi ^{\ell }$
with the superscript index $\ell =1,\ldots ,N$ labeling them. The EM fields
are related to the potentials $\varphi ,\mathbf{A}$ by the following
standard relations 
\begin{equation}
\mathbf{E}=-\nabla \varphi -\frac{1}{\mathrm{c}}\partial _{t}\mathbf{A},\ 
\mathbf{B}=\nabla \times \mathbf{A}.  \label{fia}
\end{equation}%
We introduce now the system Lagrangian $\mathcal{L}$ which is though similar
to the one introduced in \cite{BF4}, \cite{BF5} but differs from it. A
different Lagrangian is introduced here in order to get the hydrogen atom
frequency spectrum with desired precision. Namely, to every $\ell $-th
charge is assigned an adjunct potential $\left( \varphi ^{\ell },\mathbf{A}%
^{\ell }\right) $ describing its additional degrees of freedom and our
relativistic Lagrangian is defined by 
\begin{gather}
\mathcal{L}\left( \left( \varphi ,\mathbf{A}\right) ,\left\{ \psi ^{\ell
}\right\} _{\ell =1}^{N},\left\{ \left( \varphi ^{\ell },\mathbf{A}^{\ell
}\right) \right\} _{\ell =1}^{N}\right) =\frac{1}{8\pi }\left[ \left( \nabla
\varphi +\frac{1}{\mathrm{c}}\partial _{t}\mathbf{A}\right) ^{2}-\left(
\nabla \times \mathbf{A}\right) ^{2}\right]  \label{relsh} \\
+\sum_{\ell }\frac{\chi ^{2}}{2m^{\ell }}\left[ \frac{1}{\mathrm{c}^{2}}%
\left\vert \tilde{\partial}_{t}^{\ell }\psi ^{\ell }\right\vert
^{2}-\left\vert \tilde{\nabla}\psi ^{\ell }\right\vert ^{2}-\kappa _{0\ell
}^{2}\left\vert \psi ^{\ell }\right\vert ^{2}-G^{\ell }\left( \psi ^{\ell
\ast }\psi ^{\ell }\right) \right]  \notag \\
-\sum_{\ell }\frac{1}{8\pi }\left[ \left( \nabla \varphi ^{\ell }+\frac{1}{%
\mathrm{c}}\partial _{t}\mathbf{A}^{\ell }\right) ^{2}-\left( \nabla \times 
\mathbf{A}^{\ell }\right) ^{2}\right] ,  \notag
\end{gather}%
where (i) the covariant derivatives are defined by 
\begin{equation}
\tilde{\partial}_{t}^{\ell }=\partial _{t}+\frac{\mathrm{i}q^{\ell }\left(
\varphi -\varphi ^{\ell }\right) }{\chi },\ \tilde{\nabla}^{\ell }=\nabla -%
\frac{\mathrm{i}q^{\ell }\left( \mathbf{A-A}^{\ell }\right) }{\chi c},
\label{dtl}
\end{equation}%
(ii) $\psi ^{\ast }$ is complex conjugate to $\psi $; (iii) $m^{\ell }>0$ is
the $\ell $-th charge mass, $q^{\ell }$ is the value of the charge, $\kappa
_{0\ell }=\frac{m^{\ell }\mathrm{c}}{\chi }$, and $\chi >0$ is a constant
similar to the Planck constant $\hbar =\frac{h}{2\pi }$; (iv) $G^{\ell
}\left( \psi ^{\ell \ast }\psi ^{\ell }\right) $ is a nonlinearity which
will be described below.

The Euler-Lagrange field equations for the Lagrangian $\mathcal{L}$ defined
in (\ref{relsh}) include, first of all, the Maxwell equations for the EM
potentials 
\begin{gather}
\nabla \cdot \left( \frac{1}{\mathrm{c}}\partial _{t}\mathbf{A}+\nabla
\varphi \right) =-4\pi \dsum_{\ell }\rho ^{\ell },  \label{mplag14} \\
\nabla \times \left( \nabla \times \mathbf{A}\right) +\frac{1}{\mathrm{c}}%
\partial _{t}\left( \frac{1}{\mathrm{c}}\partial _{t}\mathbf{A}+\nabla
\varphi \right) =\frac{4\pi }{\mathrm{c}}\dsum_{\ell }\mathbf{J}^{\ell },
\label{mplag15}
\end{gather}%
where the charge densities and currents are defined by%
\begin{gather}
\rho ^{\ell }=-\frac{q^{\ell }\left\vert \psi ^{\ell }\right\vert ^{2}}{%
m^{\ell }\mathrm{c}^{2}}\left( \chi \func{Im}\frac{\partial _{t}\psi ^{\ell }%
}{\psi ^{\ell }}+q^{\ell }\left( \varphi -\varphi ^{\ell }\right) \right) ,
\label{mplag10} \\
\mathbf{J}^{\ell }=\frac{q^{\ell }\left\vert \psi ^{\ell }\right\vert ^{2}}{%
m^{\ell }}\left( \chi \func{Im}\frac{\nabla \psi ^{\ell }}{\psi ^{\ell }}-%
\frac{q^{\ell }\left( \mathbf{A-A}^{\ell }\right) }{\mathrm{c}}\right) .
\label{mplag11}
\end{gather}%
The field equations include similar Maxwell equations for the adjunct
potentials 
\begin{gather}
\nabla \cdot \left( \frac{1}{\mathrm{c}}\partial _{t}\mathbf{A}^{\ell
}+\nabla \varphi ^{\ell }\right) =-4\pi \rho ^{\ell },  \label{fiash} \\
\ \nabla \times \left( \nabla \times \mathbf{A}^{\ell }\right) +\frac{1}{%
\mathrm{c}}\partial _{t}\left( \frac{1}{\mathrm{c}}\partial _{t}\mathbf{A}%
^{\ell }+\nabla \varphi ^{\ell }\right) =\frac{4\pi }{\mathrm{c}}\mathbf{J}%
^{\ell },\ \ell =1,...,N,  \label{fial}
\end{gather}%
and equations for the wave functions $\psi ^{\ell }$ in the form of
nonlinear Klein-Gordon equations 
\begin{equation}
-\frac{1}{\mathrm{c}^{2}}\tilde{\partial}_{t}^{\ell }\tilde{\partial}%
_{t}^{\ell }\psi ^{\ell }+\tilde{\nabla}^{\ell 2}\psi ^{\ell }-G^{\ell
\prime }\left( \psi ^{\ell \ast }\psi ^{\ell }\right) \psi ^{\ell }-\kappa
_{0}^{2}\psi ^{\ell }=0,\ \ell =1,...,N.  \label{KG}
\end{equation}%
Note that equations (\ref{KG}) for $\psi ^{\ell }$ are coupled to the
equations for EM potentials via the covariant derivatives.

The Lagrangian and equations are Lorentz and gauge invariant, and one can
verify that defined above $\rho ^{\ell }$ and $\mathbf{J}^{\ell }$ satisfy
conservation/continuity equation 
\begin{equation*}
\frac{1}{\mathrm{c}}\partial _{t}\rho ^{\ell }+\nabla \cdot \mathbf{J}^{\ell
}=0.
\end{equation*}%
\ If we choose the Lorentz gauge 
\begin{equation}
\frac{1}{\mathrm{c}}\partial _{t}\varphi +\nabla \cdot \mathbf{A}=0,\ \frac{1%
}{\mathrm{c}}\partial _{t}\varphi ^{\ell }+\nabla \cdot \mathbf{A}^{\ell }=0
\label{Lorg}
\end{equation}%
for all potentials, equations (\ref{mplag14})-(\ref{fial}) take the form%
\begin{gather}
\nabla \varphi -\frac{1}{\mathrm{c}^{2}}\partial _{t}^{2}\varphi
=-\dsum_{\ell =1}^{N}4\pi \rho ^{\ell },  \label{filor0} \\
\frac{1}{\mathrm{c}^{2}}\partial _{t}^{2}\mathbf{A}-\nabla ^{2}\mathbf{A}=%
\frac{4\pi }{\mathrm{c}}\dsum_{\ell =1}^{N}\mathbf{J}^{\ell },  \label{alor0}
\end{gather}%
\begin{gather}
\nabla ^{2}\varphi ^{\ell }-\frac{1}{\mathrm{c}^{2}}\partial _{t}^{2}\varphi
^{\ell }=-4\pi \rho ^{\ell },  \label{filor} \\
\frac{1}{\mathrm{c}^{2}}\partial _{t}^{2}\mathbf{A}^{\ell }-\nabla ^{2}%
\mathbf{A}^{\ell }=\frac{4\pi }{\mathrm{c}}\mathbf{J}^{\ell },\ell =1,...,N,
\label{alor}
\end{gather}%
where $\rho ^{\ell },\mathbf{J}^{\ell }$ are defined by (\ref{mplag10}), (%
\ref{mplag11}). Based on the above equations we assume that always 
\begin{equation}
\left( \varphi ,\mathbf{A}\right) =\sum_{\ell }\left( \varphi ^{\ell },%
\mathbf{A}^{\ell }\right) .  \label{fiasum}
\end{equation}%
Hence, from now on \emph{we assume that in the relativistic case the
dynamics of EM\ fields and charges is determined by equations (\ref{filor}),
(\ref{alor}), (\ref{mplag10}), (\ref{mplag11}), (\ref{KG}) }and according to
(\ref{fiasum}) and (\ref{dtl}) the covariant derivatives $\tilde{\partial}%
_{t}^{\ell }$ and $\tilde{\nabla}^{\ell }$ in (\ref{KG}) involve only fields 
$\left( \varphi ^{\ell ^{\prime }},\mathbf{A}^{\ell ^{\prime }}\right) $
with $\ell ^{\prime }\neq \ell ,$%
\begin{equation}
\tilde{\partial}_{t}^{\ell }=\partial _{t}+\frac{\mathrm{i}q^{\ell }}{\chi }%
\sum_{\ell ^{\prime }\neq \ell }\varphi ^{\ell ^{\prime }},\qquad \tilde{%
\nabla}^{\ell }=\nabla -\frac{\mathrm{i}q^{\ell }}{\chi c}\sum_{\ell
^{\prime }\neq \ell }\mathbf{A}^{\ell ^{\prime }},  \label{dtexl}
\end{equation}%
implying that the adjunct potentials completely compensate the EM\
self-action for every charge and effectively there is no EM self-interaction
(in \cite{BF4}, \cite{BF5} this kind of compensation was gained by using an
additional nonlinear self-interaction).

Let us turn now to the nonlinearities $G$. We introduce the nonlinearities
here the same way as in \cite{BF4}, \cite{BF5} with an intension to have a
rest (ground) state for a single charge in the form of a localized wave
function (form factor). Notice that for a single charge evidently $N=1$, $%
\varphi -\varphi ^{\ell }=0$, $\mathbf{A-A}^{\ell }=0$ and we look for the
rest solution in the form 
\begin{equation}
\psi ^{\ell }\left( t,\mathbf{x}\right) =\mathrm{e}^{-i\omega _{0}t}\psi
\left( \mathbf{x}\right) ,\qquad \omega _{0}=\frac{m\mathrm{c}^{2}}{\chi }=%
\mathrm{c}\kappa _{0}  \label{psioml}
\end{equation}%
with $\mathbf{A}^{\ell }=0$. Substituting (\ref{psioml}) in (\ref{KG})
yields 
\begin{gather*}
-\nabla ^{2}\varphi =4\pi \left\vert \psi \right\vert ^{2}, \\
-\nabla ^{2}\psi +G^{\prime }\left( \psi ^{\ast }\psi \right) \psi =0.
\end{gather*}%
Now we choose a strictly positive, monotonically decreasing radial function $%
\mathring{\psi}$ (ground state) as a parameter for the model and determine
the nonlinearity $G^{\prime }$ from the following \emph{charge equilibrium
condition:}%
\begin{equation}
-\nabla ^{2}\mathring{\psi}+G^{\prime }\left( \left\vert \mathring{\psi}%
\right\vert ^{2}\right) \mathring{\psi}=0.  \label{nop40}
\end{equation}%
The above equation allows to determine the nonlinearity $G^{\prime }$ as
long as $\mathring{\psi}$ is a strictly positive, smooth and monotonically
decreasing function of $r=\left\vert \mathbf{x}\right\vert $. In Section \ref%
{snonlin} we consider the nonlinearity in more details providing also
examples.

Note that using the Lorentz invariance of the system one can easily obtain a
solution which represents the charge-field moving with a constant velocity $%
\mathbf{v}$ simply by applying to the rest solution $\left( \psi ^{\ell
},\varphi ,\mathbf{0}\right) $ the Lorentz transformation (see \cite{BF4}, 
\cite{BF5}).

\subsection{Non-relativistic Lagrangian and field equations\label{S:nrel}}

Our non-relativistic model describes the case of charges moving with
non-relativistic velocities and it is set as follows. Using the
frequency-shifting substitution (\ref{psioml}) with more general $\psi =\psi
_{\omega }^{\ell }\left( t,\mathbf{x}\right) $ which depends on $\left( t,%
\mathbf{x}\right) $ we observe that the second time derivative in (\ref{KG})
can be written in the form 
\begin{equation*}
-\frac{1}{\mathrm{c}^{2}}\tilde{\partial}_{t}^{\ell }\tilde{\partial}%
_{t}^{\ell }\psi _{\omega }^{\ell }=\frac{1}{\mathrm{c}^{2}}\left( \partial
_{t}+\frac{\mathrm{i}q^{\ell }}{\chi }\varphi _{\neq \ell }\right) ^{2}\psi
_{\omega }^{\ell }-2i\frac{m^{\ell }}{\chi }\left( \partial _{t}+\frac{%
\mathrm{i}q^{\ell }}{\chi }\varphi _{\neq \ell }\right) \psi _{\omega
}^{\ell }+\kappa _{0}^{2}\psi _{\omega }^{\ell },
\end{equation*}%
where%
\begin{equation}
\varphi _{\neq \ell }=\sum_{\ell ^{\prime }\neq \ell }\varphi ^{\ell
^{\prime }}.  \label{fineq}
\end{equation}%
We neglect the term with the factor $\frac{1}{\mathrm{c}^{2}}$ and
substitute $-2\mathrm{i}\frac{m^{\ell }}{\chi }\left( \partial _{t}+\frac{%
\mathrm{i}q^{\ell }}{\chi }\varphi _{\neq \ell }\right) \psi _{\omega
}^{\ell }+\kappa _{0}^{2}\psi _{\omega }^{\ell }$ for the term $-\frac{1}{%
\mathrm{c}^{2}}\tilde{\partial}_{t}^{\ell }\tilde{\partial}_{t}^{\ell }\psi
_{\omega }^{\ell }$ in (\ref{KG}). Consequently, we replace the nonlinear
Klein-Gordon equation (\ref{KG}) by the following nonlinear Schr\"{o}dinger
equation (where we denote the frequency shifted $\psi _{\omega }^{\ell }$ by 
$\psi ^{\ell }$) 
\begin{equation}
\chi \mathrm{i}\partial _{t}\psi ^{\ell }+\frac{\chi ^{2}}{2m^{\ell }}\left( 
\tilde{\nabla}^{\ell }\right) ^{2}\psi ^{\ell }-\frac{\chi ^{2}}{2m^{\ell }}%
G^{\ell \prime }\left( \psi ^{\ell \ast }\psi ^{\ell }\right) \psi ^{\ell
}-q^{\ell }\varphi _{\neq \ell }\psi ^{\ell }=0,  \label{eqp1}
\end{equation}%
where $\tilde{\nabla}^{\ell }$ is given by (\ref{dtexl}). Since the magnetic
fields generated by moving charges also have coefficient $\frac{1}{\mathrm{c}%
}$ we neglect them preserving only the external magnetic fields and replace $%
\tilde{\nabla}^{\ell }$ by the covariant gradient $\tilde{\nabla}_{\mathrm{ex%
}}^{\ell }$ defined by 
\begin{equation}
\tilde{\nabla}_{\mathrm{ex}}^{\ell }=\nabla -\frac{\mathrm{i}q^{\ell }%
\mathbf{A}_{\mathrm{ex}}}{\chi c}.  \label{nop1a}
\end{equation}%
\ So our non-relativistic Lagrangian is 
\begin{gather}
\mathcal{\hat{L}}_{0}\left( \varphi ,\left\{ \psi ^{\ell }\right\} _{\ell
=1}^{N},\left\{ \varphi ^{\ell }\right\} _{\ell =1}^{N}\right) =\frac{%
\left\vert \nabla \varphi \right\vert ^{2}}{8\pi }+\sum_{\ell }\hat{L}^{\ell
}\left( \psi ^{\ell },\psi ^{\ell \ast },\varphi \right) ,  \label{Lbet} \\
\hat{L}_{0}^{\ell }=\frac{\chi \mathrm{i}}{2}\left[ \psi ^{\ell \ast
}\partial _{t}\psi ^{\ell }-\psi ^{\ell }\partial _{t}\psi ^{\ell \ast }%
\right] -\frac{\chi ^{2}}{2m^{\ell }}\left\{ \left\vert \tilde{\nabla}_{%
\mathrm{ex}}^{\ell }\psi ^{\ell }\right\vert ^{2}+G^{\ell }\left( \psi
^{\ell \ast }\psi ^{\ell }\right) \right\} -  \notag \\
-q^{\ell }\left( \varphi +\varphi _{\mathrm{ex}}-\varphi ^{\ell }\right)
\psi ^{\ell }\psi ^{\ell \ast }-\frac{\left\vert \nabla \varphi ^{\ell
}\right\vert ^{2}}{8\pi },  \notag
\end{gather}%
where $\mathbf{A}_{\mathrm{ex}}\left( t,\mathbf{x}\right) $ and $\varphi _{%
\mathrm{ex}}\left( t,\mathbf{x}\right) $ are potentials of external EM
fields, $\psi ^{\ell \ast }$ is complex conjugate of $\psi ^{\ell }$. The
Euler-Lagrange equations for the electrostatic potentials have the form 
\begin{equation}
-\frac{1}{4\pi }\nabla ^{2}\varphi =\sum_{\ell =1}^{N}q^{\ell }\psi ^{\ell
}\psi ^{\ell \ast },  \label{eqp2}
\end{equation}%
\begin{equation}
-\frac{1}{4\pi }\nabla ^{2}\varphi ^{\ell }=q^{\ell }\psi ^{\ell }\psi
^{\ell \ast },\ell =1,...,N.  \label{eqp3}
\end{equation}%
Assuming that that $\varphi ,\varphi ^{\ell }$ vanish at infinity we obtain
a reduced version of (\ref{fiasum}) 
\begin{equation}
\varphi =\sum_{\ell }\varphi ^{\ell }.  \label{fisuml}
\end{equation}%
Similarly to the relativistic case we assume now that \emph{nonrelativistic
equations for dynamics of charges} in the external EM field with potentials $%
\varphi _{\mathrm{ex}},\mathbf{A}_{\mathrm{ex}}$ take the form%
\begin{equation}
\mathrm{i}\chi \partial _{t}\psi ^{\ell }=-\frac{\chi ^{2}}{2m^{\ell }}%
\left( \tilde{\nabla}_{\mathrm{ex}}^{\ell }\right) ^{2}\psi ^{\ell }+q^{\ell
}\left( \varphi _{\neq \ell }+\varphi _{\mathrm{ex}}\right) \psi ^{\ell }+%
\frac{\chi ^{2}}{2m^{\ell }}\left[ G_{a}^{\ell }\right] ^{\prime }\left(
\left\vert \psi ^{\ell }\right\vert ^{2}\right) \psi ^{\ell },  \label{NLSj0}
\end{equation}%
where $\varphi _{\neq \ell }$ is given by (\ref{fineq}) and $\varphi ^{\ell
} $ is determined from the equations 
\begin{equation}
\nabla ^{2}\varphi ^{\ell }=-4\pi q^{\ell }\left\vert \psi ^{\ell
}\right\vert ^{2},\ell =1,...,N.  \label{delfi}
\end{equation}%
The solution of the above equation is given by the formula 
\begin{equation}
\varphi ^{\ell }\left( t,\mathbf{x}\right) =q^{\ell }\dint_{\mathbb{R}^{3}}%
\frac{\left\vert \psi ^{\ell }\right\vert ^{2}\left( t,\mathbf{y}\right) }{%
\left\vert \mathbf{y}-\mathbf{x}\right\vert }\mathrm{d}\mathbf{y.}
\label{jco3}
\end{equation}%
The nonlinear self-interaction terms $G_{a}^{\ell }$ in (\ref{NLSj0}) are
determined through the charge equilibrium equation (\ref{nop40}) and index $%
a $ indicates the dependence on the\emph{\ size parameter} $a>0$ which we
introduce by the formula 
\begin{equation}
G_{a}^{\prime }\left( s\right) =a^{-2}G_{1}^{\prime }\left( a^{3}s\right) .
\label{totgkap}
\end{equation}%
More detailed discussion of the nonlinearity is given in the following
section.

\subsection{Nonlinearity, its basic properties and examples\label{snonlin}}

As we have already mentioned, the \emph{nonlinear self interaction function }%
$G$\emph{\ is determined from the charge equilibrium\ equation} (\ref{nop40}%
) based on the form factor (free ground state) $\mathring{\psi}$. Important
features of our nonlinearity include: (i) the boundedness or slow
subcritical growth of its derivative $G^{\prime }\left( s\right) $ for $%
s\rightarrow \infty $ with consequent boundedness from below of the energy;
(ii) slightly singular behavior about $s=0$, that is for small wave
amplitudes.

In this section we consider the construction of the function $G$, study its
properties and provide examples for which the construction of $G$ is carried
out explicitly. Throughout this section we have%
\begin{equation*}
\psi ,\mathring{\psi}\geq 0\text{ and hence }\left\vert \psi \right\vert
=\psi .
\end{equation*}%
We introduce explicitly the dependence of the free ground state $\mathring{%
\psi}$ on the size parameter $a>0$ through the following representation of
the function $\mathring{\psi}\left( r\right) $ 
\begin{equation}
\mathring{\psi}\left( r\right) =\mathring{\psi}_{a}\left( r\right) =a^{-3/2}%
\mathring{\psi}_{1}\left( a^{-1}r\right) ,  \label{nrac5}
\end{equation}%
where $\mathring{\psi}_{1}\left( \mathsf{r}\right) $ is a function of the
dimensionless variable $\mathsf{r}\geq 0$. The dependence on $a$ is chosen
so that $L^{2}$-norm $\left\Vert \mathring{\psi}_{a}\left( r\right)
\right\Vert $ does not depend on $a,$ hence the function $\mathring{\psi}%
_{a}\left( r\right) $ satisfies the charge normalization condition (\ref%
{norm10}) for every $a>0$. Obviously, definition (\ref{nrac5}) is consistent
with (\ref{nop40}) and (\ref{totgkap}). The size parameter $a$ naturally has
the dimension of length. A properly defined spatial size of $\mathring{\psi}%
_{a}$, based, for instance, on the variance, is proportional to $a$\ with a
coefficient depending on $\mathring{\psi}_{1}$. The charge equilibrium\
equation (\ref{nop40}) can be written in the following form: 
\begin{equation}
\nabla ^{2}\mathring{\psi}_{a}=G_{a}^{\prime }\left( \mathring{\psi}%
_{a}^{2}\right) \mathring{\psi}_{a}.  \label{stp}
\end{equation}%
The function $\mathring{\psi}_{a}\left( r\right) $ is assumed to be a smooth
(at least twice continuously differentiable) positive monotonically
decreasing function of $r\geq 0$ which is square integrable with weight $%
r^{2},$ we assume that its derivative $\mathring{\psi}_{a}^{\prime }\left(
r\right) $ is negative for $r>0$ and we assume it to satisfy the charge
normalization condition of the form (\ref{norm10}); such a function is
usually called in literature a ground state.

Let us look first at the case $a=1$,\ $\mathring{\psi}_{a}=\mathring{\psi}%
_{1}$,\ $\mathring{\varphi}_{a}=\mathring{\varphi}_{1}$, for which the
equation (\ref{stp}) yields the following representation for $G^{\prime }(%
\mathring{\psi}_{1}^{2})$ from (\ref{stp})%
\begin{equation}
G_{1}^{\prime }\left( \mathring{\psi}_{1}^{2}\left( r\right) \right) =\frac{%
(\nabla ^{2}\mathring{\psi}_{1})\left( r\right) }{\mathring{\psi}_{1}\left(
r\right) }.  \label{gg}
\end{equation}%
Since $\mathring{\psi}_{1}^{2}\left( r\right) $ is a \emph{monotonic}
function,\ we can find its inverse $r=r\left( \psi ^{2}\right) ,$ yielding 
\begin{equation}
G_{1}^{\prime }\left( s\right) =\frac{\nabla ^{2}\mathring{\psi}_{1}\left(
r\left( s\right) \right) }{\mathring{\psi}_{1}\left( r\left( s\right)
\right) },\ 0=\mathring{\psi}_{1}^{2}\left( \infty \right) \leq s\leq 
\mathring{\psi}_{1}^{2}\left( 0\right) .  \label{intps}
\end{equation}%
Since $\mathring{\psi}_{1}\left( r\right) $ is smooth and $\partial _{r}%
\mathring{\psi}_{1}<0$, $G^{\prime }(\left\vert \psi \right\vert ^{2})$ is
smooth for $0<\left\vert \psi \right\vert ^{2}<\mathring{\psi}_{1}^{2}\left(
0\right) $. If we do not need $G^{\prime }\left( s\right) $ to be smooth, we
extend $G^{\prime }\left( s\right) $ for $s\geq \mathring{\psi}%
_{1}^{2}\left( 0\right) $ as a constant, namely 
\begin{equation}
G_{1}^{\prime }\left( s\right) =G_{1}^{\prime }\left( \mathring{\psi}%
_{1}^{2}\left( 0\right) \right) \text{ if }s\geq \mathring{\psi}%
_{1}^{2}\left( 0\right) .  \label{intps1}
\end{equation}%
The first derivative of such an extension at $s=\mathring{\psi}%
_{1}^{2}\left( 0\right) $ has a discontinuity point. If $\mathring{\psi}%
_{a}\left( r\right) $ is a smooth function of class $C^{n}$, $n>2,$ we
always can define an extension of $G^{\prime }\left( s\right) $ for $s\geq 
\mathring{\psi}_{1}^{2}\left( 0\right) $ as a bounded function of class $%
C^{n-2}$ for all $r>0$ and 
\begin{equation}
G_{1}^{\prime }\left( s\right) =G_{1}^{\prime }\left( \mathring{\psi}%
_{1}^{2}\left( 0\right) \right) -1\text{ if }s\geq \mathring{\psi}%
_{1}^{2}\left( 0\right) +1.  \label{intps2}
\end{equation}%
Slowly growing (subcritical) functions $G^{\prime }\left( s\right) $ which
are not constant for large $s$ also can be used, see examples below.

In the case of arbitrary size parameter $a>0$ we define $G_{a}^{\prime
}\left( s\right) $ by formula (\ref{totgkap}), and this definition is
consistent with (\ref{nrac5}) and (\ref{intps}).

Let us take a look at general properties of $G^{\prime }\left( s\right) $ as
they follow from defining them relations (\ref{intps}). In the examples
below the function $G^{\prime }\left( s\right) $ is not differentiable at $%
s=0$, but if $\mathring{\psi}\left( r\right) $ decays exponentially or with
a power law the nonlinearity $g\left( \psi \right) =G^{\prime }(\left\vert
\psi \right\vert ^{2})\psi $ as it enters the field equation (\ref{NLSj0})
is differentiable for all $\psi $ including zero, hence it satisfies\ the
Lipschitz condition. For a Gaussian $\mathring{\psi}_{1}\left( r\right) $
which decays superexponentially $G^{\prime }(\left\vert \psi \right\vert
^{2})$ is unbounded at zero and $g\left( \psi \right) $ is not
differentiable at zero. Since $\mathring{\psi}\left( \left\vert \mathbf{x}%
\right\vert \right) >0$, the sign of $G_{1}^{\prime }\left( \left\vert \psi
\right\vert ^{2}\right) $ coincides with the sign of $\nabla ^{2}\mathring{%
\psi}_{1}\left( \left\vert \mathbf{x}\right\vert \right) $. At the origin $%
\mathbf{x}=\mathbf{0}$ the function $\mathring{\psi}_{1}\left( \left\vert 
\mathbf{x}\right\vert \right) $ has its maximum and, consequently, $%
G_{1}^{\prime }\left( s\right) \leq 0$ for $s$ close to $s=\mathring{\psi}%
_{1}^{2}\left( 0\right) $. The Laplacian applied to the radial function $%
\mathring{\psi}_{1}$ takes the form $\frac{1}{r}\frac{\partial ^{2}}{%
\partial r^{2}}\left( r\mathring{\psi}_{1}\left\vert \mathbf{x}\right\vert
\right) $. Consequently, if $r\mathring{\psi}_{1}\left( r\right) $ is convex
at $r=\left\vert \mathbf{x}\right\vert $ we have $\nabla ^{2}\mathring{\psi}%
_{1}\left( \left\vert \mathbf{x}\right\vert \right) \geq 0$. Since $r^{2}%
\mathring{\psi}_{1}\left( r\right) $ is integrable, we naturally assume that 
$\left\vert \mathbf{x}\right\vert \mathring{\psi}_{1}\left( \left\vert 
\mathbf{x}\right\vert \right) \rightarrow 0$ as $\left\vert \mathbf{x}%
\right\vert \rightarrow \infty $. Then if the second derivative of $r%
\mathring{\psi}_{1}\left( r\right) $ has a constant sign near infinity, it
must be non-negative as well as $G_{1}^{\prime }\left( s\right) $ for $s\ll
1 $. In the examples we give below $G_{1}^{\prime }\left( s\right) $ has
exactly one zero on the half-axis.

\textbf{Example 1.} Consider a form factor $\mathring{\psi}_{1}\left(
r\right) $ decaying as a power law, namely 
\begin{equation}
\mathring{\psi}_{1}\left( r\right) =\frac{c_{\mathrm{pw}}}{\left(
1+r^{2}\right) ^{5/4}},\   \label{expsi1}
\end{equation}%
where $c_{\mathrm{pw}}$ is the normalization factor, $c_{\mathrm{pw}%
}=3^{1/2}/\left( 4\pi \right) ^{1/2}$. This function evidently is positive
and monotonically decreasing. Let us find now $G^{\prime }\left( s\right) $
based on the relations (\ref{intps}). An elementary computation of $\nabla
^{2}\mathring{\psi}_{1}$ shows that%
\begin{gather}
G^{\prime }\left( s\right) =\frac{15s^{2/5}}{4c_{\mathrm{pw}}^{4/5}}-\frac{%
45s^{4/5}}{4c_{\mathrm{pw}}^{8/5}},  \label{exGd1} \\
G\left( s\right) =\frac{75s^{7/5}}{28c_{\mathrm{pw}}^{4/5}}-\frac{25s^{9/5}}{%
4c_{\mathrm{pw}}^{8/5}},\text{ for }0\leq s\leq c_{\mathrm{pw}}^{2}.  \notag
\end{gather}%
The extension for $s\geq c_{\mathrm{pw}}^{2}$ can be defined as a constant
or the same formula (\ref{exGd1}) can be used for all $s\geq 0$ since
corresponding nonlinearity has a subcritical growth.

If we explicitly introduce size parameter $a$ into the form factor using (%
\ref{nrac5}), we define $G_{a}^{\prime }\left( s\right) $ by (\ref{totgkap}%
). Notice that the variance of the form factor $\mathring{\psi}%
_{1}^{2}\left( \left\vert \mathbf{x}\right\vert \right) $ decaying as a
power law (\ref{expsi1}) is infinite.

\textbf{Example 2}. Now we consider an exponentially decaying form factor $%
\mathring{\psi}_{1}$ of the form 
\begin{equation}
\mathring{\psi}_{1}\left( r\right) =c_{\mathrm{e}}\mathrm{e}^{-\left(
r^{2}+1\right) ^{1/2}}\text{,\qquad }c_{\mathrm{e}}=\left( 4\pi
\int_{0}^{\infty }r^{2}\mathrm{e}^{-2\left( r^{2}+1\right) ^{1/2}}\,\mathrm{d%
}r\right) ^{-1/2}.  \label{grs}
\end{equation}%
Evidently $\mathring{\psi}_{1}\left( r\right) $ is positive and
monotonically decreasing. The dependence $r\left( s\right) $ defined by the
relation (\ref{grs})\ is as follows:\ 
\begin{equation}
r=\left[ \ln ^{2}\left( c_{\mathrm{e}}/\sqrt{s}\right) -1\right] ^{1/2},%
\text{ if }\sqrt{s}\leq \mathring{\psi}_{1}\left( 0\right) =c_{\mathrm{e}}%
\mathrm{e}^{-1}.  \label{rofp}
\end{equation}%
An elementary computation shows that 
\begin{equation}
-\frac{\nabla ^{2}\mathring{\psi}_{1}}{\mathring{\psi}_{1}}=\frac{2}{\left(
r^{2}+1\right) ^{\frac{1}{2}}}+\frac{1}{\left( r^{2}+1\right) }+\frac{1}{%
\left( r^{2}+1\right) ^{\frac{3}{2}}}-1.  \label{rofp2}
\end{equation}%
Combining (\ref{rofp}) with (\ref{rofp2}) we readily obtain the following
function for $s\leq c_{\mathrm{e}}^{2}\mathrm{e}^{-2}$ 
\begin{equation}
G_{1}^{\prime }\left( s\right) =\left[ 1-\frac{4}{\ln \left( c_{\mathrm{e}%
}^{2}/s\right) }-\frac{4}{\ln ^{2}\left( c_{\mathrm{e}}^{2}/s\right) }-\frac{%
8}{\ln ^{3}\left( c_{\mathrm{e}}^{2}/s\right) }\right] .\text{ }
\label{ggkap}
\end{equation}%
We can extend it for larger $s$ as follows:%
\begin{equation}
G_{1}^{\prime }\left( s\right) =G_{1}^{\prime }\left( c_{\mathrm{e}}^{2}%
\mathrm{e}^{-2}\right) =-3\text{ if }s\geq c_{\mathrm{e}}^{2}\mathrm{e}^{-2},
\label{ggkap1}
\end{equation}%
or we can use a smooth extension as in (\ref{intps2}). The function $%
G_{1}^{\prime }\left( s\right) $ is not differentiable at $s=0$. At the same
time the function $g\left( \psi \right) =G_{1}^{\prime }\left( \psi \left(
r\right) \right) \psi $ if we set $g\left( 0\right) =0$ is continuous and $%
g\left( \psi \right) $ is continuously differentiable with respect to $\psi
\ $at zero and $g\left( \psi \right) $ satisfies a Lipschitz condition. The
variance of the exponential form factor $\mathring{\psi}_{1}\left( r\right) $
is obviously finite. To find $G_{a}^{\prime }\left( s\right) $ for arbitrary 
$a$ we use its representation (\ref{totgkap}).

\textbf{Example 3}. Now we define a \emph{Gaussian} form factor by the
formula 
\begin{equation}
\mathring{\psi}\left( r\right) =C_{g}\mathrm{e}^{-r^{2}/2},\quad C_{g}=\frac{%
1}{\pi ^{3/4}}.  \label{Gaussp}
\end{equation}%
Such a ground state is called \emph{gausson} in \cite{Bialynicki}.
Elementary computation shows that 
\begin{equation*}
\frac{\nabla ^{2}\mathring{\psi}\left( r\right) }{\mathring{\psi}\left(
r\right) }=r^{2}-3=-\ln \left( \mathring{\psi}^{2}\left( r\right)
/C_{g}^{2}\right) -3.
\end{equation*}%
Hence, we define the nonlinearity by the formula%
\begin{equation}
G^{\prime }\left( \left\vert \psi \right\vert ^{2}\right) =-\ln \left(
\left\vert \psi \right\vert ^{2}/C_{g}^{2}\right) -3,  \label{Gaussg}
\end{equation}%
and refer to it as \emph{logarithmic nonlinearity. }The nonlinear potential
function has the form%
\begin{equation}
G\left( s\right) =\int_{0}^{s}\left( -\ln \left( s^{\prime
}/C_{g}^{2}\right) -3\right) ds^{\prime }=-s\ln s+s\left( \ln \frac{1}{\pi
^{3/2}}-2\right) .  \label{g1gauss}
\end{equation}%
\emph{\ }Dependence on the size parameter $a>0$ is given by the formula 
\begin{equation}
G_{a}^{\prime }\left( \left\vert \psi \right\vert ^{2}\right) =-a^{-2}\ln
\left( a^{3}\left\vert \psi \right\vert ^{2}/C_{g}^{2}\right) -3a^{-2}.
\label{Gpa}
\end{equation}%
Obviously $g\left( \psi \right) =G_{1}^{\prime }\left( |\psi |^{2}\right)
\psi $ is continuous for all $\psi \in \mathbb{C}$ if at zero we set $%
g\left( 0\right) $ $=0$ and is differentiable for every $\psi \neq 0$ but is
not differentiable at $\psi =0$ and does not satisfy the Lipschitz
condition. Notice also that $g\left( \psi \right) $ has a subcritical growth
as $\left\vert \psi \right\vert \rightarrow \infty .$

\section{Charges in remote interaction regimes\label{nrapr1}}

The primary focus of this section is to show that if the size parameter $%
a\rightarrow 0$ then the dynamics of centers of localized solutions is
approximated by the Newton equations with the Lorentz forces. This is done
in the spirit of the well known in the quantum mechanics Ehrenfest\ Theorem%
\emph{, }\cite[Sections 7, 23]{Schiff}. We also provide as an example
explicit wave-corpuscle solutions which have such a dynamics.

As a first step we describe basic properties of the classical solutions of (%
\ref{NLSj0}), (\ref{delfi}). The Lagrangian $\mathcal{\hat{L}}$ is gauge
invariant and every $\ell $-th charge has a 4-current $\left( \rho ^{\ell },%
\mathbf{J}^{\ell }\right) $\ defined by 
\begin{equation}
\rho ^{\ell }=q\left\vert \psi ^{\ell }\right\vert ^{2},\ \mathbf{J}^{\ell
}=\left( \frac{\chi q^{\ell }}{m^{\ell }}\func{Im}\frac{\nabla \psi ^{\ell }%
}{\psi ^{\ell }}-\frac{q^{\ell 2}\mathbf{A}_{\mathrm{ex}}}{m^{\ell }\mathrm{c%
}}\right) \left\vert \psi ^{\ell }\right\vert ^{2},  \label{jco1}
\end{equation}%
which satisfies the continuity equations $\partial _{t}\rho ^{\ell }+\nabla
\cdot \mathbf{J}^{\ell }=0$ or%
\begin{equation}
\partial _{t}\left\vert \psi ^{\ell }\right\vert ^{2}+\nabla \cdot \left( 
\frac{\chi }{m^{\ell }}\func{Im}\frac{\nabla \psi ^{\ell }}{\psi ^{\ell }}%
\left\vert \psi ^{\ell }\right\vert ^{2}-\frac{q^{\ell }}{m^{\ell }\mathrm{c}%
}\mathbf{A}_{\mathrm{ex}}\left\vert \psi ^{\ell }\right\vert ^{2}\right) =0.
\label{jco2}
\end{equation}%
Note that $\mathbf{J}^{\ell }$ defined by (\ref{jco1}) agrees with the
definition of the current (\ref{mplag11}) in the Maxwell equations.
Equations (\ref{jco2}) can be obtained by multiplying (\ref{NLSj0}) by $\psi
^{\ell \ast }$ and taking the imaginary part. Integrating the continuity
equation we see that $\left\Vert \psi ^{\ell }\right\Vert ^{2}=\mathrm{const}
$ and we impose the following normalization condition:%
\begin{equation}
\left\Vert \psi ^{\ell }\right\Vert ^{2}=\int_{\mathbb{R}^{3}}\left\vert
\psi ^{\ell }\right\vert ^{2}{}\mathrm{d}\mathbf{x}=1,\ t\geq 0,\ \ell
=1,...,N.  \label{norm10}
\end{equation}%
The motivation for this particular normalization is that this normalization
allows $\left\vert \psi ^{\ell }\right\vert ^{2}=\left\vert \psi _{a}^{\ell
}\right\vert ^{2}$ to converge to a delta-function and, in addition, for
this normalization the solution to (\ref{delfi}) given by the formula (\ref%
{jco3}) converges to the Coulomb's potential with the value $q^{\ell }$ of
the charge if $\left\vert \psi _{a}^{\ell }\right\vert ^{2}$ converges to a
delta function. The momentum density $\mathbf{P}^{\ell }$ for the Lagrangian 
$\mathcal{\hat{L}}_{0}$ in (\ref{Lbet}) is defined by the formula%
\begin{equation*}
\mathbf{P}^{\ell }=\frac{\mathrm{i}\chi }{2}\left( \psi ^{\ell }\tilde{\nabla%
}^{\ell \ast }\psi ^{\ell \ast }-\psi ^{\ell \ast }\tilde{\nabla}^{\ell
}\psi ^{\ell }\right) .
\end{equation*}%
Note that so defined momentum density $\mathbf{P}^{\ell }$ is related with
the current $\mathbf{J}^{\ell }$ in (\ref{jco1}) by the formula 
\begin{equation}
\mathbf{P}^{\ell }\left( t,\mathbf{x}\right) =\frac{m^{\ell }}{q^{\ell }}%
\mathbf{J}^{\ell }\left( t,\mathbf{x}\right) .  \label{jco5}
\end{equation}%
We introduce the total individual momenta $\mathsf{P}^{\ell }$ for $\ell $%
-th charge by%
\begin{equation}
\mathsf{P}^{\ell }=\int_{\mathbb{R}^{3}}\mathbf{P}^{\ell }\,\mathrm{d}%
\mathbf{x},  \label{peu1}
\end{equation}%
and obtain the following equations for the total individual momenta 
\begin{equation}
\frac{\mathrm{d}\mathsf{P}^{\ell }}{\mathrm{d}t}=q^{\ell }\int_{\mathbb{R}%
^{3}}\left[ \left( \dsum\nolimits_{\ell ^{\prime }\neq \ell }\mathbf{E}%
^{\ell ^{\prime }}+\mathbf{E}_{\mathrm{ex}}\right) \left\vert \psi ^{\ell
}\right\vert ^{2}+\frac{1}{\mathrm{c}}\mathbf{v}^{\ell }\times \mathbf{B}_{%
\mathrm{ex}}\right] \,\mathrm{d}\mathbf{x},  \label{peu2}
\end{equation}%
where 
\begin{equation}
\mathbf{v}^{\ell }\left( t,\mathbf{x}\right) =\frac{1}{m^{\ell }}\mathbf{P}%
^{\ell }\left( t,\mathbf{x}\right) =\frac{1}{q^{\ell }}\mathbf{J}^{\ell
}\left( t,\mathbf{x}\right) .  \label{jco6}
\end{equation}%
The external EM fields $\mathbf{E}_{\mathrm{ex}}$,$\ \mathbf{B}_{\mathrm{ex}%
} $ in (\ref{peu2}) corresponding to the potentials $\varphi _{\mathrm{ex}}$%
, $\mathbf{A}_{\mathrm{ex}}$\ are determined by standard formulas (\ref{fia}%
). Derivation of (\ref{peu2}) is\ rather elementary. Indeed, in the simplest
case where $\mathbf{A}_{\mathrm{ex}}=0$ we multiply (\ref{NLSj0}) by $\nabla
\psi ^{\ell \ast }$, take the real part and integrate the result over the
entire space using integration by parts. To obtain (\ref{peu2}) in more
involved general case one can similarly multiply (\ref{NLSj0}) by $\tilde{%
\nabla}^{\ell \ast }\psi ^{\ell \ast }$ and then integrate the result by
parts using some vector algebra manipulation.

\subsection{Newtonian mechanics as an approximation}

Let us show now that if the size parameter $a$ is small compared to the
macroscopic scale of variation of EM fields, the charge evolution can be
described approximately by the Newton equations with the Lorentz forces
similar to (\ref{pchar1}). In this subsection we give a heuristic
derivation. Conditions under which this kind of derivation is justified are
given in the next section.

We introduce the $\ell $-th charge position $\mathbf{r}^{\ell }\left(
t\right) $ and velocity $\mathsf{v}^{\ell }\left( t\right) $ as the
following spatial averages%
\begin{equation}
\mathbf{r}^{\ell }\left( t\right) =\mathbf{r}_{a}^{\ell }\left( t\right)
=\int_{\mathbb{R}^{3}}\mathbf{x}\left\vert \psi _{a}^{\ell }\left( t,\mathbf{%
x}\right) \right\vert ^{2}\,\mathrm{d}\mathbf{x},\ \mathsf{v}^{\ell }\left(
t\right) =\frac{1}{q^{\ell }}\int_{\mathbb{R}^{3}}\mathbf{J}^{\ell }\left( t,%
\mathbf{x}\right) \,\mathrm{d}\mathbf{x},  \label{peu4}
\end{equation}%
where the current density $\mathbf{J}^{\ell }$ is defined by (\ref{jco1}).
We show below that a combination of the continuity equation (\ref{jco2})
with the momentum evolution equations (\ref{peu2}) imply the following
remarkable property: the positions $\mathbf{r}^{\ell }\left( t\right) $
satisfy with a high accuracy Newton's equations of motion for the system of $%
N$ point charges if the size parameter $a$ is small enough.

Multiplying continuity equation (\ref{jco2}) by $\mathbf{x\ }$and
integrating we find the following identities%
\begin{equation}
\frac{\mathrm{d}\mathbf{r}^{\ell }\left( t\right) }{\mathrm{d}t}=\int_{%
\mathbb{R}^{3}}\mathbf{x}\partial _{t}\left\vert \psi ^{\ell }\right\vert
^{2}\,\mathrm{d}\mathbf{x}=\mathbf{-}\frac{1}{q^{\ell }}\int_{\mathbb{R}^{3}}%
\mathbf{x}\nabla \cdot \mathbf{J}^{\ell }\,\mathrm{d}\mathbf{x}=\frac{1}{%
q^{\ell }}\int_{\mathbb{R}^{3}}\mathbf{J}^{\ell }\mathrm{d}\mathbf{x}=%
\mathsf{v}^{\ell }\left( t\right) ,  \label{peu5}
\end{equation}%
showing that the positions and velocities defined by formulas (\ref{peu4})
are related exactly as in the point charge mechanics. Then integrating (\ref%
{jco5}) we obtain the following kinematic representation for the total
momentum%
\begin{equation}
\mathsf{P}^{\ell }\left( t\right) =\frac{m^{\ell }}{q^{\ell }}\int_{\mathbb{R%
}^{3}}\mathbf{J}^{\ell }\left( t,\mathbf{x}\right) \,\mathrm{d}\mathbf{x}%
=m^{\ell }\mathsf{v}^{\ell }\left( t\right) ,  \label{peu6}
\end{equation}%
which also is exactly the same as for the point charges mechanics. Relations
(\ref{peu5}) and (\ref{peu6}) yield 
\begin{equation}
m^{\ell }\frac{\mathrm{d}^{2}\mathbf{r}^{\ell }\left( t\right) }{\mathrm{d}%
^{2}t}=m^{\ell }\frac{\mathrm{d}}{\mathrm{d}t}\mathsf{v}^{\ell }\left(
t\right) =\frac{\mathrm{d}\mathsf{P}^{\ell }}{\mathrm{d}t},  \label{mp1}
\end{equation}%
and we obtain from (\ref{peu2}) the following system of equations of motion
for $N$ charges:%
\begin{equation}
m^{\ell }\frac{\mathrm{d}^{2}\mathbf{r}^{\ell }\left( t\right) }{\mathrm{d}%
^{2}t}=q^{\ell }\int_{\mathbb{R}^{3}}\left[ \left( \dsum\nolimits_{\ell
^{\prime }\neq \ell }\mathbf{E}^{\ell ^{\prime }}+\mathbf{E}_{\mathrm{ex}%
}\right) \left\vert \psi ^{\ell }\right\vert ^{2}+\frac{1}{\mathrm{c}}%
\mathbf{v}^{\ell }\times \mathbf{B}_{\mathrm{ex}}\right] \,\mathrm{d}\mathbf{%
x},\ \ell =1,...,N,  \label{Couf}
\end{equation}%
where $\mathbf{E}^{\ell ^{\prime }}\left( t,\mathbf{x}\right) =-\nabla
\varphi ^{\ell ^{\prime }}\left( t,\mathbf{x}\right) $, $\mathbf{E}_{\mathrm{%
ex}}$ and $\mathbf{B}_{\mathrm{ex}}$\ are defined by (\ref{fia}).

The derivation of the above system is analogous to the well known in quantum
mechanics \emph{Ehrenfest\ Theorem}, \cite[Sections 7, 23]{Schiff}, \cite%
{Bialynicki}. Now we give a formal derivation of Newton's law of motion. Let
us suppose that for every $\ell $-th charge density $\left\vert \psi ^{\ell
}\right\vert ^{2}$ and the corresponding current density $\mathbf{J}^{\ell }$
are localized in $a$-vicinity of the position $\mathbf{r}^{\ell }\left(
t\right) $, and that $\left\vert \mathbf{r}^{\ell }\left( t\right) -\mathbf{r%
}^{\ell ^{\prime }}\left( t\right) \right\vert \geq \gamma >0$ with $\gamma $
independent on $a$ on time interval $\left[ 0,T\right] $. Then if $%
a\rightarrow 0$ we get 
\begin{equation}
\left\vert \psi ^{\ell }\right\vert ^{2}\left( t,\mathbf{x}\right)
\rightarrow \delta \left( \mathbf{x}-\mathbf{r}^{\ell }\left( t\right)
\right) ,\ \mathbf{v}^{\ell }\left( t,x\right) =\mathbf{J}^{\ell }/q^{\ell
}\rightarrow \mathsf{v}^{\ell }\left( t\right) \delta \left( \mathbf{x}-%
\mathbf{r}^{\ell }\left( t\right) \right) ,  \label{todel}
\end{equation}%
where the coefficients before the Dirac delta-functions are determined by
the charge normalization conditions (\ref{norm10}) and relations (\ref{peu4}%
). Using potential representations (\ref{jco3}) we infer from (\ref{todel})
the convergence of the potentials $\varphi ^{\ell }$ to the corresponding
Coulomb's potentials, namely%
\begin{equation}
\varphi ^{\ell }\left( t,\mathbf{x}\right) \rightarrow \varphi _{0}^{\ell
}\left( t,\mathbf{x}\right) =\frac{q^{\ell }}{\left\vert \mathbf{x}-\mathbf{r%
}^{\ell }\right\vert },\ \nabla _{\mathbf{r}}\varphi ^{\ell }\left( t,%
\mathbf{x}\right) \rightarrow \frac{q^{\ell }\left( \mathbf{x}-\mathbf{r}%
^{\ell }\right) }{\left\vert \mathbf{x}-\mathbf{r}^{\ell }\right\vert ^{3}}%
\text{ as }a\rightarrow 0.  \label{tocou}
\end{equation}%
Hence, when passing to the limit as $a\rightarrow 0$ we can recast the
equations of motion (\ref{Couf}) as the system 
\begin{equation}
m^{\ell }\frac{\mathrm{d}^{2}\mathbf{r}^{\ell }}{\mathrm{d}t^{2}}=\mathsf{f}%
^{\ell }+\epsilon _{0},  \label{peu8}
\end{equation}%
where%
\begin{equation*}
\mathsf{f}^{\ell }=\sum_{\ell ^{\prime }\neq \ell }q^{\ell }\mathbf{E}%
_{0}^{\ell ^{\prime }}+q^{\ell }\mathbf{E}_{\mathrm{ex}}\left( \mathbf{r}%
^{\ell }\right) +\frac{1}{\mathrm{c}}\mathsf{v}^{\ell }\times \mathbf{B}_{%
\mathrm{ex}}\left( \mathbf{r}^{\ell }\right) ,\ \ell =1,...,N,
\end{equation*}%
and $\epsilon _{0}\rightarrow 0$ as $a\rightarrow 0$. Notice that the terms $%
\mathsf{f}^{\ell }$ in equations (\ref{peu8}) coincide with the Lorentz
forces and we see that the limit equations of motion obtained from (\ref%
{peu8}) coincide with \emph{Newton's equations of motion} for point charges
interacting via the Coulomb forces and with the external EM field via
corresponding Lorentz forces, namely 
\begin{equation}
m^{\ell }\frac{\mathrm{d}^{2}\mathbf{r}^{\ell }}{\mathrm{d}t^{2}}%
=-\sum_{\ell ^{\prime }\neq \ell }\frac{q^{\ell }q^{\ell ^{\prime }}\left( 
\mathbf{r}^{\ell ^{\prime }}-\mathbf{r}^{\ell }\right) }{\left\vert \mathbf{r%
}^{\ell ^{\prime }}-\mathbf{r}^{\ell }\right\vert ^{3}}+q^{\ell }\mathbf{E}_{%
\mathrm{ex}}\left( \mathbf{r}^{\ell }\right) +\frac{1}{\mathrm{c}}\mathsf{v}%
^{\ell }\times \mathbf{B}_{\mathrm{ex}}\left( \mathbf{r}^{\ell }\right) ,\
\ell =1,\ldots ,N.  \label{Newt}
\end{equation}%
The above formal derivation of Newton's equations of motion is very simple,
but if implemented with full rigor would require to impose rather high
regularity conditions on the solution and rather strong requirements on the
convergence to the delta function. Note that we use essentially the fact
that the nonlinearity $G_{a}^{\prime }$, that singularly depends on $a$ as $%
a\rightarrow 0$ according to (\ref{totgkap}), does not enter the system (\ref%
{Couf}). In the following subsection we rewrite the system (\ref{Couf}) in
an integral form which allows to pass to the limit under less restrictive
assumptions, and also study an important case of a single charge in an
external field.

\subsection{Single charge in an external EM field\label{S:singleg}}

In this section we describe assumptions under which the above derivation of
Newton's law with the Lorentz forces for localized charges can be rigorously
justified. In the case of many well separated charges every single charge
senses all other charges through the sum of their EM field, and it is this
feature of the EM interaction makes a problem of a single charge in external
EM field of special importantce.

We consider a single charge described by (\ref{NLSj0}) with $N=1$ in an
external EM field where $\varphi _{\mathrm{ex}}\left( t,x\right) $, $\mathbf{%
A}_{\mathrm{ex}}\left( t,x\right) $ are two times continuously
differentiable functions. Now index $\ell $ \ takes only one value, but we
still keep it to be consistent with our previous notation. We assume that
the first spatial derivatives of $\mathbf{E}_{\mathrm{ex}}$ and $\mathbf{B}_{%
\mathrm{ex}}$ are bounded uniformly in $\left( t,x\right) $. We also assume
that equation (\ref{Couf}) holds in a weaker sense which we describe below.

First let us recast (\ref{Couf}) in a different form. Multiplying (\ref{jco2}%
) by $\mathbf{x-r}^{\ell }\left( t\right) $ we obtain%
\begin{gather*}
m^{\ell }\partial _{t}\left( \left\vert \psi ^{\ell }\right\vert ^{2}\left( 
\mathbf{x-r}^{\ell }\left( t\right) \right) \right) +m^{\ell }\left\vert
\psi ^{\ell }\right\vert ^{2}\partial _{t}\mathbf{r}^{\ell }\left( t\right) +
\\
+\nabla \cdot \left( \left( \chi \func{Im}\frac{\nabla \psi ^{\ell }}{\psi
^{\ell }}\left\vert \psi ^{\ell }\right\vert ^{2}-\frac{q^{\ell }}{\mathrm{c}%
}\mathbf{A}_{\mathrm{ex}}\left\vert \psi ^{\ell }\right\vert ^{2}\right)
\left( \mathbf{x-r}^{\ell }\left( t\right) \right) \right) \\
=\chi \func{Im}\frac{\nabla \psi ^{\ell }}{\psi ^{\ell }}\left\vert \psi
^{\ell }\right\vert ^{2}-\frac{q^{\ell }}{\mathrm{c}}\mathbf{A}_{\mathrm{ex}%
}\left\vert \psi ^{\ell }\right\vert ^{2}.
\end{gather*}%
Note that the right-hand side of the above equation coincides with $m^{\ell }%
\mathbf{v}^{\ell }\left( t,\mathbf{x}\right) $\textbf{.} Substituting this
expression for $\mathbf{v}^{\ell }$ into (\ref{Couf}) (for a single
particle) and we get 
\begin{equation}
m^{\ell }\frac{\mathrm{d}^{2}\mathbf{r}^{\ell }\left( t\right) }{\mathrm{d}%
^{2}t}=q^{\ell }\int_{\mathbb{R}^{3}}\mathbf{E}_{\mathrm{ex}}\left\vert \psi
^{\ell }\right\vert ^{2}\mathrm{d}\mathbf{x}+\frac{q^{\ell }}{\mathrm{c}}%
\partial _{t}\mathbf{r}^{\ell }\left( t\right) \times \int_{\mathbb{R}%
^{3}}\left\vert \psi ^{\ell }\right\vert ^{2}\mathbf{B}_{\mathrm{ex}}\,%
\mathrm{d}\mathbf{x+\epsilon }_{3},  \label{rl3}
\end{equation}%
where%
\begin{gather}
\mathbf{\epsilon }_{3}=\frac{q^{\ell }}{\mathrm{c}}\int_{\mathbb{R}%
^{3}}\partial _{t}\left( \left\vert \psi ^{\ell }\right\vert ^{2}\left( 
\mathbf{x-r}^{\ell }\left( t\right) \right) \right) \times \mathbf{B}_{%
\mathrm{ex}}\,\mathrm{d}\mathbf{x}  \label{eps3} \\
\mathbf{+}\frac{q^{\ell }}{m^{\ell }\mathrm{c}}\int_{\mathbb{R}^{3}}\nabla
\cdot \left[ \left( \chi \func{Im}\frac{\nabla \psi ^{\ell }}{\psi ^{\ell }}%
\left\vert \psi ^{\ell }\right\vert ^{2}-\frac{q^{\ell }}{\mathrm{c}}\mathbf{%
A}_{\mathrm{ex}}\left\vert \psi ^{\ell }\right\vert ^{2}\right) \left( 
\mathbf{x-r}^{\ell }\left( t\right) \right) \right] \times \mathbf{B}_{%
\mathrm{ex}}\,\mathrm{d}\mathbf{x.}  \notag
\end{gather}%
Observe then that according to the charge normalization 
\begin{gather}
q^{\ell }\int_{\mathbb{R}^{3}}\mathbf{E}_{\mathrm{ex}}\left\vert \psi ^{\ell
}\right\vert ^{2}\mathrm{d}\mathbf{x}=q^{\ell }\mathbf{E}_{\mathrm{ex}%
}\left( t,\mathbf{r}^{\ell }\left( t\right) \right) +\mathbf{\epsilon }_{1},%
\text{ where}  \label{eps1} \\
\mathbf{\epsilon }_{1}=q^{\ell }\int_{\mathbb{R}^{3}}\left( \mathbf{E}_{%
\mathrm{ex}}\left( t,\mathbf{x}\right) -\mathbf{E}_{\mathrm{ex}}\left( t,%
\mathbf{r}^{\ell }\left( t\right) \right) \right) \left\vert \psi ^{\ell
}\right\vert ^{2}\mathrm{d}\mathbf{x},  \notag
\end{gather}%
and%
\begin{gather}
\int_{\mathbb{R}^{3}}\left\vert \psi ^{\ell }\right\vert ^{2}\mathbf{B}_{%
\mathrm{ex}}\mathrm{d}\mathbf{x=B}_{\mathrm{ex}}\left( t,\mathbf{r}^{\ell
}\left( t\right) \right) +\mathbf{\epsilon }_{2},\text{ where}  \label{eps2}
\\
\mathbf{\epsilon }_{2}=\int_{\mathbb{R}^{3}}\left( \mathbf{B}_{\mathrm{ex}%
}\left( t,\mathbf{x}\right) -\mathbf{B}_{\mathrm{ex}}\left( t,\mathbf{r}%
^{\ell }\left( t\right) \right) \right) \left\vert \psi ^{\ell }\right\vert
^{2}\,\mathrm{d}\mathbf{x.}  \notag
\end{gather}%
Hence (\ref{rl3}) can be written in the form 
\begin{equation}
m^{\ell }\frac{\mathrm{d}^{2}\mathbf{r}^{\ell }\left( t\right) }{\mathrm{d}%
^{2}t}=q^{\ell }\mathbf{E}_{\mathrm{ex}}\left( t,\mathbf{r}^{\ell }\left(
t\right) \right) +\mathbf{\epsilon }_{1}+\frac{q^{\ell }}{\mathrm{c}}%
\partial _{t}\mathbf{r}^{\ell }\left( t\right) \times \left[ \mathbf{B}_{%
\mathrm{ex}}\left( t,\mathbf{r}^{\ell }\left( t\right) \right) +\mathbf{%
\epsilon }_{2}\right] +\mathbf{\epsilon }_{3}.  \label{RL4}
\end{equation}

\subsubsection{Dynamics in an external electric field}

Here we consider in detail a simpler case where external magnetic field is
absent, $\mathbf{A}_{\mathrm{ex}}=0$. In this case 
\begin{equation*}
\mathbf{B}_{\mathrm{ex}}=0,\quad \mathbf{E}_{\mathrm{ex}}=-\nabla \varphi _{%
\mathrm{ex}},\quad \mathbf{\epsilon }_{2}=0,\quad \mathbf{\epsilon }_{3}=0,
\end{equation*}%
and the field equations take the form%
\begin{gather}
\mathrm{i}\chi \partial _{t}\psi ^{\ell }=-\frac{\chi ^{2}\nabla ^{2}\psi
^{\ell }}{2m^{\ell }}+q\varphi _{\mathrm{ex}}\psi ^{\ell }+\frac{\chi ^{2}}{%
2m}G_{a}^{\prime }\left( \left\vert \psi ^{\ell }\right\vert ^{2}\right)
\psi ^{\ell },  \label{NLSel} \\
\nabla ^{2}\varphi ^{\ell }=-4\pi q\left\vert \psi ^{\ell }\right\vert ^{2}.
\label{fiel}
\end{gather}%
As we already mentioned, in this section index $\ell $ takes only one value.
First, let us recall some properties of equations (\ref{NLSel}), (\ref{fiel}%
). Note that equation (\ref{NLSel}) does not involve the potential $\varphi
^{\ell }$\ and can be solved independently from equation (\ref{fiel}). If
the nonlinearity $g(\psi )=G_{a}^{\prime }\left( \left\vert \psi \right\vert
^{2}\right) \psi $ is a continuously differentiable function of $\psi $ with
uniformly bounded derivative, and $\varphi _{\mathrm{ex}}$ is a smooth
bounded function, then the local existence and uniqueness of a solution of (%
\ref{NLSel}) with a prescribed initial data in $H^{1}\left( \mathbb{R}%
^{3}\right) $ which belongs to and is bounded in $H^{1}\left( \mathbb{R}%
^{3}\right) $ is well known, see \cite[Remark 4.4.8]{Cazenave03}. This
solution is defined for all $t>0$ if (\ref{Gless}) holds for $G^{\ell }$
since the energy 
\begin{equation}
\mathsf{E}_{0}\left( \psi ^{\ell }\right) =\int \frac{\chi ^{2}}{2m^{\ell }}%
\left\{ \left\vert \nabla \psi ^{\ell }\right\vert ^{2}+G^{\ell }\left(
\left\vert \psi ^{\ell }\right\vert ^{2}\right) \right\} +q^{\ell }\varphi _{%
\mathrm{ex}}\left\vert \psi ^{\ell }\right\vert ^{2}\,\mathrm{d}x
\label{ten}
\end{equation}%
is bounded from below on the invariant set $\left\{ \psi \in
H^{1},\left\Vert \psi \right\Vert =1\right\} $. The existence and uniqueness
of solutions of the initial value problem for (\ref{NLSel}) with $\varphi _{%
\mathrm{ex}}=0$ and the logarithmic nonlinearity as in (\ref{Gaussg}) were
proven in \cite{CazenaveHaraux80}, see also \cite{Cazenave03}.

Now we will show that, under a localization assumption, the centers $\mathbf{%
r}^{\ell }\left( t\right) $ of wave function $\psi ^{\ell }$ defined by (\ref%
{peu4}) converge to trajectories governed Newton's law of motion for point
charges. To eliminate the time derivatives in equation (\ref{rl3}) we
integrate it twice in time obtaining 
\begin{equation}
\mathbf{r}^{\ell }\left( t\right) =\mathbf{r}_{0}^{\ell }+\mathbf{\dot{r}}%
_{0}^{\ell }t+\int_{0}^{t}\int_{0}^{t_{2}}\frac{q^{\ell }}{m^{\ell }}\mathbf{%
E}_{\mathrm{ex}}\left( t,\mathbf{r}^{\ell }\left( t_{1}\right) \right) \,%
\mathrm{d}t_{1}\mathrm{d}t_{2}+\mathbf{\epsilon }_{4},  \label{inte}
\end{equation}%
where%
\begin{gather}
\mathbf{\epsilon }_{4}=\frac{1}{m^{\ell }}\int_{0}^{t}\int_{0}^{t_{2}}%
\mathbf{\epsilon }_{1}\,\mathrm{d}t_{1}\mathrm{d}t_{2}  \label{eps4} \\
=\frac{1}{m^{\ell }}\int_{0}^{t}\int_{0}^{t_{2}}q^{\ell }\int_{\mathbb{R}%
^{3}}\left( \mathbf{E}_{\mathrm{ex}}\left( t,\mathbf{x}\right) -\mathbf{E}_{%
\mathrm{ex}}\left( t,\mathbf{r}^{\ell }\left( t\right) \right) \right)
\left\vert \psi ^{\ell }\right\vert ^{2}\,\mathrm{d}\mathbf{x}\mathrm{d}t_{1}%
\mathrm{d}t_{2}.  \notag
\end{gather}%
Let us denote $\mathbf{z}\left( t\right) =\mathbf{r}^{\ell }\left( t\right) -%
\mathbf{r}_{0}^{\ell }-\mathbf{\dot{r}}_{0}^{\ell }t$ and rewrite (\ref{inte}%
) in the form 
\begin{gather}
\mathbf{z}=F\left( \mathbf{z}\right) +\mathbf{\epsilon }_{4},  \label{intez}
\\
F\left( \mathbf{z}\right) =\int_{0}^{t}\int_{0}^{t_{2}}\frac{q^{\ell }}{%
m^{\ell }}\mathbf{E}_{\mathrm{ex}}\left( t,\mathbf{r}^{\ell }\left( 0\right)
+\mathbf{\dot{r}}^{\ell }\left( 0\right) t+\mathbf{z}\right) \,\mathrm{d}%
t_{1}\mathrm{d}t_{2}.  \notag
\end{gather}%
We consider now equation (\ref{inte}) on a small time interval $\left[ 0,T%
\right] $ with solutions in the space of continuous functions $C\left( \left[
0,T\right] \right) $. Since the first order spatial derivatives of $\mathbf{E%
}_{\mathrm{ex}}$ are bounded uniformly, the operator $F$ satisfies a global
Lipschitz estimate 
\begin{equation}
\left\Vert F\left( \mathbf{z}_{1}\right) -F\left( \mathbf{z}_{2}\right)
\right\Vert _{C\left( \left[ 0,T\right] \right) }\leq CT^{2}\left\Vert 
\mathbf{z}_{1}-\mathbf{z}_{2}\right\Vert _{C\left( \left[ 0,T\right] \right)
}.  \label{lip1}
\end{equation}%
In addition to that we have 
\begin{equation*}
\left\Vert F\left( \mathbf{z}\right) \right\Vert _{C\left( \left[ 0,T\right]
\right) }\leq C_{1}T^{2}\text{ if }\left\Vert \mathbf{z}\right\Vert
_{C\left( \left[ 0,T\right] \right) }\leq 2\left\vert \mathbf{r}^{\ell
}\left( 0\right) \right\vert .
\end{equation*}%
Let us take $T>0\ $so small that 
\begin{equation}
CT^{2}\leq 1/4,\qquad C_{1}T^{2}\leq 1/4,\qquad \left\vert \mathbf{\dot{r}}%
^{\ell }\left( 0\right) T\right\vert \leq \left\vert \mathbf{r}^{\ell
}\left( 0\right) \right\vert ,  \label{Tsm}
\end{equation}%
and let $\mathbf{z}_{0}$ be a solution of the equation $\mathbf{z}%
_{0}=F\left( \mathbf{z}_{0}\right) $ satisfying $\left\Vert \mathbf{z}%
_{0}\right\Vert _{C\left( \left[ 0,T\right] \right) }\leq 2\left\vert 
\mathbf{r}^{\ell }\left( 0\right) \right\vert $. Such a solution exists and
is unique by the contraction principle. Note then that the corresponding $%
\mathbf{r}\left( t\right) =\mathbf{z}_{0}\left( t\right) +\mathbf{r}%
_{0}^{\ell }+\mathbf{\dot{r}}_{0}^{\ell }t$ is a solution of (\ref{inte})
with $\mathbf{\epsilon }_{4}=0$ and consequently is a solution of the
equation 
\begin{equation}
m^{\ell }\frac{\mathrm{d}^{2}\mathbf{r}\left( t\right) }{\mathrm{d}^{2}t}%
=q^{\ell }\mathbf{E}_{\mathrm{ex}}\left( \mathbf{r}\right) ,\qquad \mathbf{r}%
\left( 0\right) =\mathbf{r}_{0}^{\ell },\frac{\mathrm{d}\mathbf{r}\left(
\left( 0\right) \right) }{\mathrm{d}t}=\mathbf{\dot{r}}_{0}^{\ell },
\label{relec}
\end{equation}%
which coincides with Newton's equation (\ref{Newt}) for a single point
charge in an external electric field. To estimate a difference between the
Newtonian trajectory $\mathbf{r}\left( t\right) $ and the charge position $%
\mathbf{r}^{\ell }\left( t\right) $ defined by (\ref{peu4}) we subtract from
(\ref{intez}) equation for $\mathbf{z}_{0}$ and obtain 
\begin{equation*}
\mathbf{z-z}_{0}=F\left( \mathbf{z}\right) -F\left( \mathbf{z}_{0}\right) +%
\mathbf{\epsilon }_{4}.
\end{equation*}%
Using (\ref{lip1}) and smallness of $T$ we obtain the following inequality 
\begin{equation}
\left\Vert \mathbf{r}^{\ell }\mathbf{-r}\right\Vert _{C\left( \left[ 0,T%
\right] \right) }=\left\Vert \mathbf{z-z}_{0}\right\Vert _{C\left( \left[ 0,T%
\right] \right) }\leq 2\left\Vert \mathbf{\epsilon }_{4}\right\Vert
_{C\left( \left[ 0,T\right] \right) }.  \label{rlr}
\end{equation}%
The above argument proves the following theorem.

\begin{theorem}
\label{Thepsel}Suppose that $\mathbf{A}_{\mathrm{ex}}\left( t,x\right) =0$
and $\varphi _{\mathrm{ex}}\left( t,x\right) $ are two times continuously
differentiable functions, and that the first spatial derivatives of $\mathbf{%
E}_{\mathrm{ex}}$ are bounded uniformly in $\left( t,x\right) $. Suppose
also that: (i) there is a family of solutions $\psi ^{\ell }=\psi _{a}$ of
the NLS (\ref{NLSel}) with $G=G_{a}$ which depend on the size parameter $a>0$%
; (ii) $\mathbf{r}^{\ell }\left( t\right) $ as defined by (\ref{peu4}) is a
continuous function of $t\in \left[ 0,T\right] $, where $T$ is a
sufficiently small fixed number which satisfies (\ref{Tsm}); (iii) (\ref%
{inte}) holds where $\mathbf{\epsilon }_{4}$ satisfies the following relation%
\begin{gather}
\lim_{a\rightarrow 0}\left\Vert \mathbf{\epsilon }_{4}\right\Vert _{C\left( %
\left[ 0,T\right] \right) }=0,\text{ where }\left\Vert \mathbf{\epsilon }%
_{4}\right\Vert _{C\left( \left[ 0,T\right] \right) }=  \label{eps4lim} \\
=\sup_{t\in \left[ 0,T\right] }\left\vert \int_{0}^{t}\int_{0}^{t_{2}}\frac{%
q^{\ell }}{m^{\ell }}\int_{\mathbb{R}^{3}}\left( \mathbf{E}_{\mathrm{ex}%
}\left( t,\mathbf{x}\right) -\mathbf{E}_{\mathrm{ex}}\left( t,\mathbf{r}%
^{\ell }\left( t\right) \right) \right) \left\vert \psi ^{\ell }\right\vert
^{2}\mathrm{d}\mathbf{x}dt_{1}dt_{2}\right\vert .  \notag
\end{gather}%
Then if $\mathbf{r}\left( t\right) $ is a solution of Newton's equation (\ref%
{relec}) we have%
\begin{equation*}
\sup_{t\in \left[ 0,T\right] }\left\vert \mathbf{r}^{\ell }\left( t\right) -%
\mathbf{r}\left( t\right) \right\vert \rightarrow 0\text{ as }a\rightarrow 0.
\end{equation*}
\end{theorem}

Note that the function $\left( \mathbf{E}_{\mathrm{ex}}\left( t,\mathbf{x}%
\right) -\mathbf{E}_{\mathrm{ex}}\left( t,\mathbf{r}^{\ell }\left( t\right)
\right) \right) =0$ at $\mathbf{x=r}^{\ell }$ and it is a bounded
differentiable function, implying that condition (\ref{eps4lim}) is a weaker
version of convergence $\left\vert \psi ^{\ell }\right\vert ^{2}\left( t,%
\mathbf{x}\right) \rightarrow \delta \left( \mathbf{x-r}^{\ell }\left(
t\right) \right) $ in $C\left( \mathbb{R}^{3}\right) ^{\ast }$\ uniformly on 
$\left[ 0,T\right] $. Since $\left\vert \psi ^{\ell }\right\vert ^{2}$ for
every $a$ satisfy normalization condition (\ref{norm10}), the convergence to 
$\delta \left( \mathbf{x-r}^{\ell }\left( t\right) \right) $ means a weak
form of localization around $\mathbf{r}^{\ell }\left( t\right) $. In Theorem %
\ref{Thcorpel} and Remark \ref{R:unia} we give an example where condition (%
\ref{eps4lim}) holds.

\subsubsection{Dynamics in a general external electromagnetic field}

Now let us consider the case when the external magnetic field described by
its vector potential $\mathbf{A}_{\mathrm{ex}}$ does not vanish. For
localized solutions of equation (\ref{NLSj0}) where $N=1$ and $\ell $ takes
one value we want to derive the convergence of $\mathbf{r}^{\ell }\left(
t\right) $ defined by (\ref{peu4}) to a solution of Newton's equation of
motion 
\begin{equation}
m\frac{\mathrm{d}^{2}\mathbf{r}}{\mathrm{d}^{2}t}=q\mathbf{E}_{\mathrm{ex}%
}\left( t,\mathbf{r}\right) +\frac{q}{\mathrm{c}}\frac{\mathrm{d}\mathbf{r}}{%
\mathrm{d}t}\times \mathbf{B}_{\mathrm{ex}}\left( t,\mathbf{r}\right) .
\label{Lor1}
\end{equation}%
Considering equation (\ref{RL4}) as a perturbation of equation (\ref{Lor1})
let us estimate the difference between their solutions. First we derive from
equation (\ref{RL4}) estimates for $\left( \partial _{t}\mathbf{r}^{\ell
}\left( t\right) \right) ^{2}$ similar to usual energy estimates.
Multiplying (\ref{RL4}) by $\partial _{t}\mathbf{r}^{\ell }\left( t\right) $
we obtain%
\begin{equation*}
m^{\ell }\partial _{t}^{2}\mathbf{r}^{\ell }\left( t\right) \partial _{t}%
\mathbf{r}^{\ell }\left( t\right) =q^{\ell }\mathbf{E}_{\mathrm{ex}}\left( t,%
\mathbf{r}^{\ell }\left( t\right) \right) \partial _{t}\mathbf{r}^{\ell
}\left( t\right) +\left( \mathbf{\epsilon }_{1}+\mathbf{\epsilon }%
_{3}\right) \partial _{t}\mathbf{r}^{\ell }\left( t\right) .
\end{equation*}%
Integration of the above equation with the respect to $t$ yields 
\begin{gather}
\frac{m^{\ell }}{2}\left[ \left( \partial _{t}\mathbf{r}^{\ell }\left(
t\right) \right) ^{2}-\left( \partial _{t}\mathbf{r}^{\ell }\left( 0\right)
\right) ^{2}\right] =  \label{RLdotsq} \\
=\int_{0}^{t}q^{\ell }\mathbf{E}_{\mathrm{ex}}\left( t,\mathbf{r}^{\ell
}\left( t_{1}\right) \right) \partial _{t}\mathbf{r}^{\ell }\left(
t_{1}\right) +\left( \mathbf{\epsilon }_{1}+\mathbf{\epsilon }_{3}\right)
\partial _{t}\mathbf{r}^{\ell }\left( t_{1}\right) \,\mathrm{d}t_{1}.  \notag
\end{gather}%
Hence%
\begin{eqnarray*}
\left( \partial _{t}\mathbf{r}^{\ell }\left( t\right) \right) ^{2} &\leq
&\left( \partial _{t}\mathbf{r}^{\ell }\left( 0\right) \right) ^{2}+2M\left[
\int_{0}^{t}\left( \partial _{t}\mathbf{r}^{\ell }\left( t_{1}\right)
\right) ^{2}\,\mathrm{d}t_{1}\right] ^{1/2} \\
&\leq &\left( \partial _{t}\mathbf{r}^{\ell }\left( 0\right) \right)
^{2}+M^{2}+\int_{0}^{t}\left( \partial _{t}\mathbf{r}^{\ell }\left(
t_{1}\right) \right) ^{2}\,\mathrm{d}t_{1},
\end{eqnarray*}%
where%
\begin{equation}
M^{2}=\int_{0}^{t}\left( \frac{q^{\ell }}{m^{\ell }}\mathbf{E}_{\mathrm{ex}%
}\left( t,\mathbf{r}^{\ell }\left( t_{1}\right) \right) +\frac{\left( 
\mathbf{\epsilon }_{1}+\mathbf{\epsilon }_{3}\right) }{m^{\ell }}\right)
^{2}\,\mathrm{d}t_{1}.  \label{RLM}
\end{equation}%
Consequently, 
\begin{equation}
\left( \partial _{t}\mathbf{r}^{\ell }\left( t\right) \right) ^{2}\leq M_{1}%
\text{ for }t\in \left[ 0,T\right] ,  \label{RLM1}
\end{equation}%
where constant $M_{1}$ depends on constants $M$ and $T$. Let $\mathbf{r}%
\left( t\right) $ be a solution of (\ref{Lor1}) with initial data 
\begin{equation}
\mathbf{r}\left( 0\right) =\mathbf{r}^{\ell }\left( 0\right) ,\ \frac{%
\mathrm{d}\mathbf{r}}{\mathrm{d}t}\left( 0\right) =\frac{\mathrm{d}\mathbf{r}%
^{\ell }}{\mathrm{d}t}\left( 0\right) .  \label{RLt0}
\end{equation}%
In a similar way we obtain inequality (\ref{RLM1}) for $\left( \partial _{t}%
\mathbf{r}\left( t\right) \right) ^{2}$ with a constant $M_{1}=M_{10}$ which
does not depend on $\mathbf{\epsilon }_{1}+\mathbf{\epsilon }_{3}$. To
estimate the difference $\mathbf{r}^{\ell }-\mathbf{r}$ we subtract (\ref%
{Lor1}) from (\ref{RL4}), multiply the difference by $\partial _{t}\left( 
\mathbf{r}^{\ell }-\mathbf{r}\right) $ and integrate the result with respect
to $t$ obtaining 
\begin{gather}
\frac{m^{\ell }}{2}\left( \partial _{t}\left( \mathbf{r}^{\ell }-\mathbf{r}%
\right) \right) ^{2}=\int_{0}^{t}q^{\ell }\left( \mathbf{E}_{\mathrm{ex}%
}\left( t_{1},\mathbf{r}^{\ell }\right) -\mathbf{E}_{\mathrm{ex}}\left( t,%
\mathbf{r}\right) \right) \partial _{t}\left( \mathbf{r}^{\ell }-\mathbf{r}%
\right) \,\mathrm{d}t_{1}  \label{RLR} \\
+\frac{q^{\ell }}{\mathrm{c}}\int_{0}^{t}\partial _{t}\mathbf{r}\times
\left( \mathbf{B}_{\mathrm{ex}}\left( t,\mathbf{r}^{\ell }\left( t\right)
\right) -\mathbf{B}_{\mathrm{ex}}\left( t,\mathbf{r}\left( t\right) \right)
\right) \partial _{t}\left( \mathbf{r}^{\ell }-\mathbf{r}\right) \,\mathrm{d}%
t_{1}  \notag \\
+\int_{0}^{t}\left( \frac{q^{\ell }}{\mathrm{c}}\partial _{t}\mathbf{r}%
^{\ell }\left( t\right) \times \mathbf{\epsilon }_{2}+\mathbf{\epsilon }_{1}+%
\mathbf{\epsilon }_{3}\right) \partial _{t}\left( \mathbf{r}^{\ell }-\mathbf{%
r}\right) \,\mathrm{d}t_{1}.  \notag
\end{gather}%
Therefore, using inequality (\ref{RLM1}) for $\left( \partial _{t}\mathbf{r}%
\left( t\right) \right) ^{2}$ and $\left( \partial _{t}\mathbf{r}^{\ell
}\left( t\right) \right) ^{2}$, we obtain 
\begin{gather}
\left( \partial _{t}\left( \mathbf{r}^{\ell }-\mathbf{r}\right) \right)
^{2}\leq \left( 2C\frac{q^{\ell }}{m^{\ell }}+\frac{q^{\ell }}{m^{\ell }%
\mathrm{c}}2CM_{10}\right) \int_{0}^{t}\left\vert \mathbf{r}^{\ell }-\mathbf{%
r}\right\vert \left\vert \partial _{t}\left( \mathbf{r}^{\ell }-\mathbf{r}%
\right) \right\vert \,\mathrm{d}t_{1}  \label{dtrm} \\
+2\int_{0}^{t}\left( \frac{q^{\ell }}{m^{\ell }\mathrm{c}}M_{1}\left\vert 
\mathbf{\epsilon }_{2}\right\vert +\left\vert \mathbf{\epsilon }_{1}+\mathbf{%
\epsilon }_{3}\right\vert \right) \left\vert \partial _{t}\left( \mathbf{r}%
^{\ell }-\mathbf{r}\right) \right\vert \,\mathrm{d}t_{1}.  \notag
\end{gather}%
Using an elementary inequality 
\begin{equation}
\left\vert \mathbf{r}^{\ell }-\mathbf{r}\right\vert =\left\vert
\int_{0}^{t}\partial _{t}\left( \mathbf{r}^{\ell }-\mathbf{r}\right) \,%
\mathrm{d}t_{1}\right\vert \leq t^{1/2}Z^{1/2}\left( t\right) ,  \label{rlrt}
\end{equation}%
where 
\begin{equation}
Z\left( t\right) =\left( \int_{0}^{t}\left( \partial _{t}\left( \mathbf{r}%
^{\ell }-\mathbf{r}\right) \right) ^{2}\,\mathrm{d}t_{1}\right) ,  \label{ZZ}
\end{equation}%
we obtain from (\ref{dtrm}) for $t\in \left[ 0,T\right] $%
\begin{gather}
\left( \partial _{t}\left( \mathbf{r}^{\ell }-\mathbf{r}\right) \right)
^{2}\leq \left( 2C\frac{q^{\ell }}{m^{\ell }}+\frac{q^{\ell }}{m^{\ell }%
\mathrm{c}}2CM_{10}\right) \int_{0}^{t}t_{1}^{1/2}Z^{1/2}\left( t_{1}\right)
\left\vert \partial _{t}\left( \mathbf{r}^{\ell }-\mathbf{r}\right)
\right\vert \,\mathrm{d}t_{1}  \label{dtrm1} \\
+2\left[ \int_{0}^{t}\left( \partial _{t}\left( \mathbf{r}^{\ell }-\mathbf{r}%
\right) \right) ^{2}\,\mathrm{d}t_{1}\right] ^{1/2}\left[ \int_{0}^{t}\left( 
\frac{q^{\ell }}{m^{\ell }\mathrm{c}}M_{1}\left\vert \mathbf{\epsilon }%
_{2}\right\vert +\left\vert \mathbf{\epsilon }_{1}+\mathbf{\epsilon }%
_{3}\right\vert \right) \,\mathrm{d}t_{1}\right] ^{1/2}  \notag \\
\leq 2^{1/2}Ct\frac{q^{\ell }}{m^{\ell }}Z\left( t\right) +2Z^{1/2}\left(
t\right) \mathbf{\epsilon }_{5},  \notag
\end{gather}%
where%
\begin{equation*}
\mathbf{\epsilon }_{5}=\left[ \int_{0}^{T}\left( \frac{q^{\ell }}{m^{\ell }%
\mathrm{c}}M_{1}\left\vert \mathbf{\epsilon }_{2}\right\vert +\left\vert 
\mathbf{\epsilon }_{1}+\mathbf{\epsilon }_{3}\right\vert \right) ^{2}\,%
\mathrm{d}t_{1}\right] ^{1/2}.
\end{equation*}%
Rewriting inequality (\ref{dtrm1}) in the form 
\begin{equation*}
\partial _{t}Z\leq 2^{1/2}Ct\frac{q^{\ell }}{m^{\ell }}Z+2Z^{1/2}\mathbf{%
\epsilon }_{5},
\end{equation*}%
we obtain the following inequality 
\begin{equation*}
Z^{1/2}\leq \frac{\mathbf{\epsilon }_{5}}{C_{2}}\left( e^{C_{2}t}-1\right) 
\text{ for }t\in \left[ 0,T\right] ,
\end{equation*}%
with a constant $C_{2}$ which does not depend on $\mathbf{\epsilon }_{5}$.
Hence $Z\left( t\right) \rightarrow 0$ as $\mathbf{\epsilon }_{5}\rightarrow
0$, and using inequalities (\ref{rlrt}) and (\ref{dtrm1}) we conclude that $%
\left\vert \mathbf{r}^{\ell }-\mathbf{r}\right\vert \rightarrow 0$ and $%
\left( \partial _{t}\left( \mathbf{r}^{\ell }-\mathbf{r}\right) \right)
^{2}\rightarrow 0$ as $\mathbf{\epsilon }_{5}\rightarrow 0$.

The above argument proves the following theorem.

\begin{theorem}
\label{Thepsmag}Suppose that $\mathbf{A}_{\mathrm{ex}}\left( t,\mathbf{x}%
\right) $ and $\varphi _{\mathrm{ex}}\left( t,\mathbf{x}\right) $ are two
times continuously differentiable functions, and that the first spatial and
time derivatives of $\mathbf{E}_{\mathrm{ex}}$ and $\mathbf{B}_{\mathrm{ex}}$
are bounded uniformly in $\left( t,\mathbf{x}\right) $ for $t\in \left[ 0,T%
\right] ,\mathbf{x}\in \mathbb{R}^{3}.$ Suppose also that (i) there is a
family of solutions $\psi ^{\ell }=\psi _{a}$ of the NLS equations (\ref%
{NLSj0}) with $N=1,$ $G=G_{a}$ which depend on the size parameter $a>0$;
(ii) $\mathbf{r}^{\ell }\left( t\right) $ and $\partial _{t}\mathbf{r}^{\ell
}\left( t\right) $\ as defined by (\ref{peu4}) are continuous functions of $%
t\in \left[ 0,T\right] $; (iii) equality (\ref{RLR}) holds, and $M$ defined
by (\ref{RLM}) is bounded uniformly in $a$, $0<a\leq a_{0}$; (iv) the
following limit relation holds 
\begin{equation}
\int_{0}^{T}\left( \frac{q^{\ell }}{m^{\ell }\mathrm{c}}M_{1}\left\vert 
\mathbf{\epsilon }_{2}\right\vert +\left\vert \mathbf{\epsilon }_{1}+\mathbf{%
\epsilon }_{3}\right\vert \right) ^{2}dt_{1}\rightarrow 0\text{ as }%
a\rightarrow 0,  \label{M1lim}
\end{equation}%
where $\mathbf{\epsilon }_{1},\mathbf{\epsilon }_{2},\mathbf{\epsilon }_{3}$
are defined by (\ref{eps1}), (\ref{eps2}), (\ref{eps3}), $M_{1}$ is a
constant which depends on $M$.\newline
Then if $\mathbf{r}\left( t\right) $ satisfies Newton's equation (\ref{Lor1}%
) we have 
\begin{equation*}
\sup_{t\in \left[ 0,T\right] }\left\vert \mathbf{r}^{\ell }\left( t\right) -%
\mathbf{r}\left( t\right) \right\vert \rightarrow 0,\sup_{t\in \left[ 0,T%
\right] }\left\vert \partial _{t}\mathbf{r}^{\ell }\left( t\right) -\partial
_{t}\mathbf{r}\left( t\right) \right\vert \rightarrow 0\text{ as }%
a\rightarrow 0.
\end{equation*}
\end{theorem}

\begin{remark}
Note that the derivation of Newton's law of motion for charge centers in
Theorems \ref{Thepsel} and \ref{Thepsmag} does not depend explicitly on the
particular nonlinearity $G$ and the value of $\chi >0$. Analysis of the
hydrogen atom model in Section \ref{S:nonleig} shows that $\chi =\hbar $ is
a natural choice.
\end{remark}

\begin{remark}
Theorems similar to Theorems \ref{Thepsel} and \ref{Thepsmag} for a system
of interacting charges can be analogously formulated and proven. Among
additional conditions one has to assume the condition of non-collision, that
is the trajectories of Newton's system (\ref{Newt}) have to be separated on
the time interval $\left[ 0,T\right] $: $\left\vert \mathbf{r}^{\ell }\left(
t\right) -\mathbf{r}^{\ell ^{\prime }}\left( t\right) \right\vert \geq d>0$
if $\ell \neq \ell ^{\prime }$ with a constant $d$ which does not depend on $%
a$. Note that one cannot assume that the electrostatic potentials $\varphi
^{\ell ^{\prime }}$ have uniformly bounded gradients uniformly in $a\ $since
they converge to the Coulomb potentials. But looking at (\ref{eps4lim})
where $\mathbf{E}_{\mathrm{ex}}$ is replaced by $-\nabla \varphi _{\mathrm{ex%
}}-\nabla \varphi _{\neq \ell }$ we observe that uniform boundedness of $%
\nabla \varphi ^{\ell ^{\prime }}$ is required only in a vicinity of $%
\mathbf{r}^{\ell }\left( t\right) $ with a fixed radius $d/2$, whereas
outside the vicinity a natural assumption is convergence to zero of the
integral of $\left\vert \psi ^{\ell }\right\vert ^{2}\left\vert \nabla
\varphi _{\neq \ell }\right\vert $.
\end{remark}

\subsubsection{Exact wave-corpuscle solutions: accelerating solitons\label%
{exel}}

Here we consider the field equations (\ref{NLSj0}), (\ref{delfi}) for a
single charge, omitting the index $\ell $, and present a family of their
exact solutions in the form of accelerating solitons (wave-corpuscles). Such
solutions provide an example for which conditions of Theorem \ref{Thepsel}
hold. We assume a purely electric external EM field, i.e. when $\mathbf{A}_{%
\mathrm{ex}}=0$, $\mathbf{E}_{\mathrm{ex}}\left( t,\mathbf{x}\right)
=-\nabla \varphi _{\mathrm{ex}}\left( t,\mathbf{x}\right) $. For the purely
electric external field the field equations (\ref{NLSj0}), (\ref{delfi})
take the form (\ref{NLSel}). We define the\emph{\ wave-corpuscle} $\psi
,\varphi $ by the following formula: 
\begin{gather}
\psi \left( t,\mathbf{x}\right) =\mathrm{e}^{\mathrm{i}S/\chi }\hat{\psi}%
,\qquad S=m\mathbf{v}\cdot \left( \mathbf{x}-\mathbf{r}\right) +s_{\mathrm{p}%
}\left( t\right) ,  \label{psil0} \\
\hat{\psi}=\mathring{\psi}\left( \left\vert \mathbf{x}-\mathbf{r}\right\vert
\right) ,\qquad \varphi =\mathring{\varphi}\left( \left\vert \mathbf{x}-%
\mathbf{r}\right\vert \right) ,\qquad \mathbf{r}=\mathbf{r}\left( t\right) .
\notag
\end{gather}%
In the above formula $\mathring{\psi}$ is the form factor satisfying (\ref%
{stp}), $\mathring{\varphi}$ is a radial function determined from (\ref%
{delfi}).

We refer to the function $\mathbf{r}\left( t\right) $\ as wave-corpuscle
center.\emph{\ }Since\emph{\ }$\hat{\psi}$ is center-symmetric, this
definition agrees with more general definition (\ref{peu4}). Note that in a
simpler case when the external fields $\varphi _{\mathrm{ex}}$ and $\mathbf{A%
}_{\mathrm{ex}}$ vanish, a simpler solution of (\ref{NLSel}) and (\ref{Lor1}%
) is provided by (\ref{psil0}) with $\mathbf{r}\left( t\right) =\mathbf{r}%
_{0}+\mathbf{v}t$ with a constant velocity $\mathbf{v}$. In this case the
wave-corpuscle\emph{\ }solution (\ref{psil0}) of the field equations (\ref%
{NLSel})-(\ref{fiel}) can be obtained from the rest solution $\mathring{\psi}%
,\mathring{\varphi}$ by certain Galilean-gauge transformations. Solutions of
a similar form are known in the theory of nonlinear Schr\"{o}dinger
equations, see \cite{Sulem} and references therein. For the particular case
of the logarithmic nonlinearity solutions of the form (\ref{psil0}) were
found in \cite{Bialynicki} in the form of accelerating gaussons.

\begin{theorem}
\label{Thcorpel}Suppose that $\varphi _{\mathrm{ex}}\left( t,\mathbf{x}%
\right) $ is a continuous function which is linear with respect to $\mathbf{x%
}$\textbf{.} Then $\psi $ defined by (\ref{psil0}) is an exact solution to (%
\ref{NLSel}), provided that $\mathbf{r}\left( t\right) $ is determined from
the equation 
\begin{equation}
m\frac{\mathrm{d}^{2}\mathbf{r}\left( t\right) }{\mathrm{d}t^{2}}=q\mathbf{E}%
_{\mathrm{ex}}\left( t,\mathbf{r}\right) ,  \label{nlaw}
\end{equation}%
and $\mathbf{v}\left( t\right) ,s_{\mathrm{p}}\left( t\right) $ are
determined by formulas%
\begin{equation}
\mathbf{v}=\frac{\mathrm{d}r}{\mathrm{d}t}\mathbf{,\qquad }s_{\mathrm{p}%
}=\dint_{0}^{t}\left( \frac{m\mathbf{v}^{2}}{2}-q\varphi _{\mathrm{ex}%
}\left( t,\mathbf{r}\left( t\right) \right) \right) \,\mathrm{d}t^{\prime }.
\label{eel}
\end{equation}
\end{theorem}

\begin{proof}
If the potential $\varphi _{\mathrm{ex}}\left( t,\mathbf{x}\right) $ is
linear in $\mathbf{x}$ then for any given trajectory $\mathbf{r}\left(
t\right) $ we can write 
\begin{equation}
\varphi _{\mathrm{ex}}\left( t,\mathbf{x}\right) =\varphi _{0,\mathrm{ex}%
}\left( t\right) +\varphi _{0,\mathrm{ex}}^{\prime }\left( t\right) \cdot
\left( \mathbf{x}-\mathbf{r}\left( t\right) \right) ,  \label{elst}
\end{equation}%
where%
\begin{equation*}
\varphi _{0,\mathrm{ex}}^{\prime }\left( t\right) =\nabla _{\mathbf{x}%
}\varphi _{\mathrm{ex}}\left( \mathbf{r}\left( t\right) ,t\right) ,\mathbf{%
\qquad }\varphi _{0,\mathrm{ex}}\left( t\right) =\varphi _{\mathrm{ex}%
}\left( t,\mathbf{r}\left( t\right) \right) .
\end{equation*}%
Observe that the representation (\ref{psil0}) implies 
\begin{equation}
\partial _{t}\psi =\exp \left( \mathrm{i}\frac{S}{\chi }\right) \left\{ %
\left[ \frac{\mathrm{i}m}{\chi }\left( \mathbf{\dot{v}}\cdot \left( \mathbf{x%
}-\mathbf{r}\right) -\mathbf{v}\cdot \mathbf{\dot{r}}\right) +\frac{\mathrm{i%
}\dot{s}_{\mathrm{p}}}{\chi }\right] \hat{\psi}-\mathbf{\dot{r}}\cdot \nabla 
\hat{\psi}\right\} ,  \label{dtpp}
\end{equation}%
and by Leibnitz formula we have 
\begin{equation}
\nabla ^{2}\psi =\exp \left( \mathrm{i}\frac{S}{\chi }\right) \left[ \left( 
\frac{\mathrm{i}m\mathbf{v}}{\chi }\right) ^{2}\hat{\psi}+2\frac{\mathrm{i}m%
}{\chi }\mathbf{v}\cdot \nabla \hat{\psi}+\nabla ^{2}\hat{\psi}\right] .
\label{del2p}
\end{equation}%
Substituting the expression (\ref{psil0}) into the field equations (\ref%
{NLSel}) we obtain the following equation for functions $\mathbf{v}$, $%
\mathbf{r}$, $s_{\mathrm{p}}$: 
\begin{gather}
\left[ -m\mathbf{\dot{v}}\cdot \left( \mathbf{x}-\mathbf{r}\right) -\mathbf{v%
}\cdot \mathbf{\dot{r}}-\dot{s}_{\mathrm{p}}\right] \hat{\psi}-\mathrm{i}%
\chi \mathbf{\dot{r}}\cdot \nabla \hat{\psi}  \label{eq2} \\
-\frac{m}{2}\mathbf{v}^{2}\hat{\psi}+\mathrm{i}\chi \mathbf{v}\cdot \nabla 
\hat{\psi}+\frac{\chi ^{2}}{2m}\nabla ^{2}\hat{\psi}-q\varphi _{\mathrm{ex}}%
\hat{\psi}-\frac{\chi ^{2}}{2m}G^{\prime }\hat{\psi}=0.  \notag
\end{gather}%
Using the charge equilibrium equation (\ref{nop40}) we eliminate the
nonlinearity $G$ and the Laplacian $\nabla ^{2}$ in the above equation (\ref%
{eq2}) and obtain the following equation equivalent to it: 
\begin{equation}
-\left\{ m\left[ \mathbf{\dot{v}}\cdot \left( \mathbf{x}-\mathbf{r}\right) -%
\mathbf{v}\cdot \mathbf{\dot{r}}\right] +\frac{m}{2}\mathbf{v}^{2}+\dot{s}_{%
\mathrm{p}}+q\varphi _{\mathrm{ex}}\right\} \hat{\psi}-\mathrm{i}\chi \left( 
\mathbf{\dot{r}-v}\right) \nabla \hat{\psi}=0.  \label{eq3}
\end{equation}%
Now, we equate to zero the coefficients before $\nabla \hat{\psi}$\ and $%
\hat{\psi}$ in that equation, resulting in two equations: 
\begin{equation}
\mathbf{v}=\mathbf{\dot{r}},\ m\left[ \mathbf{\dot{v}}\cdot \left( \mathbf{x}%
-\mathbf{r}\right) -\mathbf{v}\cdot \mathbf{\dot{r}}\right] +\frac{m}{2}%
\mathbf{v}^{2}+\dot{s}_{\mathrm{p}}+q\varphi _{\mathrm{ex}}=0,  \label{ra}
\end{equation}%
where, in view of the representation (\ref{elst}), the second equation in (%
\ref{ra}) can be recast as 
\begin{equation}
m\left[ \mathbf{\dot{v}}\cdot \left( \mathbf{x}-\mathbf{r}\right) -\mathbf{v}%
\cdot \mathbf{\dot{r}}\right] +\dot{s}_{\mathrm{p}}+\frac{m\mathbf{v}^{2}}{2}%
+q\left[ \varphi _{0,\mathrm{ex}}+\varphi _{0,\mathrm{ex}}^{\prime }\cdot
\left( \mathbf{x}-\mathbf{r}\right) \right] =0.  \label{fir}
\end{equation}%
Then we equate to zero the coefficient before $\left( \mathbf{x}-\mathbf{r}%
\right) $ and the remaining coefficient and obtain the following pair of
equations:%
\begin{equation}
m\mathbf{\dot{v}}=-q\varphi _{0,\mathrm{ex}}^{\prime }\left( t\right)
,\qquad \dot{s}_{\mathrm{p}}-m\mathbf{v}\cdot \mathbf{\dot{r}}+\frac{m%
\mathbf{v}^{2}}{2}+q\varphi _{0,\mathrm{ex}}\left( t\right) =0.  \label{V1a}
\end{equation}%
Based on the first equation (\ref{ra}) and the equations (\ref{V1a}) we
conclude that the wave-corpuscle defined by the formula (\ref{psil0}) and
equations (\ref{nlaw}), (\ref{eel}) is indeed an exact solution to the field
equations (\ref{NLSel}).
\end{proof}

\begin{remark}
\label{R:unia}The above construction does not depend on the nonlinearity $%
G^{\prime }=G_{a}^{\prime }$ as long as (\ref{nop40}) is satisfied. It is
uniform with respect to $a>0$, and the dependence on $a$ in (\ref{psil0}) is
only through $\mathring{\psi}\left( \left\vert \mathbf{x}-\mathbf{r}%
\right\vert \right) =a^{-3/2}\mathring{\psi}_{1}\left( a^{-1}\left\vert 
\mathbf{x}-\mathbf{r}\right\vert \right) $. Obviously, if $\psi \left( t,%
\mathbf{x}\right) $ is defined by (\ref{psil0}) then $\left\vert \psi \left(
t,\mathbf{x}\right) \right\vert ^{2}\rightarrow \delta \left( \mathbf{x}-%
\mathbf{r}\right) $ as $a\rightarrow 0$ and condition (\ref{eps4lim}) is
fulfilled.
\end{remark}

\begin{remark}
The form (\ref{psil0}) of exact solutions constructed in Theorem \ref%
{Thcorpel} is the same as the WKB ansatz in the quasi-classical approach.
The trajectories of the charges centers coincide with trajectories that can
be found by applying well-known quasiclassical asymptotics to solutions of (%
\ref{NLSj0}) if one neglects the nonlinearity. Note though that there are
two important effects of the nonlinearity not presented in the standard
quasiclassical approach. First of all, due to the nonlinearity the charge
preserves its shape in the course of evolution whereas in the linear model
any wavepacket disperses over time. Second of all, the quasiclassical
asymptotic expansions produce infinite asymptotic series which provide for a
formal solution, whereas the properly introduced nonlinearity as in (\ref%
{nop40}), (\ref{gg}) allows one to obtain an exact solution. For a treatment
of mathematical aspects of the approach to nonlinear wave mechanics based on
the WKB asymptotic expansions we refer the reader to \cite{Komech05} and
references therein.
\end{remark}

Similarly one can consider a single charge in an external EM field which can
have nonzero magnetic component. In this case the wave-corpuscle is again
defined by relations (\ref{psil0}) but equations (\ref{nlaw}) are replaced
by (\ref{Lor1}), see \cite{BF4} for details.

\section{Multiharmonic solutions for a system of many charges}

In this section we consider a regime of close interaction which models a
system of bound charges. This regime differs significantly from the regime
of remote interaction considered in Section \ref{nrapr1}. More specifically,
we seek special form solutions to the field equations (\ref{NLSj0}), (\ref%
{delfi}) for which (i) the potentials $\varphi ^{\ell }$ are time
independent; (ii) every wave functions $\psi ^{\ell }$ depends on time
harmonically through gauge factor $\mathrm{e}^{-\mathrm{i}\omega _{\ell }t}$
with possibly different values of frequencies $\omega _{\ell }$ for
different $\ell $. We refer to such solutions as multiharmonic. Note that
multiharmonic solutions naturally arise in the analysis of nonlinear
dispersive equations and systems, see \cite{BF1}.

Let us consider the field equations (\ref{NLSj0}) and (\ref{delfi}) without
external fields, namely%
\begin{equation}
i\chi \partial _{t}\psi ^{\ell }+\frac{\chi ^{2}}{2m^{\ell }}\nabla ^{2}\psi
^{\ell }-q^{\ell }\varphi _{\neq \ell }\psi ^{\ell }=\frac{\chi ^{2}}{%
2m^{\ell }}G_{\ell }^{\prime }\left( \left\vert \psi ^{\ell }\right\vert
^{2}\right) \psi ^{\ell },\quad \ell =1,...,N,  \label{nrs1}
\end{equation}%
with $\varphi _{\neq \ell }$ defined by (\ref{fineq}) 
\begin{equation}
\varphi _{\neq \ell }=\sum_{\ell ^{\prime }\neq \ell }\varphi ^{\ell
},\qquad \frac{1}{4\pi }\nabla ^{2}\varphi ^{\ell }=-q^{\ell }\left\vert
\psi ^{\ell }\right\vert ^{2}.  \label{nrs2}
\end{equation}%
The total energy of the system is given by the formula 
\begin{gather}
\mathcal{E}=\sum_{\ell }\mathcal{E}_{\ell },  \label{Esum} \\
\mathcal{E}_{\ell }=\frac{1}{2}\int q^{\ell }\left\vert \psi ^{\ell
}\right\vert ^{2}\varphi _{\neq \ell }\,\mathrm{d}\mathbf{x}+\int \frac{\chi
^{2}}{2m^{\ell }}\left\{ \left\vert \nabla \psi ^{\ell }\right\vert
^{2}+G^{\ell }\left( \left\vert \psi ^{\ell }\right\vert ^{2}\right)
\right\} \,\mathrm{d}\mathbf{x},  \notag
\end{gather}%
where $\varphi _{\neq \ell }$ is defined in (\ref{nrs2}), $\varphi ^{\ell
^{\prime }}$ are given by (\ref{jco3}). The energy $\mathcal{E}$ is a
conserved quantity for the system (\ref{nrs1}), (\ref{nrs2}), \ it can be
derived in a standard way from the Lagrangian (\ref{Lbet}) taking in account
equations (\ref{nrs1}), (\ref{nrs2}). Collecting terms in the expression for 
$\mathcal{E}$ which explicitly involve $\psi ^{\ell }$ with a given $\ell $
we introduce $\ell -$th charge energy in the system's field by the formula 
\begin{equation}
\mathsf{E}_{0\ell }=\int q^{\ell }\left\vert \psi ^{\ell }\right\vert
^{2}\varphi _{\neq \ell }\,\mathrm{d}\mathbf{x}+\int \frac{\chi ^{2}}{%
2m^{\ell }}\left\{ \left\vert \nabla \psi ^{\ell }\right\vert ^{2}+G^{\ell
}\left( \left\vert \psi ^{\ell }\right\vert ^{2}\right) \right\} \,\mathrm{d}%
\mathbf{x}.  \label{E0L}
\end{equation}%
Here and below integrals with respect to spatial variables are over $\mathbb{%
R}^{3}.$ We call so defined $\mathsf{E}_{0\ell }$ the $\ell $ -th particle
energy in system field. Note that for a single particle $\mathsf{E}_{0\ell }=%
\mathcal{E}_{\ell }=\mathcal{E}$, in the general case $N\geq 2$ \ the total
energy $\mathcal{E}$ does not equal the sum of $\mathsf{E}_{0\ell },$ see
for example (\ref{Epe}) for the case of two interacting charges; in fact the
difference between the sum of $\mathsf{E}_{0\ell }$ and the total energy $%
\mathcal{E}$ coincides with the total energy of EM fields. In accordance
with (\ref{norm10}) we assume that $\psi ^{\ell }\in \Xi $ where 
\begin{equation}
\Xi =\left\{ \psi \in H^{1}\left( \mathbb{R}^{3}\right) :\left\Vert \psi
\right\Vert ^{2}=\int \left\vert \psi \right\vert ^{2}\,\mathrm{d}\mathbf{x}%
=1\right\} .  \label{Ksi0}
\end{equation}

\begin{theorem}
\label{Th:Ebound}Let $G_{\ell }(s)$ be a function of class $C^{1}$ which is
subcritical, namely there exists $\delta >0$ and $C$ such that 
\begin{equation}
\left\vert \int G_{\ell }\left( \left\vert \psi \right\vert ^{2}\right) 
\mathrm{d}\mathbf{x}\right\vert \leq C+C\left\Vert \psi \right\Vert
_{H^{1}}^{2-\delta }\text{ for all }\psi \in \Xi ,\quad \ell =1,...,N.
\label{Gless}
\end{equation}%
Then the functional $\mathcal{E}$ is bounded from below on the set $\Xi ^{N}$
and boundedness of $\mathcal{E}$ implies boundedness of $\left\Vert \psi
^{\ell }\right\Vert _{H^{1}}$, $\ell =1,...,N$. Boundedness of $\mathcal{E}$
on $\Xi ^{N}$ is equivalent to boundedness of all $\mathsf{E}_{0\ell },$ $%
\ell =1,...,N.$
\end{theorem}

\begin{proof}
Note that according to (\ref{jco3})%
\begin{equation*}
\int \left\vert \psi ^{\ell }\right\vert ^{2}\varphi ^{\ell ^{\prime }}%
\mathrm{d}\mathbf{x}=\int \int \left\vert \psi ^{\ell }\left( \mathbf{x}%
\right) \right\vert ^{2}\frac{1}{\left\vert \mathbf{x}-\mathbf{y}\right\vert 
}\left\vert \psi ^{\ell ^{\prime }}\left( \mathbf{y}\right) \right\vert
^{2}\,\mathrm{d}\mathbf{x}\mathrm{d}\mathbf{y}.
\end{equation*}%
Splitting the domain of integration into $\left\vert \mathbf{x}-\mathbf{y}%
\right\vert \geq \theta $ and $\left\vert \mathbf{x}-\mathbf{y}\right\vert
<\theta $ with arbitrary small $\theta $ and using the Sobolev imbedding
theorem we obtain the following inequality%
\begin{gather}
\int \left\vert \psi ^{\ell }\left( \mathbf{x}\right) \right\vert ^{2}\frac{1%
}{\left\vert \mathbf{x}-\mathbf{y}\right\vert }|\psi ^{\ell ^{\prime
}}\left( \mathbf{y}\right) |^{2}\,\mathrm{d}\mathbf{x}\mathrm{d}\mathbf{y}%
\leq  \label{escou} \\
\leq C_{\epsilon }||\psi ^{\ell }||^{2}||\psi ^{\ell ^{\prime
}}||^{2}+\epsilon ||\psi ^{\ell }||_{H^{1}}||\psi ^{\ell ^{\prime
}}||_{H^{1}}\left\Vert \psi ^{\ell }\right\Vert ||\psi ^{\ell ^{\prime }}||,
\notag
\end{gather}%
where $\epsilon $ can be chosen arbitrary small. From this inequality and (%
\ref{Gless}) we obtain the statement of the theorem.
\end{proof}

Let us consider now multiharmonic solutions to the system (\ref{nrs1}), (\ref%
{nrs2}), that is solutions of the form 
\begin{equation}
\psi ^{\ell }\left( t,\mathbf{x}\right) =\mathrm{e}^{-\mathrm{i}\omega
_{\ell }t}\psi _{\ell }\left( \mathbf{x}\right) ,\qquad \varphi ^{\ell
}\left( t,\mathbf{x}\right) =\varphi _{\ell }\left( \mathbf{x}\right) .
\label{psiup}
\end{equation}%
Note that we use lower index for $\psi _{\ell }\left( \mathbf{x}\right) $ to
differ it from $\psi ^{\ell }\left( t,\mathbf{x}\right) $. Then the
functions $\psi _{\ell }\left( \mathbf{x}\right) $ satisfy the following
nonlinear eigenvalue problem 
\begin{equation}
\chi \omega _{\ell }\psi _{\ell }+\frac{\chi ^{2}}{2m^{\ell }}\nabla
^{2}\psi _{\ell }-q^{\ell }\varphi _{\neq \ell }\psi _{\ell }-\frac{\chi ^{2}%
}{2m^{\ell }}G_{_{\ell }}^{\prime }\left( \left\vert \psi _{\ell
}\right\vert ^{2}\right) \psi _{\ell }=0,  \label{nep1}
\end{equation}%
where $\varphi _{\neq \ell }$ is defined by (\ref{fineq}), $\ell =1,...,N$.

\subsection{ Planck-Einstein frequency-energy relation and the logarithmic
nonlinearity}

The nonlinear eigenvalue problem (\ref{nep1}), (\ref{nrs2}) may have many
solutions, and every solution $\left\{ \psi _{\ell }\right\} _{\ell =1}^{N}$
determines a set of frequencies $\left\{ \omega _{\ell }\right\} _{\ell
=1}^{N}$ and energies $\left\{ \mathsf{E}_{0\ell }\right\} _{\ell =1}^{N}$.
Recall that the fundamental Planck-Einstein frequency-energy relation reads%
\begin{equation}
E=\hbar \omega .  \label{PE}
\end{equation}%
We would like to find if possible a nonlinearity having the following
property: for any two solutions $\left\{ \psi _{\ell }\right\} _{\ell =1}^{N}
$, $\left\{ \psi _{\ell }^{\prime }\right\} _{\ell =1}^{N}$ with the
corresponding frequencies $\left\{ \omega _{\ell }\right\} _{\ell =1}^{N}$, $%
\left\{ \omega _{\ell }^{\prime }\right\} _{\ell =1}^{N}$ and energies $%
\left\{ \mathsf{E}_{0\ell }\right\} _{\ell =1}^{N}$, $\left\{ \mathsf{E}%
_{0\ell }^{\prime }\right\} _{\ell =1}^{N}$ the following \emph{%
Planck-Einstein frequency-energy relation} holds:%
\begin{equation}
\chi \left( \omega _{\ell }-\omega _{\ell }^{\prime }\right) =\mathsf{E}%
_{0\ell }-\mathsf{E}_{0\ell }^{\prime },\qquad \ell =1,...,N.  \label{PEl}
\end{equation}%
We want the above relation (\ref{PEl}) to hold for any regular multiharmonic
solution of a system of the form (\ref{nep1}), (\ref{nrs2}). Let us consider
first the frequencies $\omega _{\ell }$ which can be determined from the
equations as follows. Multiplying the $\ell $-th equation (\ref{nep1}) by $%
\psi _{\ell }^{\ast }$, integrating the result with respect to the space
variable and using the charge normalization condition we obtain 
\begin{equation}
\chi \omega _{\ell }=\int \left[ \frac{\chi ^{2}}{2m^{\ell }}\left\vert
\nabla \psi _{\ell }\right\vert ^{2}\,\mathrm{d}\mathbf{y}+q^{\ell }\varphi
_{\neq \ell }\left\vert \psi _{\ell }\right\vert ^{2}\right] \,\mathrm{d}%
\mathbf{y}+\int \frac{\chi ^{2}}{2m^{\ell }}G_{\ell }^{\prime }\left(
\left\vert \psi _{\ell }\right\vert ^{2}\right) \left\vert \psi _{\ell
}\right\vert ^{2}\mathrm{d}\mathbf{y}.  \label{omee}
\end{equation}%
Comparing the above with (\ref{E0L}) we see that 
\begin{equation}
\chi \omega _{\ell }-\mathsf{E}_{0\ell }=\frac{\chi ^{2}}{2m^{\ell }}\int %
\left[ G_{\ell }^{\prime }\left( \left\vert \psi _{\ell }\right\vert
^{2}\right) \left\vert \psi _{\ell }\right\vert ^{2}-G_{\ell }\left(
\left\vert \psi _{\ell }\right\vert ^{2}\right) \right] \,\mathrm{d}\mathbf{y%
}.  \label{omee1}
\end{equation}%
For two solutions$\left\{ \psi _{\ell }\right\} _{\ell =1}^{N},\left\{ \psi
_{\ell }^{\prime }\right\} _{\ell =1}^{N}$ we obtain the following relations
linking their frequencies and energies to the nonlinearities: 
\begin{gather}
\chi \left( \omega _{\ell }-\omega _{\ell }^{\prime }\right) -\left( \mathsf{%
E}_{0\ell }-\mathsf{E}_{0\ell }^{\prime }\right) =  \label{conom} \\
=\frac{\chi ^{2}}{2m^{\ell }}\int G_{\ell }\left( \left\vert \psi _{\ell
}^{\prime }\right\vert ^{2}\right) -G_{\ell }^{\prime }\left( \left\vert
\psi _{\ell }^{\prime }\right\vert ^{2}\right) \left\vert \psi _{\ell
}^{\prime }\right\vert ^{2}\,\mathrm{d}\mathbf{y}-  \notag \\
-\frac{\chi ^{2}}{2m^{\ell }}\int G_{\ell }\left( \left\vert \psi _{\ell
}\right\vert ^{2}\right) -G_{\ell }^{\prime }\left( \left\vert \psi _{\ell
}\right\vert ^{2}\right) \left\vert \psi _{\ell }\right\vert ^{2}\,\mathrm{d}%
\mathbf{y}.  \notag
\end{gather}%
To find a sufficient condition for (\ref{PEl}) to hold, we take into account
the charge normalization condition $\left\Vert \psi _{\ell }\right\Vert =1$
and observe that it is sufficient to impose the following condition 
\begin{equation}
\int \left[ G_{\ell }\left( \left\vert \psi _{\ell }\right\vert ^{2}\right)
-G_{\ell }^{\prime }\left( \left\vert \psi _{\ell }\right\vert ^{2}\right)
\left\vert \psi _{\ell }\right\vert ^{2}\right] \,\mathrm{d}\mathbf{y}=%
\mathrm{const}  \label{gcon}
\end{equation}%
for every $\left\vert \psi _{\ell }\right\vert ^{2}$ with $\left\Vert \psi
_{\ell }\right\Vert =1,$ where the constant may depend on $\ell .$ This
condition is fulfilled if the following differential equation with
independent variable $s=\left\vert \psi \right\vert ^{2}$ holds:%
\begin{equation}
s\frac{d}{ds}G_{\ell }\left( s\right) -G_{\ell }\left( s\right) =K_{G}s
\label{gkg}
\end{equation}%
with $K_{G}$ that depends only on $\ell $. Then (\ref{omee1}) and the
normalization condition imply 
\begin{equation}
\chi \omega _{\ell }-\mathsf{E}_{0\ell }=\frac{\chi ^{2}}{2m^{\ell }}\int %
\left[ G_{\ell }^{\prime }\left( \left\vert \psi _{\ell }\right\vert
^{2}\right) \left\vert \psi _{\ell }\right\vert ^{2}-G_{\ell }\left(
\left\vert \psi _{\ell }\right\vert ^{2}\right) \right] \,\mathrm{d}\mathbf{y%
}=-\frac{\chi ^{2}}{2m^{\ell }}K_{G}  \label{omminec}
\end{equation}%
and hence (\ref{PEl}) is fulfilled.

Solving the differential equation (\ref{gkg}) we obtain the following
explicit formula for the \emph{logarithmic nonlinearity} 
\begin{equation*}
G_{\ell }\left( s\right) =K_{G}s\ln s+Cs.
\end{equation*}%
Comparing with (\ref{Gpa}) and (\ref{g1gauss}) we find that \emph{if }$%
K_{G}<0$\emph{\ this is exactly the nonlinearity (\ref{Gpa}) which
corresponds to the Gaussian factor. }$\ $According to (\ref{g1gauss}) the
above formula takes the form 
\begin{equation*}
G_{\ell }\left( s\right) =G_{\ell ,a}\left( s\right) =-\frac{1}{\left(
a^{\ell }\right) ^{2}}s\ln s+\frac{1}{\left( a^{\ell }\right) ^{2}}s\left(
\ln \frac{1}{\pi ^{3/2}}-2-3\ln a\right) 
\end{equation*}%
where $a^{\ell }$ is the size parameter for $\ell $-th charge. If $K_{G}=0$
then $G_{\ell }\left( \left\vert \psi _{\ell }\right\vert ^{2}\right) $ is
quadratic and equation (\ref{nep1}) turns into the linear Schr\"{o}dinger
equation for which fulfillment of the Planck-Einstein relation is a
well-known fundamental property. It is absolutely remarkable that the
logarithmic nonlinearity which is singled out by the fulfillment of the
Planck-Einstein relation has a second crucial property, namely it allows a
localized soliton solution, namely the Gaussian one in (\ref{Gaussp}). Note
that above argument can be repeated literally if equation (\ref{nep1})
involves external time-independent electric field as in (\ref{NLSj0}), in
this case $\varphi _{\neq \ell }$ in the definition (\ref{E0L}) of $\mathsf{E%
}_{0\ell }$ has to be replaced by $\varphi _{\neq \ell }+\varphi _{\mathrm{ex%
}}.$

Let us study in more detail the situation with the logarithmic nonlinearity
singled out by the Planck-Einstein relation. Formula (\ref{omminec}) for the
Gaussian wave function takes the form 
\begin{equation}
\mathsf{E}_{0\ell }=\chi \omega _{\ell }+\frac{\chi ^{2}}{2\left( a^{\ell
}\right) ^{2}m^{\ell }}.  \label{ommine}
\end{equation}

\begin{theorem}
\label{Th:Eboundlog}Let every $G_{\ell }(s)$, $\ell =1,...,N$, be the
logarithmic function with the derivative defined by (\ref{Gpa}) with $%
a=a^{\ell }$. Then the functional $\mathcal{E}$ defined by (\ref{Esum}) is
bounded from below on the set $\Xi ^{N}$. The boundedness of $\mathcal{E}$
or boundedness of all $\left\{ \mathsf{E}_{0\ell }\right\} _{\ell =1}^{N}$ \
implies the boundedness of $\left\Vert \psi _{\ell }\right\Vert _{H^{1}}$ $\ 
$and $\left\Vert \psi _{\ell }\right\Vert _{X}$,\ where the space $X$ is
defined by (\ref{Xspace}), for every component of $\left\{ \psi _{\ell
}\right\} _{\ell =1}^{N}$.
\end{theorem}

\begin{proof}
The proof of the boundedness from below is similar to the proof of Theorem %
\ref{Th:Ebound}: we use (\ref{escou}), and instead of (\ref{Gless}) we use (%
\ref{Weis}). From the boundedness of $\mathcal{E}$ or $\left\{ \mathsf{E}%
_{0\ell }\right\} _{\ell =1}^{N}$ and (\ref{Weis}) we infer the boundedness
in $H^{1},$ and from the boundedness of $\int G_{\ell }\left( \left\vert
\psi _{\ell }\right\vert ^{2}\right) \,\mathrm{d}\mathbf{x}$ we infer the
boundedness of $\left\Vert \psi ^{\ell }\right\Vert _{X}$.
\end{proof}

\begin{theorem}
\label{Th:EinPlanck}Let $G_{\ell }(s),$ $\ell =1,...,N$ be the logarithmic
functions with derivatives defined by (\ref{Gpa}) with $a=a^{\ell }.$ Assume
that we have a set of solutions $\left\{ \psi _{\ell }^{\sigma }\right\}
_{\ell =1}^{N}\in \Xi ^{N}$, $\sigma \in \Sigma $ of the nonlinear
eigenvalue problem (\ref{nep1}), (\ref{nrs2}) \ with the corresponding
frequencies $\left\{ \omega _{\ell }^{\sigma }\right\} _{\ell =1}^{N}$ and
finite energies $\left\{ \mathsf{E}_{0\ell }^{\sigma }\right\} _{\ell
=1}^{N} $. Then for any two solutions $\psi _{\ell }^{\sigma }=\psi _{\ell }$
and $\psi _{\ell }^{\sigma _{1}}=\psi _{\ell }^{\prime }$ with $\sigma
,\sigma _{1}\in \Sigma $ \ the Planck-Einstein relation (\ref{PEl}) is
fulfilled.
\end{theorem}

\begin{proof}
According to Theorem \ref{Th:Eboundlog} the boundedness of $\mathsf{E}%
_{0\ell }^{\sigma }$ for every given $\sigma $ implies that functions $\psi
_{\ell }^{\sigma }$ \ are bounded in $H^{1}$ and in $X$. \ Let us consider
the left-hand side of (\ref{nep1}) and denote it by $F\left( \psi _{\ell
}\right) $. Similarly to (\ref{escou}) we obtain that 
\begin{equation*}
\left\vert \int \psi _{\ell }f\varphi _{\ell ^{\prime }}\,\mathrm{d}\mathbf{x%
}\right\vert \leq C\left\Vert \psi _{\ell }\right\Vert _{H^{1}}\left\Vert
f\right\Vert _{H^{1}},
\end{equation*}%
where $C$ does not depend on $f$. Hence multiplication by $\varphi _{\ell
^{\prime }}$\ is a bounded operator from $H^{1}$ to $H^{-1}$. Using this
fact to obtain continuous dependence of the term $\varphi _{\neq \ell }\psi
_{\ell }$ on $\psi _{\ell }$ and using Lemma 9.3.3 in \cite{Cazenave03} for
remaining terms in $F\left( \psi _{\ell }\right) $, we conclude that the
functional $F\left( \psi _{\ell }\right) $\ depends continuously in space $%
W^{\ast }$ defined in (\ref{Wspace}) on $\psi _{\ell }\in H^{1}\cap X$ and
it is bounded in $W^{\ast }$. Since 
\begin{equation*}
\int F\left( \psi _{\ell }\right) \psi _{\ell }d\mathbf{x}=\chi \omega
_{\ell }\int \,\left\vert \psi _{\ell }\right\vert ^{2}\,\mathrm{d}\mathbf{x}%
-\mathsf{E}_{0\ell }+\frac{\chi ^{2}}{2\left( a^{\ell }\right) ^{2}m^{\ell }}%
\int \,\left\vert \psi _{\ell }\right\vert ^{2}\,\mathrm{d}\mathbf{x}
\end{equation*}
\ (\ref{omee1}) holds for smooth rapidly decaying functions $\psi _{\ell }$,
we can use in a standard way that smooth functions with compact support are
dense in $H^{1}\cap X$ \ and obtain  (\ref{omee1}) written in the form 
\begin{equation*}
\chi \omega _{\ell }^{\sigma }-\mathsf{E}_{0\ell }^{\sigma }=-\frac{\chi ^{2}%
}{2\left( a^{\ell }\right) ^{2}m^{\ell }}\int \,\left\vert \psi _{\ell
}^{\sigma }\right\vert ^{2}\,\mathrm{d}\mathbf{x}
\end{equation*}%
for functions $\psi _{\ell }^{\sigma }$ from $H^{1}\cap X.$ The right-hand
side does not depend on $\sigma $ thanks to the normalization condition (\ref%
{norm10}), and we obtain (\ref{PEl}).
\end{proof}

\emph{We would like to stress remarkable universality of the Planck-Einstein
relation (\ref{PEl}) proven above for multiharmonic solutions and the fact
that it holds for every individual particle in a system of many interacting
particles.} Namely, the Planck-Einstein relation holds with the same
coefficient $\chi $ for an arbitrary pair of solutions of (\ref{nep1}), (\ref%
{nrs2}) and for an arbitrary component of those solutions, and the validity
of this relation does not depend on the number $N$ which equals the number
of interacting particles. Observe also that the derivation of (\ref{omminec}%
) and (\ref{ommine}) is based only on properties of the nonlinear terms as
in (\ref{conom}) and can be extended to more general systems than (\ref{nrs1}%
), (\ref{nrs2}). In particular, the systems may involve linear terms with
potentials which explicitly depend on \textbf{$x$} as in (\ref{parbet}).

The connection between the logarithmic nonlinearity and the Planck-Einstein
frequency-energy relation was discovered in a different setting by
Bialynicki-Birula and Mycielski in \cite{Bialynicki}. Note though that in 
\cite{Bialynicki} a system of $N$ particles is described as in the quantum
mechanics by a single wave function $\psi $ over $3N$-dimensional
configuration space, whereas in our model every of $N$ particles has its own
wave function $\psi _{\ell }$ depending on $3$ spatial variables. This is
why our approach naturally leads to the study of \emph{multi-harmonic
solutions} for interacting particles for which every individual wave
function $\psi _{\ell }$ can be associated with its individual frequency $%
\omega _{\ell }$. Another signficant difference between our approach and the
nonlinear mechanics in \cite{Bialynicki} is the way the nonlinearity enters
the Lagrangian and the field equations for $N\geq 2$ particles. Namely, in
our approach every individual particle has its own nonlinearity whereas in 
\cite{Bialynicki} there is a single nonlinearity for the entire system of $N$
particles.

\subsection{Gaussian shape as a global minimum of energy}

As one can see from the previous section the logarithmic nonlinearity
defined by (\ref{Gaussg}) is singled out by the requirement that the
Planck-Einstein frequency-energy relation holds exactly. Let us look further
into the properties of the logarithmic nonlinearity.

We already observed that the equilibrium equation with the logarithmic
nonlinearity admits the Gaussian function as an exact solution. This
solution is a critical point of the energy, and we show below that the
Gaussian function provides a unique (modulo translations) minimum of the
energy functional on functions subjected to the charge normalization
condition. This fact was already found in \cite{Bialynicki}.

Let us consider the energy $\mathcal{E}\left( \psi \right) =\mathsf{E}%
_{0}\left( \psi \right) $ restricted to one particle with $\varphi _{\mathrm{%
ex}}=0$ and $G^{\ell }$ defined for $a=1$ by (\ref{g1gauss}), namely 
\begin{equation}
\mathcal{E}\left( \psi \right) =\frac{\chi ^{2}}{2m^{\ell }}\int \left\{
\left\vert \nabla \psi \right\vert ^{2}-\left\vert \psi \right\vert ^{2}\ln
\left\vert \psi \right\vert ^{2}+\left\vert \psi \right\vert ^{2}\left( \ln 
\frac{1}{\pi ^{3/2}}-2\right) \right\} \,\mathrm{d}\mathbf{x}.  \label{ten2}
\end{equation}%
Of course, the case $a>0$ can be considered similarly. We consider this
functional on a set $\Xi $ defined by (\ref{Ksi0}). In the following theorem
we show based on the logarithmic Sobolev inequality that the Gaussian
function $\psi _{g}=C_{g}e^{-\left\vert \mathbf{x}\right\vert ^{2}/2}$
provides the global minimum of the energy, and that all the global minima
belong to the set 
\begin{equation}
\Omega =\left\{ \psi :\psi =\mathrm{e}^{i\theta }\psi _{g}\left( \mathbf{%
\cdot }-\mathbf{r}\right) ,\quad \mathbf{r}\in \mathbb{R}^{3},\quad \theta
\in \mathbb{R}\right\}  \label{Gam}
\end{equation}%
obtained from $\psi _{g}$ by spatial translations and gauge factor
multiplication. Another proof of this statement is given in \cite{Cazenave83}%
.

\begin{theorem}
\label{Thglmin}Let $\psi \in \Xi $. Then, if $a=1$ 
\begin{equation}
\mathcal{E}\left( \psi \right) \geq \mathcal{E}\left( \psi _{g}\right) =%
\frac{\chi ^{2}}{2m^{\ell }}  \label{E0gr}
\end{equation}%
and the equality holds only for $\psi \in \Omega $.
\end{theorem}

\begin{proof}
For simplicity we set here $\frac{\chi ^{2}}{m^{\ell }}=1$. The value of the
energy $\mathcal{E}\left( \psi \right) $ on $\psi _{g}$ can be found
explicitly: 
\begin{equation}
\mathcal{E}\left( \psi _{g}\right) =\frac{1}{2}.  \label{epsig}
\end{equation}%
Note that the Euler equation for constrained critical points of $\mathcal{E}%
\left( \psi \right) $ has the form 
\begin{equation}
\lambda \psi +\nabla ^{2}\psi =\left( -\ln \left( \left\vert \psi
\right\vert ^{2}\right) +\left( \ln \frac{1}{\pi ^{3/2}}-3\right) \right)
\psi ,  \label{poh1}
\end{equation}%
and $\psi _{g}$ satisfies this equation. Now we show that $\psi _{g}$
provides the global minimum of $\mathcal{E}\left( \psi \right) $ defined by (%
\ref{ten2}) for real-valued $\psi $. We will use the following Euclidean
logarithmic Sobolev inequality 
\begin{equation}
\int_{\mathbb{R}^{d}}\psi ^{2}\ln \psi ^{2}\,\mathrm{d}\mathbf{x}\leq \frac{d%
}{2}\ln \left[ \frac{2}{\pi d\mathrm{e}}\int_{\mathbb{R}^{d}}\left\vert
\nabla \psi \right\vert ^{2}\,\mathrm{d}\mathbf{x}\right]   \label{Weis}
\end{equation}%
for functions from $H^{1}\left( \mathbb{R}^{d}\right) $ where equality holds
only for Gaussian functions (see \cite{DelPinoDolbeault03}). Therefore, 
\begin{gather}
\mathcal{E}\left( \psi \right) \geq   \label{egr} \\
\frac{1}{2}\left\{ \int \left[ \left\vert \nabla \psi \right\vert
^{2}+\left\vert \psi \right\vert ^{2}\left( \ln \frac{1}{\pi ^{3/2}}%
-2\right) \right] \,\mathrm{d}\mathbf{x}-\frac{3}{2}\ln \left[ \frac{2}{3\pi 
\mathrm{e}}\int_{\mathbb{R}^{d}}\left\vert \nabla \psi \right\vert ^{2}\,%
\mathrm{d}\mathbf{x}\right] \right\}   \notag \\
=\frac{1}{2}\left\{ \int \left\vert \nabla \psi \right\vert ^{2}\,\mathrm{d}%
\mathbf{x}-\frac{1}{2}-\frac{3}{2}\ln \left[ \frac{2}{3}\int_{\mathbb{R}%
^{d}}\left\vert \nabla \psi \right\vert ^{2}\,\mathrm{d}\mathbf{x}\right]
\right\} .  \notag
\end{gather}%
To find $\int_{\mathbb{R}^{d}}\left\vert \nabla \psi \right\vert ^{2}\,%
\mathrm{d}\mathbf{x}$ we consider the Euler equation (\ref{poh1}) for the
minimizer $\psi .$ From Pohozhaev identity applied to\ (\ref{poh1}) we
obtain: 
\begin{equation}
\frac{1}{2}\int \frac{1}{2}G\left( \left\vert \Psi \right\vert ^{2}\right) \,%
\mathrm{d}\mathbf{x}=\lambda \left\Vert \psi \right\Vert ^{2}-\frac{1}{6}%
\int \left\vert \nabla \Psi \right\vert ^{2}\,\mathrm{d}\mathbf{x},
\label{mulPoh}
\end{equation}%
where%
\begin{equation*}
G\left( \left\vert \psi \right\vert ^{2}\right) =-\left\vert \psi
\right\vert ^{2}\ln \left\vert \psi \right\vert ^{2}+\left\vert \psi
\right\vert ^{2}\left( \ln \frac{1}{\pi ^{3/2}}-2\right) .
\end{equation*}%
Multiplying equation (\ref{poh1}) by $\psi $ and integrating the result we
get 
\begin{equation}
\lambda \left\Vert \psi \right\Vert ^{2}-\mathcal{E}\left( \psi \right) =-%
\frac{1}{2}\int \left\vert \psi \right\vert ^{2}\,\mathrm{d}\mathbf{x}=-%
\frac{1}{2}.  \label{poh3}
\end{equation}%
Adding (\ref{mulPoh}) and (\ref{poh3}) we obtain 
\begin{equation*}
-\frac{\chi ^{2}}{2m^{\ell }}\int \left\vert \nabla \psi \right\vert ^{2}\,%
\mathrm{d}\mathbf{x}=-\frac{1}{2}-\frac{1}{6}\int \left\vert \nabla \Psi
\right\vert ^{2}\,\mathrm{d}\mathbf{x},
\end{equation*}%
and hence%
\begin{equation*}
\int \left\vert \nabla \psi \right\vert ^{2}\,\mathrm{d}\mathbf{x}=\frac{3}{2%
}.
\end{equation*}%
Substitution of the above expression into inequality (\ref{egr}) implies $%
\mathcal{E}\left( \psi \right) \geq \frac{1}{2}$ where the equality holds
only for Gaussian functions. Since $\psi $ is a minimizer, the inequality
holds for any real $\psi \in \Xi $.\ Let us consider now a complex minimizer 
$\psi =u+\mathrm{i}v$ such that $\mathcal{E}\left( \psi \right) \leq \frac{1%
}{2}$. Notice that the following elementary inequality holds 
\begin{equation*}
\left\vert \nabla \left( u^{2}+v^{2}\right) ^{1/2}\right\vert ^{2}\leq
\left\vert \nabla u\right\vert ^{2}+\left\vert \nabla v\right\vert ^{2}
\end{equation*}%
if $u^{2}+v^{2}\neq 0$ and the equality is possible only if $v\nabla
u=u\nabla v$. The above inequality implies in particular that $\left(
u^{2}+v^{2}\right) ^{1/2}\in H^{1}$. Applying once more the Euclidean
logarithmic Sobolev inequality for real functions we get%
\begin{equation*}
\frac{1}{2}\geq \mathcal{E}\left( \psi \right) \geq \mathcal{E}\left( \left(
u^{2}+v^{2}\right) ^{1/2}\right) .
\end{equation*}%
For such a minimizer $\left( u^{2}+v^{2}\right) ^{1/2}$ must coincide with a
Gaussian function and equality $v\nabla u=u\nabla v$ must hold a.e.. Since 
\begin{equation*}
\nabla \ln u=\frac{\nabla u}{u}=\frac{\nabla v}{v}=\nabla \ln v
\end{equation*}%
for non-zero $u$ and $v$, we conclude that $\ln u-\ln v$ is a constant and
the complex minimizer $\psi $ is obtained from a Gaussian function via
multiplication by a gauge factor $\mathrm{e}^{\mathrm{i}\theta }$.
\end{proof}

The following theorem directly follows from results of Cazenave and Lions 
\cite{CazenaveLions82}.

\begin{theorem}
\label{Thstrictmin}Let $\Omega _{\epsilon }$ be $\epsilon $ -neighborhood in 
$H^{1}$ norm of $\Omega $ 
\begin{equation}
\Omega _{\epsilon }=\left\{ \psi \in \Xi :\inf_{v\in \Omega }\left\Vert \psi
-v\right\Vert _{H^{1}}\leq \epsilon \right\} ,  \label{Gamep}
\end{equation}%
and $M_{\delta }$ be a sublevel set 
\begin{equation}
M_{\delta }=\left\{ \psi \in \Xi :\mathcal{E}\left( \psi \right) \leq 
\mathcal{E}\left( \psi _{g}\right) +\delta \right\} .  \label{Omdel}
\end{equation}%
Then for any $\epsilon >0$ there exists $\delta >0$ such that $M_{\delta
}\subset \Omega _{\epsilon }$.
\end{theorem}

\begin{proof}
Assume the contrary, namely that there exists $\epsilon >0$ and a sequence $%
\psi _{j}\in \Xi $ such that $\mathcal{E}\left( \psi _{j}\right) \rightarrow 
$ $\mathcal{E}\left( \psi _{g}\right) $ and $\inf_{v\in \Omega }\left\Vert
\psi _{j}-v\right\Vert _{H^{1}}\geq \epsilon $. According to Theorem II.1
and Remark II.3 in \cite{CazenaveLions82} a subsequence $\psi _{j}\left(
\cdot -\mathbf{r}_{j}\right) $ is relatively compact in $H^{1}$. Hence there
is a subsequence that converges in $H^{1}$ to a global minimizer which
belongs to $\Omega $. This contradicts the assumption $\epsilon >0$.
\end{proof}

Let us consider now one-particle equation (\ref{nrs1}) with logarithmic
nonlinearity in the absence of external fields (a free particle) 
\begin{equation}
\mathrm{i}\chi \partial _{t}\psi =-\frac{\chi ^{2}}{2m}\nabla ^{2}\psi +%
\frac{\chi ^{2}}{2m}\left( -\ln \left( \left\vert \psi \right\vert
^{2}/C_{g}^{2}\right) -3\right) \psi .  \label{NLSfree}
\end{equation}%
The existence and the uniqueness of solutions to the initial value problem
for this equation is proven in \cite{CazenaveHaraux80}. Since the logarithm
has a singularity at $\psi =0$, to describe classes of functions for which
the problem is well-posed we need to introduce special spaces. Following
Cazenave \cite{Cazenave03} we introduce functions 
\begin{eqnarray}
A\left( \left\vert \psi \right\vert \right) &=&-\left\vert \psi \right\vert
^{2}\ln \left\vert \psi \right\vert ^{2}\text{ if }0\leq \left\vert \psi
\right\vert \leq \mathrm{e}^{-3},  \label{Apsi} \\
A\left( \left\vert \psi \right\vert \right) &=&3\left\vert \psi \right\vert
^{2}+4\mathrm{e}^{-3}\left\vert \psi \right\vert -\mathrm{e}^{-3}\text{\ if }%
\left\vert \psi \right\vert \geq \mathrm{e}^{-3},  \notag
\end{eqnarray}%
\begin{equation}
B\left( \left\vert \psi \right\vert \right) =\left\vert \psi \right\vert
^{2}\ln \left\vert \psi \right\vert ^{2}+A\left( \left\vert \psi \right\vert
\right) .  \label{Bpsi}
\end{equation}%
So defined $A$ is a \emph{convex} $C^{1}$ function which is $C^{2}$ for $%
\left\vert \psi \right\vert \neq 0$. The function $B\left( \left\vert \psi
\right\vert \right) $ vanishes for small $\left\vert \psi \right\vert $ and $%
\psi B\left( \left\vert \psi \right\vert \right) /\left\vert \psi
\right\vert ^{2}$ is of class $C^{1}$ and satisfies uniform Lipschitz
condition. Let $A^{\ast }$ be the convex conjugate function of $A,$ it is
also convex of class $C^{1}$ and positive except at zero. We introduce
Banach spaces 
\begin{gather}
X=\left\{ \psi \in L^{1}\left( \mathbb{R}^{3}\right) :A\left( \left\vert
\psi \right\vert \right) \in L^{1}\left( \mathbb{R}^{3}\right) \right\} ,
\label{Xspace} \\
X^{\prime }=\left\{ \psi \in L^{1}\left( \mathbb{R}^{3}\right) :A^{\ast
}\left( \left\vert \psi \right\vert \right) \in L^{1}\left( \mathbb{R}%
^{3}\right) \right\} .  \notag
\end{gather}%
Let 
\begin{equation}
W=H^{1}\left( \mathbb{R}^{3}\right) \cap X,\ W^{\ast }=H^{1}\left( \mathbb{R}%
^{3}\right) +X^{\prime }.  \label{Wspace}
\end{equation}%
Note that $M_{\delta }$ is bounded in $W$. By Lemma 9.3.2 from \cite%
{Cazenave03} the function $\psi \rightarrow \psi \ln \left\vert \psi
\right\vert ^{2}$ is continuous from $X$ to $X^{\prime }$ and is bounded on
bounded sets.

According to Cazenave \cite{Cazenave83}, \cite{Cazenave03}, if $\psi _{0}\in
W$ there exists unique solution $\psi \in C\left( \mathbb{R},W\right) \cap
C^{1}\left( \mathbb{R},W^{\ast }\right) $ of (\ref{NLSfree}) with initial
data $\psi \left( 0\right) =\psi _{0}$. Properties of solutions of (\ref%
{NLSfree}) are described in \cite{Cazenave83}, \cite{Cazenave03}, in
particular $\Xi \cap W$ and $M_{\delta }$ are invariant sets. According to 
\cite{CazenaveLions82} the solution $\psi _{g}$ is \emph{orbitally stable},
namely invariance of $M_{\delta }$ and Theorem \ref{Thstrictmin} imply that
if $\psi \left( 0,\cdot \right) \in M_{\delta }$ then $\psi \left(
0,t\right) \in \Omega _{\epsilon }$ for $t\geq 0$. Consequently, small
initial perturbations of the Gaussian shape do not cause its large
perturbations in the course of time evolution given by (\ref{NLSfree}) but
may cause considerable spatial shifts of the entire wave function. Such a
time evolution of charges in electric field is described in Theorem \ref%
{Thepsel}. Note that if the potential $\varphi _{\mathrm{ex}}\left(
t,x\right) $ is linear as in Theorem \ref{Thcorpel}, one can rewrite
equation (\ref{NLSel}) in a moving frame with origin $\mathbf{r}\left(
t\right) $ and corresponding phase correction for $\psi $ as in (\ref{psil0}%
) (see \cite{BF4} for details) and obtain an equivalent equation of the form
(\ref{NLSel}) with $\varphi _{\mathrm{ex}}\left( t,\mathbf{x}\right) =0$,
namely (\ref{NLSfree}). Wave-corpuscle solution (\ref{psil0}) after the
change of variables turns into $\psi _{g}$. Therefore, using Theorem \ref%
{Thstrictmin} we conclude that the wave-corpuscle solutions of (\ref{NLSel})
constructed in Theorem \ref{Thcorpel} are orbitally stable.

\section{Two particle hydrogen-like system}

In this section we consider a particular case of multiharmonic solutions of
the system (\ref{nep1}) with $N=2$ which models a bound proton-electron
system. For this system we discuss general properties and provide a
heuristic analysis of the coupling between proton and electron. Using this
analysis as a motivation, we then write equations and energy functional for
one charge (electron) in the Coulomb field of proton. We study this problem
in more detail along the lines of \cite{BerestyckiLions83I}, \cite%
{BerestyckiLions83II}.

\subsection{Electron-proton system}

To model states of a bound proton-electron system, the hydrogen atom, we
consider the multiharmonic solutions of (\ref{NLSj0}), (\ref{delfi})
described by the system (\ref{nep1}), (\ref{nrs2}) with $N=2$, where indices 
$\ell $\ take two values $\ell =1$ and $\ell =2$. The charges have opposite
values $q_{1}=-q_{2}=-q$. For brevity we introduce notation 
\begin{equation}
\Phi _{1}=\varphi _{1}/q_{1},\qquad \Phi _{2}=\varphi _{2}/q_{2},
\label{fifi}
\end{equation}%
\begin{equation}
a_{1}=\frac{\chi ^{2}}{q^{2}m_{1}},\qquad a_{2}=\frac{\chi ^{2}}{q^{2}m_{2}}.
\label{a1a2}
\end{equation}%
The quantity $a_{1}$ turns into \emph{the Bohr radius} if $\chi $ coincides
with Planck constant $\hbar $, and $m_{1},q$ are the electron mass and
charge respectively. Using the above notation we rewrite the two-particle
system (\ref{nep1}), (\ref{nrs2}) in the form of the following \emph{%
nonlinear eigenvalue problem} 
\begin{equation}
\frac{\chi }{q^{2}}\omega _{1}\psi _{1}+\frac{a_{1}}{2}\nabla ^{2}\psi
_{1}+\Phi _{2}\psi _{1}=\frac{a_{1}}{2}G_{1}^{\prime }\left( \left\vert \psi
_{1}\right\vert ^{2}\right) \psi _{1},  \label{H1}
\end{equation}%
\begin{equation}
\frac{\chi }{q^{2}}\omega _{2}\psi _{2}+\frac{a_{2}}{2}\nabla ^{2}\psi
_{2}+\Phi _{1}\psi _{2}=\frac{a_{2}}{2}G_{2}^{\prime }\left( \left\vert \psi
_{2}\right\vert ^{2}\right) \psi _{2},  \label{H2}
\end{equation}%
\begin{equation}
\nabla ^{2}\Phi _{1}=-4\pi \left\vert \psi _{1}\right\vert ^{2},\qquad
\nabla ^{2}\Phi _{2}=-4\pi \left\vert \psi _{2}\right\vert ^{2}.
\label{v1v2}
\end{equation}%
The functions $\psi _{1}$ and $\psi _{2}$ are respectively the wave
functions for the electron and the proton. As always, we look for solutions $%
\psi _{\ell }\in \Xi $ with $\Xi $\ defined by (\ref{Ksi0}). The
nonlinearities $G_{1}^{\prime },$ $G_{2}^{\prime }$ are assumed to be
logarithmic as defined by (\ref{Gpa}) where the size parameter $a=a^{\ell }$
is different for electron and proton. One can similarly consider other
nonlinearities but in this paper we stay with the logarithmic ones.

In accordance with (\ref{Esum})\ we introduce the energy functional by the
following formula 
\begin{gather}
\mathcal{E}\left( \psi _{1},\psi _{2}\right)   \label{Epe} \\
=q^{2}\int \left[ \frac{a_{1}}{2}\left\vert \nabla \psi _{1}\right\vert ^{2}+%
\frac{a_{1}}{2}G_{1}\left( \left\vert \psi _{1}\right\vert ^{2}\right) -4\pi
\left\vert \psi _{1}\right\vert ^{2}\left( -\nabla ^{2}\right)
^{-1}\left\vert \psi _{2}\right\vert ^{2}\right] \,\mathrm{d}\mathbf{x} 
\notag \\
+q^{2}\int \left[ \frac{a_{2}}{2}\left\vert \nabla \psi _{2}\right\vert ^{2}+%
\frac{a_{2}}{2}G_{2}\left( \left\vert \psi _{2}\right\vert ^{2}\right) %
\right] \,\mathrm{d}\mathbf{x,}  \notag
\end{gather}%
where $\left( -\nabla ^{2}\right) ^{-1}\left\vert \psi _{\ell }\right\vert
^{2}$ is defined by (\ref{jco3}). Equations (\ref{H1}), (\ref{H2}), (\ref%
{v1v2}) can be derived from the Euler equations for a critical point of $%
\mathcal{E}\left( \psi _{1},\psi _{2}\right) $\ with the constraint $\left(
\psi _{1},\psi _{2}\right) \in \Xi ^{2}$, if we set in (\ref{v1v2}) $\Phi
_{2}=4\pi \left[ \left( -\nabla ^{2}\right) ^{-1}\left\vert \psi
_{2}\right\vert ^{2}\right] ,$ $\Phi _{1}=4\pi \left( \left( -\nabla
^{2}\right) ^{-1}\left\vert \psi _{2}\right\vert ^{2}\right) $, the
frequencies $\omega _{1},\omega _{2}$ being proportional to the Lagrange
multipliers. Using (\ref{jco3}) one can see that 
\begin{equation*}
\int \left\vert \psi _{1}\right\vert ^{2}\left( -\nabla ^{2}\right)
^{-1}\left\vert \psi _{2}\right\vert ^{2}\mathrm{d}\mathbf{x=}\int
\left\vert \psi _{2}\right\vert ^{2}\left( -\nabla ^{2}\right)
^{-1}\left\vert \psi _{1}\right\vert ^{2}\mathrm{d}\mathbf{x}
\end{equation*}%
hence the coupling term in (\ref{Epe}) is symmetric with respect to $\psi
_{1},\psi _{2}.$ Note also that the first term in formula (\ref{Epe})\
coincides with the energy $\mathsf{E}_{01}$ of the first charge in the
system field as given by (\ref{E0L}) with $\ell =1,$ and the second term
with the expression for the energy $\mathcal{E}_{2}\left( \psi _{2}\right) $
of a free second particle given by (\ref{ten2}). Obviously, $\mathcal{E}%
\left( \psi _{1},\psi _{2}\right) $ also can be written as a sum of energy $%
\mathsf{E}_{02}$ of the second charge in the system field plus the free
energy of the first charge.

According to Theorem \ref{Th:Eboundlog} the energy $\mathcal{E}\left( \psi
_{1},\psi _{2}\right) $ is bounded from below.

\begin{remark}
Note that system (\ref{H1})-(\ref{v1v2}) is similar to the Hartree equations
studied in \cite{Lions87}, though it differs from it because of the presence
of the nonlinearities $G_{\ell }$.
\end{remark}

\subsection{Reduction to one charge in the Coulomb field\label{S:redone}}

In this subsection we give a motivation for replacing the problem of
determination of critical values of the energy functional $\mathcal{E}$
given by (\ref{Epe}) for a system of two charges by a simpler problem for a
single charge similarly to the Born-Oppenheimer approximation in quantum
mechanics. To this end we use two changes of variables in two equations, $%
\mathbf{x}=a_{1}\mathbf{y}_{1}$ in (\ref{H1}) and $\mathbf{x}=a_{2}\mathbf{y}%
_{2}$ in (\ref{H2}), where $a_{\ell }$ are defined by (\ref{a1a2}), namely 
\begin{equation}
\mathbf{x}=a_{\ell }\mathbf{y}_{\ell },\qquad \ell =1,2,  \label{not121}
\end{equation}%
and rescale the fields as follows:%
\begin{equation}
\Phi _{\ell }\left( \mathbf{x}\right) =\frac{\phi _{\ell }\left( \mathbf{y}%
_{\ell }\right) }{a_{\ell }},\qquad \psi _{\ell }\left( \mathbf{x}\right) =%
\frac{1}{a_{\ell }^{3/2}}\Psi _{\ell }\left( \mathbf{y}_{\ell }\right) ,\ell
=1,2.  \label{fipsi}
\end{equation}%
Hence (\ref{H1}), (\ref{H2}) takes the form of the following \emph{nonlinear
hydrogen system}%
\begin{equation}
\frac{\chi }{q^{2}}\omega _{1}\Psi _{1}+\frac{1}{2a_{1}}\nabla _{\mathbf{y}%
_{1}}^{2}\Psi _{1}+\frac{1}{a_{2}}\phi _{2}\left( \frac{a_{1}}{a_{2}}\mathbf{%
y}_{1}\right) \Psi _{1}=\frac{1}{2a_{1}}G_{1}^{\prime }\left( \left\vert
\Psi _{1}\right\vert ^{2}\right) \Psi _{1},  \label{nh1}
\end{equation}%
\begin{equation}
\frac{\chi }{q^{2}}\omega _{2}\Psi _{2}+\frac{1}{2a_{2}}\nabla _{\mathbf{y}%
_{2}}^{2}\Psi _{2}+\frac{1}{a_{1}}\phi _{1}\left( \frac{a_{2}}{a_{1}}\mathbf{%
y}_{2}\right) \Psi _{2}=\frac{1}{2a_{2}}G_{2}^{\prime }\left( \left\vert
\Psi _{2}\right\vert ^{2}\right) \Psi _{2},  \label{nh2}
\end{equation}%
\begin{equation}
\nabla _{\mathbf{y}_{1}}^{2}\phi _{1}=-4\pi \left\vert \Psi _{1}\right\vert
^{2},\qquad \nabla _{\mathbf{y}_{2}}^{2}\phi _{2}=-4\pi \left\vert \Psi
_{2}\right\vert ^{2}.  \label{vv2}
\end{equation}%
Here we use the same letter $G$ to determine the function of rescaled
variables. Note that the electron/proton mass ratio $\frac{m_{1}}{m_{2}}%
\simeq \frac{1}{1837}$ is small, therefore the parameter 
\begin{equation}
b=\frac{a_{2}}{a_{1}}=\frac{m_{1}}{m_{2}}\ll 1  \label{baa}
\end{equation}%
is small too. We rewrite (\ref{nh1}), (\ref{nh2}) as follows: 
\begin{equation}
\frac{\chi a_{1}}{q^{2}}\omega _{1}\Psi _{1}+\frac{1}{2}\nabla ^{2}\Psi _{1}+%
\frac{1}{b}\phi _{2}\left( \frac{\mathbf{y}}{b}\right) \Psi _{1}=\frac{1}{2}%
G_{1}^{\prime }\left( \left\vert \Psi _{1}\right\vert ^{2}\right) \Psi _{1},
\label{eqh1}
\end{equation}%
\begin{equation}
\frac{\chi a_{2}}{q^{2}}\omega _{2}\Psi _{2}+\frac{1}{2}\nabla ^{2}\Psi
_{2}+b\phi _{1}\left( b\mathbf{y}\right) \Psi _{2}=\frac{1}{2}G_{2}^{\prime
}\left( \left\vert \Psi _{2}\right\vert ^{2}\right) \Psi _{2}.  \label{eqh2}
\end{equation}

We would like to show that if we are interested in electron frequencies $%
\omega _{1}$ which correspond to lower energy levels of $\mathcal{E}\left(
\psi _{1},\psi _{2}\right) $ then, in the case $b\rightarrow 0$, it is
reasonable to replace system (\ref{eqh1}), (\ref{eqh2}) by one equation with
the Coulomb potential. The discrete spectrum of the classical linear
non-relativistic Schr\"{o}dinger operator for hydrogen atom with the Coulomb
potential can be recovered (if multiplicities are not taken into account)
from radial eigenfunctions. Therefore here we restrict ourselves to the
radial solutions of (\ref{eqh1}), (\ref{eqh2}) and restrict (\ref{Epe}) to
radial functions. Consider a radial solution $\left( \Psi _{1},\Psi
_{2}\right) $ of (\ref{eqh1}), (\ref{eqh2}). Equation for the electron
frequency $\omega _{1}$ given by (\ref{omee}) with $\ell =1$ takes the form 
\begin{equation}
\frac{\chi a_{1}}{q^{2}}\omega _{1}=\int \frac{1}{2}G_{1}^{\prime }\left(
\left\vert \Psi _{1}\right\vert ^{2}\right) \left\vert \Psi _{1}\right\vert
^{2}-\frac{1}{2}\left\vert \nabla \Psi _{1}\right\vert ^{2}-\left( \frac{1}{b%
}\phi _{2}\left( \frac{\mathbf{y}}{b}\right) -\frac{1}{\left\vert \mathbf{y}%
\right\vert }\right) \left\vert \Psi _{1}\right\vert ^{2}-\frac{1}{%
\left\vert \mathbf{y}\right\vert }\left\vert \Psi _{1}\right\vert ^{2}\,%
\mathrm{d}\mathbf{y}  \label{elen}
\end{equation}%
with the proton potential $\phi _{2}$. Then we would like to replace the
potential $\phi _{2}$ in (\ref{elen}) by the Coulomb potential $\frac{1}{%
\left\vert \mathbf{y}\right\vert }$ as an approximation. To show that for
small $b$ one may expect the resulting perturbation of eigenvalues to be
small, we use the following representation for the radial potential $\phi
\left( r\right) =4\pi \left( -\nabla ^{2}\right) ^{-1}\left\vert \Psi
\right\vert ^{2}$: 
\begin{equation}
\phi \left( r\right) =\frac{1}{r}\left[ 1-4\pi \int_{r}^{\infty }\left(
r_{1}-r\right) r_{1}\left\vert \Psi \left( r_{1}\right) \right\vert ^{2}\,%
\mathrm{d}r_{1}\right] .  \label{ficr1}
\end{equation}%
The term which we expect to be small is 
\begin{gather}
D_{\mathrm{prot}}=-\int \left( \frac{1}{b}\phi _{2}\left( \frac{1}{b}\mathbf{%
y}_{1}\right) -\frac{1}{\left\vert \mathbf{y}_{1}\right\vert }\right)
\left\vert \Psi _{1}\right\vert ^{2}\,\mathrm{d}\mathbf{y}_{1}=
\label{dprot0} \\
=4\pi \int_{0}^{\infty }\frac{4\pi }{r}\left\vert \Psi _{1}\left( r\right)
\right\vert ^{2}r^{2}\int_{r/b}^{\infty }\left( r_{1}-r/b\right)
r_{1}\left\vert \Psi _{2}\left( r_{1}\right) \right\vert ^{2}\,\mathrm{d}%
r_{1}\mathrm{d}r.  \notag
\end{gather}%
Changing the order of integration and using substitution $r/b=r_{2}$ we
obtain 
\begin{gather*}
D_{\mathrm{prot}}=\left( 4\pi \right) ^{2}b^{2}\int_{0}^{\infty
}\int_{r_{2}}^{\infty }\left\vert \Psi _{1}\left( br_{2}\right) \right\vert
^{2}r_{2}\left( r_{1}-r_{2}\right) r_{1}\left\vert \Psi _{2}\left(
r_{1}\right) \right\vert ^{2}\,\mathrm{d}r_{1}dr_{2} \\
=\left( 4\pi \right) ^{2}b^{2}\int_{0}^{\infty }\left\vert \Psi _{2}\left(
r_{1}\right) \right\vert ^{2}r_{1}\int_{0}^{r_{1}}\left\vert \Psi _{1}\left(
br_{2}\right) \right\vert ^{2}r_{2}\left( r_{1}-r_{2}\right) \,dr_{2}\,%
\mathrm{d}r_{1}.
\end{gather*}%
Hence 
\begin{equation}
\left\vert D_{\mathrm{prot}}\right\vert \leq \frac{\left( 4\pi \right) ^{2}}{%
6}b^{2}\max_{r}\left\vert \Psi _{1}\left( r\right) \right\vert
^{2}\int_{0}^{\infty }\left\vert \Psi _{2}\left( r_{1}\right) \right\vert
^{2}r_{1}^{4}\mathrm{d}r_{1}.  \label{dprot}
\end{equation}%
The right-hand side of (\ref{dprot}) involves the variance of $\Psi _{2}$
and $\left\Vert \Psi _{1}\right\Vert _{L^{\infty }}.$ To show that $%
\left\vert D_{\mathrm{prot}}\right\vert $\ is small it is sufficient to show
that the variance of $\Psi _{2}$ and $\left\Vert \Psi _{1}\right\Vert
_{L^{\infty }}$ are bounded uniformly for small $b$. Note that $\Psi
_{1},\Psi _{2}$ in (\ref{dprot}) are not arbitrary radial functions, but
special\ solutions of (\ref{eqh1}), (\ref{eqh2}), (\ref{vv2}) which
correspond to lower critical energy levels of the energy functional $%
\mathcal{E}=\mathsf{E}_{01}+\mathsf{E}_{02}$. Dependence of the\ coupling
between (\ref{eqh1}) and (\ref{eqh2}) on $b$ comes through the positive
potentials $\frac{1}{b}\phi _{2}\left( \frac{1}{b}\mathbf{y}\right) \ $and $%
b\phi _{1}\left( b\mathbf{y}\right) $ in (\ref{eqh1}) and (\ref{eqh2})
respectively, according to (\ref{ficr1}) these potentials are bounded by $%
\frac{1}{\left\vert \mathbf{y}\right\vert }$ uniformly for $b>0$. Using the
uniform boundedness of the potentials and estimate (\ref{C01}) we conclude
that energy and $H^{1}$-norms of $\Psi _{1}\ $and $\Psi _{2}$ are bounded
uniformly in $b.$ As in the proof of Lemma \ref{ThconvLp} below, we conclude
that radial $\Psi _{1}$ and $\Psi _{2}$ are bounded in $L^{5},$ they also
satisfy (\ref{str}). From boundedness in $L^{5}$ we derive that $\frac{1}{%
\left\vert \mathbf{y}\right\vert }\Psi _{1}$ is bounded in $L^{7/4}$ and so
does $G_{1}^{\prime }\left( \left\vert \Psi _{1}\right\vert ^{2}\right) \Psi
_{1}$. Note also that since the energy is bounded from below, the negative
values of $\omega =\frac{\chi a_{1}}{q^{2}}\omega _{1}$ in (\ref{eqh1})
which correspond to lower energy levels are bounded. Hence, the term 
\begin{equation*}
\frac{1}{b}\phi _{2}\left( \frac{1}{b}\mathbf{y}\right) \Psi _{1}-\frac{1}{2}%
G_{1}^{\prime }\left( \left\vert \Psi _{1}\right\vert ^{2}\right) \Psi
_{1}+\omega \Psi _{1}
\end{equation*}%
in (\ref{eqh1}) is bounded in $L^{7/4},$ therefore the solution $\Psi _{1}$\
of the elliptic equation (\ref{eqh1}) belongs to Sobolev space $W^{7/4,2}$
and by Sobolev imbedding\ is bounded in $L^{\infty }.$ As for $\Psi _{2}$,
from (\ref{str}) using the comparison argument as in the proof of Lemma \ref%
{L:xi0} we derive exponential decay of solutions of the equation (\ref{eqh2}%
) for large $r$ uniformly in $b$ and conclude that the variance of $\Psi
_{2} $ is uniformly bounded for small $b.$ Hence, for such solutions $%
\left\vert D_{\mathrm{prot}}\right\vert $ is small for small $b$. The
smallness of $b$ in (\ref{dprot}) (recall that $b\simeq \frac{1}{1837}$ for
proton-electron system) gives us a motivation to set $b=0$ in (\ref{eqh2})
and replace $\frac{1}{b}\phi _{2}\left( \frac{1}{b}\mathbf{y}\right) $ by $%
\frac{1}{\left\vert \mathbf{y}\right\vert }$ in (\ref{eqh1}).

So, we replace the problem of finding frequencies $\omega _{1}$ by formula (%
\ref{elen}) based on critical points of $\mathcal{E}\left( \psi _{1},\psi
_{2}\right) $ which correspond to lower energy levels of $\mathcal{E}\left(
\psi _{1},\psi _{2}\right) $ by the problem of finding critical points,
lower critical levels and corresponding frequencies $\omega _{1}$ (which for
the logarithmic nonlinearity are expressed in terms of the critical levels
by (\ref{ommine}))\ for the following energy functional with the Coulomb
potential: 
\begin{equation}
\mathcal{E}_{\mathrm{Cb}}\left( \Psi _{1}\right) =\frac{q^{2}}{a_{1}\chi }%
\int_{\mathbb{R}^{3}}\left[ \frac{1}{2}\left\vert \nabla \Psi
_{1}\right\vert ^{2}+\frac{1}{2}G_{1}\left( \left\vert \Psi _{1}\right\vert
^{2}\right) -\frac{1}{\left\vert \mathbf{y}\right\vert }\left\vert \Psi
_{1}\right\vert ^{2}\right] \,\mathrm{d}\mathbf{y}.  \label{E10}
\end{equation}%
This functional involves only $\Psi _{1}$. In quantum mechanics a somewhat
similar reduction (in a completely different setting) from proton-electron
system to a single equation with the Coulomb potential is made via
Born-Oppenheimer approximation. Additional motivation for the reduction is
given in Remark \ref{R:prel}.

\section{ Variational problem for a charge in the Coulomb field}

In this section we consider in detail \emph{radial} critical points of the
functional $\mathcal{E}_{\mathrm{Cb}}\left( \Psi _{1}\right) $ defined by (%
\ref{E10}) and establish its basic properties. In this and following section
we look for \emph{real solutions} $\Psi =\Psi ^{\ast }$ and all function
spaces involve only real functions. In what follows we explicitly take into
account the dependence of the nonlinearity on the size parameter $a=a^{1}$
(or on related parameter $\kappa $), namely 
\begin{gather*}
G_{/\kappa }^{\prime }\left( \left\vert \Psi \right\vert ^{2}\right) =\kappa
^{2}G_{1}^{\prime }\left( \kappa ^{-3}\left\vert \Psi \right\vert
^{2}\right) , \\
G_{/\kappa }\left( \left\vert \Psi \right\vert ^{2}\right) =\kappa
^{5}G_{1}\left( \kappa ^{-3}\left\vert \Psi \right\vert ^{2}\right) ,\qquad
\kappa =\frac{a_{1}}{a},
\end{gather*}%
where, for consistency with the notation (\ref{totgkap}) where $%
a=a_{1}/\kappa ,$ we use notation $/\kappa \ $to denote the dependence of
rescaled $G$ on $\kappa $. Note that the dependence on $a_{1}$ and $a=a^{1}$
is factored out after the rescaling, and that $G_{/\kappa }$ in the above
formula depends only on their ratio $\kappa .$ We consider the case of small 
$\kappa $, namely 
\begin{equation}
\kappa =\frac{a_{1}}{a}\ll 1,  \label{kapaa}
\end{equation}%
when the electron size parameter $a$ is much larger than the Bohr radius $%
a_{1}$. To treat technical difficulties related to the singularity of
logarithm at zero, together with the logarithmic nonlinearity given by (\ref%
{Gpa}) we use a regularized logarithmic nonlinearity defined as follows: 
\begin{equation}
G_{/\kappa ,\xi }^{\prime }\left( \left\vert \psi \right\vert ^{2}\right)
=-\kappa ^{2}\ln _{+,\xi }\left( \kappa ^{-3}\left\vert \psi \right\vert
^{2}/C_{g}^{2}\right) -3\kappa ^{2},  \label{Gpkap}
\end{equation}%
where $\xi \leq 0$,%
\begin{equation}
\ln _{+,\xi }\left( s\right) =\max \left( \ln \left( s\right) ,\xi \right) 
\text{ \ for }s>0,\qquad \ln _{+,\xi }\left( 0\right) =\xi .  \label{lnplus}
\end{equation}%
For $\xi =-\infty $ we set $\ln _{+,-\infty }\left( s\right) =\ln \left(
s\right) $. The function $\ln _{+,\xi }\left( s\right) $ with finite $\xi $
is bounded for bounded $s$ and satisfies Lipschitz condition for $s\geq 0.$
Obviously, 
\begin{equation}
-\ln _{+,\xi }\left( s\right) \leq -\ln _{+,\xi ^{\prime }}\left( s\right) 
\text{ \ for }\xi >\xi ^{\prime }\geq -\infty .  \label{Gksi1}
\end{equation}%
Now we describe the basic properties of $G_{/\kappa ,\xi }^{\prime }$ and
its integral $G_{/\kappa ,\xi }$%
\begin{equation}
G_{1,\xi }\left( s\right) =\int_{0}^{s}G_{1,\xi }^{\prime }\left( s^{\prime
}\right) ds^{\prime },\qquad G_{/\kappa ,\xi }\left( s\right) =-\kappa
^{5}G_{1,\xi }\left( \kappa ^{-3}s\right) .  \label{G1ksi}
\end{equation}

\begin{proposition}
\label{Pr:1} The function $G_{/\kappa ,\xi }$ with $-\infty \leq \xi \leq 0$
can be written in the form

\begin{equation}
G_{/\kappa ,\xi }\left( s\right) =-\kappa ^{2}s\left( \ln _{+,\xi }\left(
\kappa ^{-3}s/C_{g}^{2}\right) +2+C_{g}^{2}\mathrm{e}^{\xi -\ln _{+,\xi
}\kappa ^{-3}s}\right)  \label{G1xikap}
\end{equation}%
where%
\begin{equation}
\mathrm{e}^{\xi -\ln _{+,\xi }s}\leq \min \left( 1,\frac{\mathrm{e}^{\xi }}{s%
}\right) \leq 1  \label{exi}
\end{equation}%
and satisfies the inequalities 
\begin{equation}
G_{/\kappa ,\xi }\left( s\right) \leq G_{/\kappa ,\xi ^{\prime }}\left(
s\right) \leq G_{/\kappa ,-\infty }\left( s\right) =-\kappa ^{2}s\ln \left(
s\right) -\kappa ^{2}s\left( \ln \kappa ^{-3}/C_{g}^{2}+2\right) ,
\label{G1xi1}
\end{equation}%
for $0\geq \xi \geq \xi ^{\prime }\geq -\infty ,$ 
\begin{equation}
\left\vert G_{1,\xi }\left( \left\vert \Psi \right\vert ^{2}\right)
-G_{1,-\infty }\left( \left\vert \Psi \right\vert ^{2}\right) \right\vert
\leq C\mathrm{e}^{\xi /4}\left\vert \Psi \right\vert .  \label{g11p}
\end{equation}
\end{proposition}

\begin{proof}
Elementary computation shows that for$\ s\geq C_{g}^{2}\mathrm{e}^{\xi }$ 
\begin{equation}
G_{1,\xi }\left( s\right) =-s\ln \left( s/C_{g}^{2}\right) -s\left( C_{g}^{2}%
\mathrm{e}^{\xi -\ln s}+2\right)  \label{G1xip}
\end{equation}%
and 
\begin{equation}
G_{1,\xi }\left( s\right) =-\xi s-3s\text{ \ for \ \ \ \ \ }s\leq \mathrm{e}%
^{\xi }C_{g}^{2}.  \label{G1xim}
\end{equation}%
Hence 
\begin{equation}
G_{1,\xi }\left( s\right) =-s\left( \ln _{+,\xi }\left( s/C_{g}^{2}\right)
+2+C_{g}^{2}\mathrm{e}^{\xi -\ln _{+,\xi }s}\right)  \label{G1xi}
\end{equation}%
and (\ref{G1xikap}) holds. Using (\ref{Gksi1}) we derive (\ref{G1xi1}) from (%
\ref{G1ksi}). Obviously%
\begin{equation*}
G_{1,\xi }\left( s\right) -G_{1,-\infty }\left( s\right) =-\mathrm{e}^{\xi }%
\text{ for }s\geq \mathrm{e}^{\xi }C_{g}^{2},
\end{equation*}%
and hence 
\begin{equation}
\left\vert G_{1,\xi }\left( \left\vert \Psi \right\vert ^{2}\right)
-G_{1,-\infty }\left( \left\vert \Psi \right\vert ^{2}\right) \right\vert
\leq \mathrm{e}^{\xi /2}\left\vert \Psi \right\vert /C_{g}\text{ \ for }%
\left\vert \Psi \right\vert ^{2}\geq \mathrm{e}^{\xi }C_{g}^{2}.  \label{g1g}
\end{equation}%
For $\left\vert \Psi \right\vert ^{2}\leq \mathrm{e}^{\xi }C_{g}^{2}$ 
\begin{gather}
\left\vert G_{1,\xi }\left( \left\vert \Psi \right\vert ^{2}\right)
-G_{1,-\infty }\left( \left\vert \Psi \right\vert ^{2}\right) \right\vert
=\left\vert \Psi \right\vert ^{2}\left\vert \left( \xi -\ln \left\vert \Psi
\right\vert ^{2}/C_{g}^{2}\right) \right\vert  \label{g2g} \\
\leq \mathrm{e}^{\xi /4}C_{g}\left\vert \Psi \right\vert ^{3/2}\left\vert
\ln \left\vert \Psi \right\vert ^{2}/C_{g}^{2}\right\vert \text{ }\leq C_{1}%
\mathrm{e}^{\xi /4}C_{g}\left\vert \Psi \right\vert .  \notag
\end{gather}%
>From (\ref{g1g}) and (\ref{g2g}) we obtain (\ref{g11p}).
\end{proof}

The nonlinear eigenvalue problem with the Coulomb potential has the form
similar to (\ref{eqh1}):%
\begin{equation}
\omega \Psi +\frac{1}{2}\nabla ^{2}\Psi +\frac{1}{\left\vert \mathbf{x}%
\right\vert }\Psi =\frac{1}{2}G_{/\kappa ,\xi }^{\prime }\left( \left\vert
\Psi \right\vert ^{2}\right) \Psi ,  \label{parbet}
\end{equation}%
where the dimensionless spectral parameter $\omega =\omega _{\kappa ,\xi }$
is given by the formula 
\begin{equation}
\omega =\frac{\chi a_{1}}{q^{2}}\omega ^{1}.  \label{omom1}
\end{equation}%
Note that equation (\ref{parbet}) can be obtained as the Euler equation by
variation of the energy functional (\ref{E10}) which we write in the form%
\begin{equation}
\mathcal{E}_{\kappa ,\xi }\left( \Psi \right) =\mathcal{E}_{0}\left( \Psi
\right) +\mathcal{G}_{\kappa ,\xi }\left( \Psi \right) ,  \label{Ekapbet}
\end{equation}%
where $\mathcal{E}_{0}$ is the quadratic energy functional 
\begin{equation}
\mathcal{E}_{0}=\int \left( \frac{1}{2}\left\vert \nabla \Psi \right\vert
^{2}-\frac{1}{\left\vert \mathbf{x}\right\vert }\left\vert \Psi \right\vert
^{2}\right) \,\mathrm{d}\mathbf{x},
\end{equation}%
and $\mathcal{G}_{\kappa ,\xi }$ is the nonlinear functional 
\begin{equation}
\mathcal{G}_{\kappa ,\xi }\left( \Psi \right) =-\frac{\kappa ^{2}}{2}\int
\left( \ln _{+,\xi }\left( \kappa ^{-3}\left\vert \Psi \right\vert
^{2}/C_{g}^{2}\right) +2+C_{g}^{2}\mathrm{e}^{\xi -\ln _{+,\xi }\kappa
^{-3}\left\vert \Psi \right\vert ^{2}}\right) \left\vert \Psi \right\vert
^{2}\,\mathrm{d}\mathbf{x}.  \label{E1bet}
\end{equation}

As always we assume the charge normalization constraint 
\begin{equation}
\left\Vert \Psi \right\Vert ^{2}=\int \left\vert \Psi \right\vert ^{2}\,%
\mathrm{d}\mathbf{x}=1  \label{norp}
\end{equation}%
and denote similarly to (\ref{Ksi0}) the set of radial functions (which
depend only on $\left\vert \mathbf{x}\right\vert $) by 
\begin{equation}
\Xi _{\mathrm{rad}}=\left\{ \Psi \in H_{\mathrm{rad}}^{1}\left( \mathbb{R}%
^{3}\right) :\left\Vert \Psi \right\Vert ^{2}=1\right\} .  \label{Ksi}
\end{equation}%
Obviously, for $\Psi \in \Xi \cap X$ where $X$ is defined by (\ref{Xspace}) 
\begin{equation}
\mathcal{G}_{\kappa }\left( \Psi \right) =\mathcal{G}_{\kappa ,-\infty
}\left( \Psi \right) =-\frac{\kappa ^{2}}{2}\int \left( \left\vert \Psi
\right\vert ^{2}\ln \left\vert \Psi \right\vert ^{2}\right) \,\mathrm{d}%
\mathbf{x}+\frac{\kappa ^{2}}{2}\left( \ln \frac{1}{\pi ^{3/2}}-2+3\ln
\kappa \right) .  \label{Gkaps}
\end{equation}%
The spectral parameter $\omega $ is the Lagrange multiplier and it relates
with corresponding critical energy levels $E_{n}^{\kappa ,\xi }$ of $%
\mathcal{E}_{\kappa ,\xi }$ on $\Xi \cap X$ by formula (\ref{ommine}) which
takes the form 
\begin{equation}
E_{n}^{\kappa ,\xi }=\omega _{\kappa ,\xi ,n}+\frac{\kappa ^{2}}{2}.
\label{eom}
\end{equation}%
Let us introduce for $\kappa >0,\xi \leq 0,\Psi \in \mathbb{C}$ the function 
\begin{equation}
g_{\kappa ,\xi }\left( \Psi \right) =G_{/\kappa ,\xi }^{\prime }(\left\vert
\Psi \right\vert ^{2})\Psi =\left( -\kappa ^{2}\ln _{+,\xi }\left( \kappa
^{-3}\left\vert \Psi \right\vert ^{2}/C_{g}^{2}\right) -3\kappa ^{2}\right)
\Psi .\text{ }  \label{gkapp}
\end{equation}%
General properties of functional $\mathcal{E}_{\kappa ,\xi }\left( \Psi
\right) $ with regularized logarithm are described in the following
statements.

\begin{lemma}
\label{L:bound}\ The functional $\mathcal{E}_{\kappa ,\xi }\left( \Psi
\right) $ defined by (\ref{Ekapbet}), (\ref{G1xi}) and (\ref{E1bet}) with $%
-\infty <\xi \leq 0$ on $\Xi _{\mathrm{rad}}$ has the following properties:
(i)\ it is bounded from below on $\Xi $ uniformly in $\xi .$ (ii) If $\Psi
\in \Xi $, $\kappa \leq 1$ and $\mathcal{E}_{\kappa ,\xi }\left( \Psi
\right) \leq C$ then $\left\Vert \Psi \right\Vert _{H^{1}}\leq C^{\prime }$
where $C^{\prime }$ depends only on $C.$ (iii) The functional $\mathcal{E}%
_{\kappa ,\xi }\left( \Psi \right) $ is of class $C^{1}$ with respect to $%
H^{1}$ norm on $\Xi _{\mathrm{rad}}.$ (iv) 
\begin{equation*}
\mathcal{G}_{\kappa ,\xi }\left( \Psi \right) \geq \mathcal{G}_{\kappa
,0}\left( \Psi \right) \geq -C^{\prime \prime }\kappa ^{2}\left(
1+\left\vert \ln \kappa \right\vert \right) \text{ if\ }\left\Vert \Psi
\right\Vert _{H^{1}}\leq C^{\prime }.
\end{equation*}
\end{lemma}

\begin{proof}
Boundedness from below follows from (\ref{egr}), (\ref{G1xi1}) and
inequality 
\begin{equation}
\int_{\mathbb{R}^{3}}\frac{1}{|\mathbf{x}|}\left\vert \Psi \right\vert ^{2}\,%
\mathrm{d}\mathbf{x}\leq C_{0}\left\Vert \Psi \right\Vert \left\Vert \nabla
\Psi \right\Vert .  \label{C01}
\end{equation}%
>From boundedness of $\mathcal{E}_{\kappa ,\xi }\left( \Psi \right) $, (\ref%
{egr}) and (\ref{G1xi1}) we obtain boundedness of $\left\Vert \nabla \Psi
\right\Vert $ which on $\Xi $ implies boundedness in $H^{1}$. To prove (iii)
observe that by (\ref{Gpkap})%
\begin{equation*}
g_{\kappa ,\xi }\left( \Psi \right) \ =\kappa ^{2}\left( -\xi -3\right) \Psi
+g_{\kappa ,\xi }^{1}\left( \left\vert \Psi \right\vert ^{2}\right)
\end{equation*}%
where $g_{\kappa ,\xi }^{1}\left( \left\vert \Psi \right\vert ^{2}\right) $
is identically zero for small $\left\vert \Psi \right\vert $\ and it has
less then quadratic growth for $\left\vert \Psi \right\vert \rightarrow
\infty $. According to Theorem A.VI in \cite{BerestyckiLions83I} then $%
\mathcal{G}_{\kappa ,\xi }\left( \Psi \right) +\kappa ^{2}\left( \xi
+3\right) \left\Vert \Psi \right\Vert ^{2}$ is of class $C^{1}$, and hence $%
\mathcal{G}_{\kappa ,\xi }\left( \Psi \right) $ is of class $C^{1}$ too. To
obtain (iv) we use (\ref{G1xi1}) and observe that $\mathcal{G}_{\kappa
,0}\left( \Psi \right) $ can be estimated in terms of $\left\Vert \Psi
\right\Vert _{L^{6}}$ which, in turn, can be estimated by $\left\Vert \Psi
\right\Vert _{H^{1}}$.
\end{proof}

\begin{lemma}
\label{ThconvLp}Let a sequence of radial functions $\Psi _{j}\in \Xi _{%
\mathrm{rad}}$\ satisfy (\ref{enbound}). Then: (i) a subsequence of $\Psi
_{j}$ converges in $L^{p}\left( \mathbb{R}^{3}\right) $ and almost
everywhere to $\Psi _{\infty }$ if $2<p<6;$ (ii) the functions $\Psi
_{j}\left( \mathbf{x}\right) \rightarrow 0$ as $\left\vert \mathbf{x}%
\right\vert \rightarrow \infty $ \ uniformly in $j$.
\end{lemma}

\begin{proof}
The estimate (\ref{enbound}) together with the normalization condition imply
the uniform boundedness of $\Psi _{j}$ in the Sobolev space $H^{1}\left( 
\mathbb{R}^{3}\right) $. According to Theorem A.I' of \cite%
{BerestyckiLions83I}, the space of radial functions $H_{\mathrm{rad}%
}^{1}\left( \mathbb{R}^{3}\right) $ is compactly imbedded into $L^{p}\left( 
\mathbb{R}^{3}\right) $ if $2<p<6$. Hence, a subsequence $\Psi _{j}\left( 
\mathbf{x}\right) \rightarrow \Psi _{\infty }\left( \mathbf{x}\right) $
strongly in $L^{p}\left( \mathbb{R}^{3}\right) $ and almost everywhere.
Statement (ii) follows from Radial Lemma A.II in \cite{BerestyckiLions83I}.
\end{proof}

The functional $\mathcal{E}_{\kappa ,\xi }\left( \Psi \right) $ with
regularized logarithm satisfies the Palais-Smale condition similar to
condition (P-S$^{+}$) in \cite{BerestyckiLions83II}:

\begin{theorem}
\label{ThPS} Let $\xi >-\infty $. Let a sequence $\Psi _{j}\in \Xi _{\mathrm{%
rad}}$ have the following properties: 
\begin{equation}
\mathcal{E}_{\kappa ,\xi }\left( \Psi _{j}\right) \leq -\beta ,\qquad \beta
>0,  \label{enbound}
\end{equation}%
and 
\begin{equation}
F\Psi _{j}=\omega _{j}\Psi _{j}+\frac{1}{2}\nabla ^{2}\Psi _{j}+\frac{1}{%
\left\vert \mathbf{x}\right\vert }\Psi _{j}-\frac{1}{2}g_{\kappa ,\xi
}\left( \Psi _{j}\right) \rightarrow 0  \label{convneg}
\end{equation}%
strongly in $H^{-1}\left( \mathbb{R}^{3}\right) $. Then the sequence
contains a subsequence which converges strongly in $H_{\mathrm{rad}%
}^{1}\left( \mathbb{R}^{3}\right) $.
\end{theorem}

\begin{proof}
The proof is similar to treatment of radial solutions in \cite%
{BerestyckiLions83I}, \cite{Lions87}. We only sketch main steps. Since (\ref%
{C01}) holds, (\ref{enbound}), (\ref{G1xi1}) and (\ref{Weis}) imply
boundedness of $\Psi _{j}$ in $H^{1}\left( \mathbb{R}^{3}\right) $. Using
Lemma \ref{ThconvLp} we conclude that we can choose a subsequence such that $%
\Psi _{j}\rightarrow \Psi _{\infty }$ in $L^{5}\left( \mathbb{R}^{3}\right) $
and $L^{8/3}\left( \mathbb{R}^{3}\right) ,$ it converges weakly in $%
H^{-1}\left( \mathbb{R}^{3}\right) $ and almost everywhere in $\mathbb{R}%
^{3}.$ Multiplying (\ref{convneg}) by $\Psi _{j}$ in $L^{2}\left( \mathbb{R}%
^{3}\right) =H$ with inner product $\left\langle \cdot ,\cdot \right\rangle $
we obtain 
\begin{gather}
\omega _{j}\left\langle \Psi _{j},\Psi _{j}\right\rangle -\frac{1}{2}%
\left\langle \nabla \Psi _{j},\nabla \Psi _{j}\right\rangle -\frac{1}{2}%
\left\langle \left( G_{/\kappa ,\xi }^{\prime }\left\vert \Psi
_{j}\right\vert ^{2}\right) \Psi _{j},\Psi _{j}\right\rangle  \label{brapsi}
\\
=\left\langle F\Psi _{j},\Psi _{j}\right\rangle -\left\langle \frac{1}{%
\left\vert \mathbf{x}\right\vert }\Psi _{j},\Psi _{j}\right\rangle .  \notag
\end{gather}%
We rewrite (\ref{brapsi}) in the form 
\begin{equation*}
\omega _{j}\left\langle \Psi _{j},\Psi _{j}\right\rangle =\left\langle F\Psi
_{j},\Psi _{j}\right\rangle +\mathcal{E}_{\kappa ,\xi }\left( \Psi
_{j}\right) -\int G_{/\kappa ,\xi }(\left\vert \Psi \right\vert
^{2})-G_{/\kappa ,\xi }^{\prime }(\left\vert \Psi \right\vert
^{2})\left\vert \Psi \right\vert ^{2}\,\mathrm{d}\mathbf{x.}
\end{equation*}%
Using (\ref{exi}) we obtain for $\Psi \in \Xi _{\mathrm{rad}}$ 
\begin{equation}
0<\int G_{/\kappa ,\xi }(\left\vert \Psi \right\vert ^{2})-G_{/\kappa ,\xi
}^{\prime }(\left\vert \Psi \right\vert ^{2})\left\vert \Psi \right\vert
^{2}\,\mathrm{d}\mathbf{x}\leq \frac{\kappa ^{2}}{2}.  \label{omminx}
\end{equation}%
Since $\left\langle F\Psi _{j},\Psi _{j}\right\rangle \rightarrow 0,$ (\ref%
{enbound}) and (\ref{omminx})\ imply that we can choose a subsequence $\Psi
_{j}$ with $\omega _{j}\rightarrow \omega _{\infty }$ where $\omega _{\infty
}\leq -\beta $. Multiplying (\ref{convneg}) by a smooth test function with
compact support and passing to the limit we obtain in standard way that $%
\Psi _{\infty }\in H^{1}\left( \mathbb{R}^{3}\right) $ satisfies (\ref%
{parbet}) with $\omega =\omega _{\infty }.$

Now we want to prove that $\Psi _{\infty }\in \Xi $ and convergence in $H_{%
\mathrm{rad}}^{1}\left( \mathbb{R}^{3}\right) $ is strong. We rewrite (\ref%
{brapsi}) in the form 
\begin{gather}
\omega _{j}\left\langle \Psi _{j},\Psi _{j}\right\rangle -\frac{1}{2}%
\left\langle \nabla \Psi _{j},\nabla \Psi _{j}\right\rangle -\frac{1}{2}%
\left\langle \left( G_{/\kappa ,\xi }\left( \left\vert \Psi _{j}\right\vert
^{2}\right) -\kappa ^{2}\xi \right) \Psi _{j},\Psi _{j}\right\rangle
\label{brapsi1} \\
+\kappa ^{2}\xi \left\langle \Psi _{j},\Psi _{j}\right\rangle =\left\langle
F\Psi _{j},\Psi _{j}\right\rangle -\left\langle \frac{1}{\left\vert \mathbf{x%
}\right\vert }\Psi _{j},\Psi _{j}\right\rangle .  \notag
\end{gather}%
We have $\left\langle F\Psi _{j},\Psi _{j}\right\rangle \rightarrow 0$, 
\begin{equation*}
\omega _{j}\left\langle \Psi _{j},\Psi _{j}\right\rangle +\kappa ^{2}\xi
\left\langle \Psi _{j},\Psi _{j}\right\rangle =\omega _{j}+\kappa ^{2}\xi
\rightarrow \omega _{\infty }+\kappa ^{2}\xi .
\end{equation*}
Since $G_{\kappa ,\xi }\left( \left\vert \Psi \right\vert ^{2}\right)
-\kappa ^{2}\xi \ $is identically zero for small $\left\vert \Psi
\right\vert ^{2}$ and grows slower than $\left\vert \Psi \right\vert ^{3}$
as $\left\vert \Psi \right\vert \rightarrow \infty $, and $\left\vert \Psi
_{j}\right\vert ^{5}$ are converging in $L^{1}\left( \mathbb{R}^{3}\right) $%
,\ we can apply the compactness Lemma of Strauss (see Theorem A.I in \cite%
{BerestyckiLions83I}) and obtain%
\begin{equation*}
\left\langle \left( G_{/\kappa ,\xi }\left( \left\vert \Psi _{j}\right\vert
^{2}\right) -\kappa ^{2}\xi \right) \Psi _{j},\Psi _{j}\right\rangle
\rightarrow \left\langle \left( G_{/\kappa ,\xi }\left( \left\vert \Psi
_{\infty }\right\vert ^{2}\right) -\kappa ^{2}\xi \right) \Psi _{\infty
},\Psi _{\infty }\right\rangle .
\end{equation*}%
Using the H\"{o}lder inequality we obtain that the quadratic mapping $\Psi
\rightarrow \frac{1}{\left\vert \mathbf{x}\right\vert }\left\vert \Psi
\right\vert ^{2}$ is continuous from $L^{5}\left( \mathbb{R}^{3}\right) \cap
L^{8/3}\left( \mathbb{R}^{3}\right) $ into $L^{1}\left( \mathbb{R}%
^{3}\right) .$ Therefore%
\begin{equation*}
\left\langle \frac{1}{\left\vert \mathbf{x}\right\vert }\Psi _{j},\Psi
_{j}\right\rangle \rightarrow \left\langle \frac{1}{\left\vert \mathbf{x}%
\right\vert }\Psi _{\infty },\Psi _{\infty }\right\rangle .
\end{equation*}%
We denote 
\begin{equation*}
M_{0}=\lim_{j\rightarrow \infty }\inf \left\langle \Psi _{j},\Psi
_{j}\right\rangle =1,\qquad M_{1}=\lim_{j\rightarrow \infty }\inf
\left\langle \nabla \Psi _{j},\nabla \Psi _{j}\right\rangle .
\end{equation*}%
>From (\ref{brapsi1}) we obtain passing to the limit 
\begin{gather}
\left( \omega _{\infty }+\kappa ^{2}\xi \right) M_{0}-\frac{1}{2}M_{1}-\frac{%
1}{2}\left\langle \left( G_{/\kappa ,\xi }\left( \left\vert \Psi _{\infty
}\right\vert ^{2}\right) -\kappa ^{2}\xi \right) \Psi _{\infty },\Psi
_{\infty }\right\rangle  \label{brapsi2} \\
=-\left\langle \frac{1}{\left\vert \mathbf{x}\right\vert }\Psi _{\infty
},\Psi _{\infty }\right\rangle .  \notag
\end{gather}%
Multiplying (\ref{parbet}) with $\omega =\omega _{\infty }$ by $\Psi $ in $H$
we obtain 
\begin{gather*}
\omega _{\infty }\left\langle \Psi _{\infty },\Psi _{\infty }\right\rangle -%
\frac{1}{2}\left\langle \nabla \Psi _{\infty },\nabla \Psi _{\infty
}\right\rangle -\frac{1}{2}\left\langle \left( G_{/\kappa ,\xi }\left(
\left\vert \Psi _{\infty }\right\vert ^{2}\right) -\kappa ^{2}\xi \right)
\Psi _{\infty },\Psi _{\infty }\right\rangle \\
+\kappa ^{2}\xi \left\langle \Psi _{\infty },\Psi _{\infty }\right\rangle
=\left\langle F\Psi _{\infty },\Psi _{\infty }\right\rangle -\left\langle 
\frac{1}{\left\vert \mathbf{x}\right\vert }\Psi _{\infty },\Psi _{\infty
}\right\rangle .
\end{gather*}%
Comparing with (\ref{brapsi2}) we obtain 
\begin{equation}
\left( \omega _{\infty }+\kappa ^{2}\xi \right) M_{0}-\frac{1}{2}%
M_{1}=\left( \omega _{\infty }+\kappa ^{2}\xi \right) \left\langle \Psi
_{\infty },\Psi _{\infty }\right\rangle -\frac{1}{2}\left\langle \nabla \Psi
_{\infty },\nabla \Psi _{\infty }\right\rangle .  \label{brapsi3}
\end{equation}%
>From weak convergence in $H_{\mathrm{rad}}^{1}\left( \mathbb{R}^{3}\right) $
we infer that $M_{0}\geq \left\Vert \Psi _{\infty }\right\Vert ^{2}$, $%
M_{1}\geq \left\Vert \nabla \Psi _{\infty }\right\Vert ^{2}$. Since $\left(
\omega _{\infty }+\kappa ^{2}\xi \right) <0,$ (\ref{brapsi3}) is possible
only if $M_{0}=\left\vert \Psi _{\infty }\right\vert ^{2}$, $%
M_{1}=\left\vert \nabla \Psi _{\infty }\right\vert ^{2}$. These equalities
imply that the weak convergence of the subsequence in $H^{1}$ to $\Psi
_{\infty }$ is strong.
\end{proof}

\subsection{Nonlinear eigenvalues for a charge in the Coulomb field\label%
{S:nonleig}}

In this section we prove that if $\kappa $ is small enough, the nonlinear
eigenvalue problem (\ref{parbet}) with logarithmic nonlinearity has
solutions with eigenvalues which are close to the eigenvalues of the
corresponding linear problem. We look for\emph{\ real-valued} solutions. As
a first step we prove existence of such solutions of the problem with
regularized logarithm $\ln _{+,\xi }$. As the second step we pass to the
limit as $\xi \rightarrow -\infty $ and obtain solutions of the eigenvalue
problem with the original logarithmic nonlinearity.

Consider the linear Schr\"{o}dinger operator $\mathcal{O}$ with the Coulomb
potential which corresponds to $\mathcal{E}_{0}\left( \Psi \right) $, that
is 
\begin{equation}
\mathcal{O}\Psi =-\frac{1}{2}\nabla ^{2}\Psi -\frac{1}{\left\vert \mathbf{x}%
\right\vert }\Psi .  \label{OC}
\end{equation}%
Note that $\mathcal{O}$ has the following well known negative eigenvalues 
\begin{equation}
E_{n}^{0}=\omega _{0,n}=-\frac{1}{2n^{2}},\qquad n=1,2,...  \label{om0n}
\end{equation}%
which coincide with the negative eigenvalues of the operator $\mathcal{O}_{%
\mathrm{rad}}$ obtained by restriction of $\mathcal{O}$ to radial functions
(without counting their multiplicity) and are equal to the negative critical
energy levels of $\mathcal{E}_{0}\left( \Psi \right) $. Every eigenvalue of
the radial problem 
\begin{equation*}
\mathcal{O}_{\mathrm{rad}}\Psi _{n}^{0}=\omega _{0,n}\Psi _{n}^{0}
\end{equation*}%
is simple and corresponding eigenfunctions $\Psi _{n}^{0}\left( r\right) $
given by well-known explicit formulas decay exponentially as $r\rightarrow
\infty $, for example the linear ground state is given by the formula 
\begin{equation*}
\Psi _{1}^{0}\left( \mathbf{x}\right) =\frac{1}{\pi ^{1/2}}\mathrm{e}%
^{-\left\vert \mathbf{x}\right\vert },\qquad E_{1}^{0}=-\frac{1}{2}.
\end{equation*}%
Let us denote by $L_{n}^{0}$ an eigenspace of $\mathcal{O}_{\mathrm{rad}}$
which corresponds to the eigenvalue $E_{n}^{0}$ and consists of radial
functions, by $L_{n}^{-}\subset H_{\mathrm{rad}}^{1}$ a finite-dimensional
invariant subspace of $\mathcal{O}_{\mathrm{rad}}$ which corresponds to
eigenvalues $\lambda \leq E_{n}^{0}$ and by $L_{n}^{+}\subset H_{\mathrm{rad}%
}^{1}\left( \mathbb{R}^{3}\right) $ invariant infinite-dimensional subspace
in $H_{\mathrm{rad}}^{1}\left( \mathbb{R}^{3}\right) $ of $\mathcal{O}_{%
\mathrm{rad}}$ which is orthogonal in $L^{2}$ to $L_{n}^{-}$. We have an
orthogonal decomposition 
\begin{equation*}
H_{\mathrm{rad}}^{1}\left( \mathbb{R}^{3}\right)
=L_{n}^{-}+L_{n}^{+}=L_{n-1}^{-}+L_{n}^{0}+L_{n}^{+}.
\end{equation*}%
We look for eigenvalues $\omega =\omega _{\kappa ,\xi ,n}<0$ of the
nonlinear eigenvalue problem (\ref{parbet}) which can be considered as a
perturbation of the eigenvalues $E_{n}^{0}$. The ground state of $\mathcal{E}%
_{\kappa ,-\infty }\left( \Psi \right) $ and the corresponding energy level $%
\omega _{\kappa ,-\infty ,1}$\ was found in \cite{Bialynicki}, it is given
by the formula 
\begin{equation}
\Psi _{\kappa ,-\infty ,1}\left( \mathbf{x}\right) =\kappa ^{3/2}C_{g}%
\mathrm{e}^{-\left\vert \kappa \mathbf{x}\right\vert ^{2}/2}C\mathrm{e}%
^{-\left\vert \mathbf{x}\right\vert },\qquad \omega _{\kappa ,1}=-\frac{1}{2}%
-\frac{1}{2}\kappa ^{2}\ln C^{2},  \label{psi1k}
\end{equation}%
where 
\begin{equation*}
C^{-2}=\frac{4}{\pi ^{1/2}}\left[ \frac{3}{4}\sqrt{\pi }\mathrm{e}-\frac{3}{4%
}\sqrt{\pi }\func{erf}\left( 1\right) \mathrm{e}-\frac{1}{2}\right] ,\qquad 
\frac{1}{2}\ln C^{2}\simeq 1.868.
\end{equation*}%
Hence the nonlinear correction for the ground level is of order $\kappa ^{2}$%
.

Since the Palais-Smale condition is satisfied according to Theorem \ref{ThPS}%
, the existence of an infinite sequence of discrete negative energy values
of $\mathcal{E}_{\kappa ,\xi }\left( \Psi \right) $ can be proven exactly
along the lines of \cite{BerestyckiLions83I}. But we want to show, in
addition, that the eigenvalues of the nonlinear problem with small $\kappa $
are close to the eigenvalues of the linear problem (this is not surprising,
since all the eigenvalues of the linear problem for radial functions are
simple). Therefore we modify the construction of critical points of the
functional $\mathcal{E}_{\kappa ,\xi }$ on $\Xi $.

We will use the following deformation lemma which is a minor modification
and reformulation in our notation of Lemma 5 from \cite{BerestyckiLions83I}
(with essentially the same proof).

\begin{lemma}
\label{L:defor}Suppose that the functional $\mathcal{E}_{\kappa ,\xi }\in
C^{1}\left( \Xi _{\mathrm{rad}},\mathbb{R}\right) $ satisfies Palais-Smale
condition as in Theorem \ref{ThPS}. Suppose also that $E^{\prime \prime }<0$%
\ and that the segment $E^{\prime }\leq E\leq E^{\prime \prime }\ $does not
contain critical values of $\mathcal{E}_{\kappa ,\xi }$. Let $\bar{\epsilon}%
>0$ be arbitrary small number. Then there exists $0<\epsilon <\bar{\epsilon}$
and a mapping $\eta \in C\left( \Xi _{\mathrm{rad}},\Xi _{\mathrm{rad}%
}\right) $ such that

\begin{enumerate}
\item[(i)] $\eta \left( u\right) =u$ for $u\in \Xi _{\mathrm{rad}}$ with $%
\mathcal{E}_{\kappa ,\xi }\left( u\right) \leq E^{\prime }-\bar{\epsilon}$
or $\mathcal{E}_{\kappa ,\xi }\left( u\right) \geq E^{\prime \prime }+\bar{%
\epsilon}.$

\item[(ii)] $\eta $ is a homeomorphism $\Xi _{\mathrm{rad}}\rightarrow \Xi _{%
\mathrm{rad}}$ and it is odd if $\mathcal{E}_{\kappa ,\xi }$ is even.

\item[(iii)] $\mathcal{E}_{\kappa ,\xi }\left( \eta \left( u\right) \right)
\leq \mathcal{E}_{\kappa ,\xi }\left( u\right) .$

\item[(iv)] $\eta \left( \left\{ u\in \Xi _{\mathrm{rad}}:\mathcal{E}%
_{\kappa ,\xi }\left( u\right) \leq E+\epsilon \right\} \right) \subset
\left\{ u\in \Xi _{\mathrm{rad}}:\mathcal{E}_{\kappa ,\xi }\left( u\right)
\leq E-\epsilon \right\} $ for $E\in \left[ E^{\prime },E^{\prime \prime }%
\right] $.
\end{enumerate}
\end{lemma}

To prove existence of critical points of $\mathcal{E}_{\kappa ,\xi }\left(
\Psi \right) $ on the infinite-dimensional sphere $\Xi _{\mathrm{rad}}$\
with critical values $E_{n}^{\kappa ,\xi }$ near $E_{n}^{0}$, $n=2,3...$
when $\kappa $ is small, we slightly modify the construction from \cite%
{BerestyckiLions83I}. As a first step we define finite-dimensional domain $%
B=B_{n}\subset L_{n}^{-}$\ which lies in $\Xi _{\mathrm{rad}}$\ near
eigenfunction $\Psi _{n}^{0}$ of $\mathcal{O}_{\mathrm{rad}}$ as follows: 
\begin{equation}
B_{n}=\left\{ \Psi :\Psi =\left( 1-R^{2}\right) ^{1/2}\Psi _{n}^{0}+v,\ v\in
L_{n-1}^{-},\ \left\Vert v\right\Vert =R,\ 0\leq R\leq R_{0}\right\} ,
\label{Bn}
\end{equation}%
where $R_{0}<1/2$ is a small fixed number, for simplicity we fix $R_{0}=1/3.$
The domain $B_{n}$ is diffeomorphic to a $n-1$ dimensional ball. The
boundary of $B_{n}$ is the $n-2$ dimensional sphere 
\begin{equation}
\partial B_{n}=\left\{ \Psi :\Psi =\left( 1-R_{0}^{2}\right) ^{1/2}\Psi
_{n}^{0}+v,v\in L_{n-1}^{-},\ \left\Vert v\right\Vert =R_{0}\right\} .
\label{dBn}
\end{equation}%
We consider continuous functions $\Gamma $ defined on $B_{n}$ with values in 
$\Xi $. We consider the class $\{\Gamma \}_{n}$ of continuous mappings $%
\Gamma :B_{n}\rightarrow \Xi _{\mathrm{rad}}$ such that $\Gamma $\
restricted to $\partial B_{n}$ is identical, 
\begin{equation}
\Gamma \left( \Psi \right) =\Psi \text{ if }\Psi \in \partial B_{n}.
\label{Gamdb}
\end{equation}%
The $n$-th min-max energy level $E_{\kappa ,\xi ,n}$ is defined as follows: 
\begin{equation}
E_{\kappa ,\xi ,n}=\inf_{\Gamma \in \{\Gamma \}_{n}}\max_{\Psi \in B_{n}}%
\mathcal{E}_{\kappa ,\xi }\left( \Gamma \left( \Psi \right) \right) .
\label{omkapn}
\end{equation}

\begin{lemma}
\label{L:seg}Let $\xi \leq 0,$ let $n=2,3,...$., $0<\kappa \leq 1$. Then $%
E_{\kappa ,\xi ,n}$ defined by (\ref{omkapn}) lies in the interval 
\begin{equation}
I_{n}=\left[ E^{\prime },E^{\prime \prime }\right] ,\text{ with }E^{\prime
}=E_{n}^{0}-C_{3n}\kappa ^{2}\left( 1+\left\vert \ln \kappa \right\vert
\right) ,\qquad E^{\prime \prime }=E_{n}^{0}+C_{1n}\kappa ^{2},  \label{Ipm}
\end{equation}%
where $C_{1n}$ and $C_{3n}$ depend only on $n$.
\end{lemma}

\begin{proof}
Since we can take a particular function $\Gamma \left( \Psi \right) =\Gamma
_{0}\left( \Psi \right) =\Psi _{n}^{0}+v$ for $\Psi \in B_{n}$ we have 
\begin{equation}
E_{\kappa ,\xi ,n}\leq \max_{\Psi \in B_{n}}\mathcal{E}_{\kappa ,\xi }\left(
\Psi \right) \leq \mathcal{E}_{0}\left( \Psi _{n}^{0}\right) +\max_{B_{n}}%
\mathcal{G}_{\kappa ,\xi }\left( \Psi \right) =E_{n}^{0}+\max_{B_{n}}%
\mathcal{G}_{\kappa ,\xi }\left( \Psi \right) .  \label{omup}
\end{equation}%
Note that $B_{n}$ lies in a ball in a finite-dimensional subspace $L_{n}^{-}$
in $H^{1}$, the basis of $L_{n}^{-}$ consists of exponentially decaying
functions with bounded first derivatives, hence $B_{n}$ is compact in $%
H^{1}\left( \mathbb{R}^{3}\right) $ and in $L^{p}\left( \mathbb{R}%
^{3}\right) $ for every $p\geq 1.$ Therefore 
\begin{equation*}
\left\Vert \Psi \right\Vert _{H^{1}}+\left\Vert \Psi \right\Vert
_{L^{3}}+\left\Vert \Psi \right\Vert _{L^{4/3}}\leq C_{0,n}\text{ for }\Psi
\in B_{n}.
\end{equation*}%
Hence, using (\ref{G1xi1}) and power estimates for $\Psi ^{2}\ln \Psi ^{2}$
we conclude that 
\begin{equation*}
\mathcal{G}_{\kappa ,\xi }\left( \Psi \right) \leq \mathcal{G}_{\kappa
,-\infty }\left( \Psi \right) \leq C\kappa ^{2}\left( \left\Vert \Psi
\right\Vert _{L^{3}}^{3}+\left\Vert \Psi \right\Vert _{L^{4/3}}^{4/3}\right)
-\kappa ^{2}\left( \ln \kappa ^{-3}/C_{g}^{2}+2\right) \left\Vert \Psi
\right\Vert ^{2}
\end{equation*}%
and we obtain 
\begin{equation}
\max_{B_{n}}\mathcal{G}_{\kappa ,\xi }\left( \Psi \right) \leq C_{1n}\kappa
^{2},  \label{maxgkap}
\end{equation}%
where $C_{1n}$ depends only on $n$. Using $\Gamma _{0}\left( \Psi \right) $
and the estimate (\ref{omup}) of $\mathcal{E}_{\kappa ,\xi }\left( \Psi
\right) $ we observe that 
\begin{equation}
E_{\kappa ,\xi ,n}=\inf_{\Gamma \in \left\{ \Gamma \right\} _{n},\ \mathcal{E%
}_{\kappa ,\xi }\left( \Gamma \left( B_{n}\right) \right) \leq
C_{2n}}\max_{\Psi \in B_{n}}\mathcal{E}_{\kappa ,\xi }\left( \Gamma \left(
\Psi \right) \right) ,  \label{omkapn1}
\end{equation}%
where $C_{2n}=C_{1n}\kappa ^{2}+E_{n}^{0}$. Inequality (\ref{maxgkap})
provides $E^{\prime \prime }$ in (\ref{Ipm}).

Now we estimate $E_{\kappa ,\xi ,n}$ from below. Mappings $\Gamma \in
\left\{ \Gamma \right\} _{n}$ which satisfy (\ref{Gamdb}) have the following
topological property: the set $\Gamma \left( B_{n}\right) $ must intersect
the subspace $L_{n-1}^{+}$. Indeed, assume that the contrary were true and
for some $\Gamma $ there were not intersection of $\Gamma \left( \Psi
\right) $ with $L_{n-1}^{+}$, that is $\left\Vert \Pi _{n-1}^{-}\Gamma
\left( \Psi \right) \right\Vert \neq 0$ for all $\Psi \in B_{n}$ where $\Pi
_{n-1}^{-}$\ is the orthoprojection onto the finite-dimensional subspace $%
L_{n-1}^{-}$. Then we could consider the following function on the ball $%
\left\Vert v\right\Vert \leq R_{0}$ in $L_{n-1}^{-}$: 
\begin{equation*}
\Gamma _{1}\left( v\right) =\frac{R_{0}}{N\left( v\right) }\Pi
_{n-1}^{-}\Gamma \left( v\right) ,
\end{equation*}%
where $N\left( v\right) =\left\Vert \Pi _{n-1}^{-}\left( \Gamma \left(
v\right) \right) \right\Vert .$ This function is continuous on the ball $%
\left\{ \left\Vert v\right\Vert \leq R_{0}\right\} $ (since the denominator $%
N\left( v\right) $ does not vanish) and $\Gamma _{1}\left( v\right) =v$ if $%
\left\Vert v\right\Vert =R_{0}$, and also it has the property $\left\Vert
\Gamma _{1}\left( v\right) \right\Vert =R_{0}$ if $\left\Vert v\right\Vert
\leq R_{0}$. A function with such properties is called a retraction of the
ball $\left\{ \left\Vert v\right\Vert \leq R_{0}\right\} $ in the $\left(
n-1\right) $-dimensional space $L_{n-1}^{-}$ onto its boundary$\left\{
\left\Vert v\right\Vert =R_{0}\right\} $. It is a well-known fact from the
algebraic topology that such a retraction cannot exist. Therefore for every $%
\Gamma $ we have $\left\Vert \Pi _{n-1}^{-}\Gamma \left( \Psi _{0}\right)
\right\Vert =0$ for some$\Psi _{0}\in B$. Therefore there exists $\Psi
_{0}\in B_{n}$ such that 
\begin{equation}
\Gamma \left( \Psi _{0}\right) =\left( 1-R^{2}\right) ^{1/2}\Psi
_{n}^{0}+w,\quad w\in L_{n}^{+},\quad \left\Vert w\right\Vert ^{2}=R^{2}\leq
1.  \label{Gampn}
\end{equation}%
To obtain estimate of $E_{\kappa ,\xi ,n}$ from below note that (\ref{Gampn}%
) implies 
\begin{equation*}
\mathcal{E}_{0}\left( \Gamma \left( \Psi _{0}\right) \right) =\left(
1-R^{2}\right) \mathcal{E}_{0}\left( \Psi _{n}^{0}\right) +\mathcal{E}%
_{0}\left( w\right) \geq \left( 1-R^{2}\right) E_{n}^{0}+E_{n}^{0}\left\Vert
w\right\Vert ^{2}\geq E_{n}^{0}.
\end{equation*}%
Therefore, we obtain for $\Gamma \in \left\{ \Gamma \right\} _{n}$, $%
\mathcal{E}_{\kappa ,\xi }\left( \Gamma \left( B_{n}\right) \right) \leq
C_{2n}$ 
\begin{equation*}
\mathcal{E}_{\kappa ,\xi }\left( \Gamma \left( \Psi _{0}\right) \right) \geq
E_{n}^{0}+\min_{\Psi \in \Xi ,\mathcal{E}_{\kappa ,\xi }\left( u\right) \leq
C_{2n}}\mathcal{G}_{\kappa ,\xi }\left( u\right) .
\end{equation*}%
Inequality $\mathcal{E}_{\kappa ,\xi }\left( \Gamma \left( B_{n}\right)
\right) \leq C_{2n}$ by Lemma \ref{L:bound} implies the boundedness $\Gamma
\left( B_{n}\right) $ in $H^{1}$. Using (iv) in Lemma \ref{L:bound} we
conclude that 
\begin{equation}
E_{\kappa ,\xi ,n}\geq E_{n}^{0}-\kappa ^{2}C_{3n}\left( 1+\left\vert \ln
\kappa \right\vert \right) .  \label{ekapln}
\end{equation}%
Combining (\ref{ekapln}), (\ref{omup}) and (\ref{maxgkap}) we obtain (\ref%
{Ipm}).
\end{proof}

Now we prove the existence of solutions of eigenvalue problem with
regularized logarithm.

\begin{theorem}
\label{Th:spectrum0} Let $-\infty <\xi \leq 0.$ Let integer $n\geq 2$ and $%
\kappa \leq 1$ be so small that 
\begin{equation}
\frac{1}{9}\left( E_{n-1}^{0}-E_{n}^{0}\right) +C_{1n}\kappa
^{2}+C_{3n}\kappa ^{2}\left( 1+\left\vert \ln \kappa \right\vert \right) <0,
\label{emine}
\end{equation}%
\ where $E_{n}^{0}$ are defined by (\ref{om0n}) and $C_{1n}$ is the constant
from (\ref{maxgkap}). Then

\begin{enumerate}
\item[(i)] the interval $I_{n}$ defined by (\ref{Ipm}) contains a critical
point $E_{n}^{\kappa ,\xi }$ of $\mathcal{E}_{\kappa ,\xi };$

\item[(ii)] there exists a solution $\Psi _{n}\in \Xi _{\mathrm{rad}}$ of (%
\ref{parbet}) with eigenvalue $\omega =\omega _{\kappa ,\xi ,n}$ such that $%
\left\vert E_{n}^{\kappa ,\xi }-\omega _{\kappa ,\xi ,n}\right\vert \leq 
\frac{\kappa ^{2}}{2}$ which satisfies 
\begin{equation}
\left\vert \omega _{\kappa ,\xi ,n}-\omega _{0,n}\right\vert \leq
C_{n}\kappa ^{2}\left( 1+\left\vert \ln \kappa \right\vert \right) .
\label{ommin}
\end{equation}
\end{enumerate}
\end{theorem}

\begin{proof}
We follow here the lines of \cite{BerestyckiLions83I} with minor
modifications. Point (ii) follows from point (i) and (\ref{omminx}). To
prove (i) assume the contrary, namely let $n$ and $\kappa $ satisfy all the
conditions and assume that that the interval $I_{n}$ defined by (\ref{Ipm})
does not contain critical points of $\mathcal{E}_{\kappa ,\xi }$. We apply
then Lemma \ref{L:defor} where we take 
\begin{equation*}
\bar{\epsilon}\leq \frac{1}{2}\left\vert \frac{1}{9}\left(
E_{n-1}^{0}-E_{n}^{0}\right) +C_{1n}\kappa ^{2}+C_{3n}\kappa ^{2}\left(
1+\left\vert \ln \kappa \right\vert \right) \right\vert ,
\end{equation*}%
with $E^{\prime }=E_{n}^{0}-\kappa ^{2}C_{3n}\left( 1+\left\vert \ln \kappa
\right\vert \right) $ and find a mapping $\eta :\Xi _{\mathrm{rad}%
}\rightarrow \Xi _{\mathrm{rad}}$. By Lemma \ref{L:seg} the value $E_{\kappa
,\xi ,n}$ defined by (\ref{omkapn}) or (\ref{omkapn1}) lies inside $I_{n}$.
By point (iv) of Lemma \ref{L:defor} 
\begin{equation*}
\eta \left( \left\{ u\in \Xi _{\mathrm{rad}}:\mathcal{E}_{\kappa ,\xi
}\left( u\right) \leq E_{\kappa ,\xi ,n}+\epsilon \right\} \right) \subset
\left\{ u\in \Xi _{\mathrm{rad}}:\mathcal{E}_{\kappa ,\xi }\left( u\right)
\leq E_{\kappa ,\xi ,n}-\epsilon \right\} .
\end{equation*}%
We can find $\Gamma \in \left\{ \Gamma \right\} _{n}$ such that $%
\max_{B_{n}}\Gamma \left( \Psi \right) \leq E_{\kappa ,\xi ,n}+\epsilon /2.$
If we consider $\Gamma ^{\prime }\left( \Psi \right) =\eta \left( \Gamma
\left( \Psi \right) \right) $ then we get $\max_{B_{n}}\Gamma ^{\prime
}\left( \Psi \right) \leq E_{\kappa ,\xi ,n}-\epsilon /2.$ To infer from
this inequality a contradiction with definition (\ref{omkapn}) we have to
check that $\Gamma ^{\prime }\in \left\{ \Gamma \right\} _{n},$ namely to
check (\ref{Gamdb}). Note that on $\partial B_{n}$%
\begin{gather*}
\mathcal{E}_{0}\left( \left( 1-R_{0}^{2}\right) ^{1/2}\Psi _{n}^{0}+v\right)
\\
=\left( 1-R_{0}^{2}\right) E_{n}^{0}+\mathcal{E}_{0}\left( v\right) \leq
E_{n}^{0}+R_{0}^{2}\left( E_{n-1}^{0}-E_{n}^{0}\right) ,
\end{gather*}%
and hence%
\begin{gather}
\mathcal{E}_{\kappa ,\xi }\left( \left( 1-R_{0}^{2}\right) ^{1/2}\Psi
_{2}^{0}+v\right) \\
\leq E_{n}^{0}+R_{0}^{2}\left( E_{n-1}^{0}-E_{n}^{0}\right) +\mathcal{G}%
_{\kappa ,\xi }\left( \left( 1-R_{0}^{2}\right) ^{1/2}\Psi _{n}^{0}+v\right)
.  \notag
\end{gather}%
Therefore 
\begin{equation}
\left. \mathcal{E}_{\kappa ,\xi }\right\vert _{\partial B_{n}}\leq
E_{n}^{0}+R_{0}^{2}\left( E_{n-1}^{0}-E_{n}^{0}\right) +\max_{\partial
B_{n}}\left( \mathcal{G}_{\kappa ,\xi }\left( \Psi \right) \right) ,
\label{Ekdb}
\end{equation}%
and using (\ref{maxgkap}) and (\ref{emine}), since $R_{0}^{2}=1/9$, we
conclude that 
\begin{equation}
\left. \mathcal{E}_{\kappa ,\xi }\right\vert _{\partial B_{n}}\leq
E_{n}^{0}+R_{0}^{2}\left( E_{n-1}^{0}-E_{n}^{0}\right) \ +C_{1n}\kappa
^{2}\leq E^{\prime }-\bar{\epsilon}.
\end{equation}%
implying $\eta \left( \Psi \right) =\Psi $ on $\partial B_{n}$ by point (i)
of Lemma \ref{L:defor}. Hence, $\Gamma ^{\prime }\in \left\{ \Gamma \right\}
_{n}$ and we obtain a contradiction with (\ref{omkapn}). So we conclude that 
$I_{n}$ contains a critical point $\Psi ^{\prime }$. Since $\mathcal{E}%
_{\kappa ,\xi }$ is differentiable at $\Psi ^{\prime }$, this critical point
provides a solution to (\ref{parbet}), and (\ref{ommin}) follows from (\ref%
{eom}).
\end{proof}

Now we show that solutions of (\ref{parbet}) decay exponentially as $%
r=\left\vert \mathbf{x}\right\vert \rightarrow \infty $.

\begin{lemma}
\label{L:xi0}Let $n\leq n_{0}.$ Let $0<\kappa \leq \kappa _{0}$ where $%
\kappa _{0}$ is small enough for conditions of Theorem \ref{Th:spectrum0} to
be fulfilled for $n\leq n_{0}$. Let $\bar{\xi}\leq 0$ satisfy 
\begin{equation}
1-\frac{\kappa ^{2}}{4}\left\vert \bar{\xi}\right\vert +\frac{3\kappa ^{2}}{2%
}<0.  \label{xi0}
\end{equation}%
Let $p=\frac{\kappa }{3}\left\vert \bar{\xi}\right\vert ^{1/2}.$ Let $\Psi
_{n}\in \Xi _{\mathrm{rad}}$ be a solution of (\ref{parbet}) with eigenvalue 
$\omega =\omega _{\kappa ,\xi ,n}$ which is described in Theorem \ref%
{Th:spectrum0} with $\xi \leq $ $\bar{\xi}$. Then there exist positive
constants $R_{1}$ and $C$ which depend only on $n_{0},\kappa $ and $\bar{\xi}%
,$ such that for $\xi \leq \bar{\xi},$ $n\leq n_{0}$%
\begin{equation}
\left\vert \Psi _{n}\left( r\right) \right\vert \leq C^{\prime \prime }%
\mathrm{e}^{-pr}\text{ for }r\geq R_{1}.  \label{psiexp}
\end{equation}
\end{lemma}

\begin{proof}
Since $\left\Vert \Psi _{n}\right\Vert _{H^{1}}$ are bounded uniformly in $%
\xi \leq 0$ and $\kappa \leq 1$\ we can use the following result of Strauss
(see Lemma A.II in \cite{BerestyckiLions83I})%
\begin{equation}
\left\vert \Psi _{n}\left( r\right) \right\vert \leq C_{0}^{\prime
}r^{-1}\left\Vert \Psi _{n}\right\Vert _{H^{1}}\text{\ for \ }r\geq \bar{R},
\label{str}
\end{equation}%
where $C_{0}^{\prime }$ and $\bar{R}$ are absolute constants, $\bar{R}\geq 1$%
. We apply this inequality to radial solutions $\Psi =\Psi _{n}$ of (\ref%
{parbet}) with $\omega =\omega _{\kappa ,\xi ,n}<0\ $and obtain 
\begin{equation}
\left\vert \Psi _{n}\left( r\right) \right\vert \leq C_{0}r^{-1}\text{ for\ }%
r\geq \bar{R},  \label{r10}
\end{equation}%
where, according to Lemma \ref{L:bound}, the constant $C_{0}$ is bounded
uniformly in $\kappa \leq 1,\xi \leq 0$ and $n$. We write (\ref{parbet}) in
the form%
\begin{equation}
\frac{1}{2}\nabla ^{2}\Psi +V_{\kappa ,\xi }\Psi =0,  \label{omv}
\end{equation}%
where%
\begin{equation*}
V_{\kappa ,\xi }\left( \mathbf{x}\right) =\frac{1}{\left\vert \mathbf{x}%
\right\vert }-\frac{1}{2}G_{/\kappa ,\xi }^{\prime }\left( \left\vert \Psi
\left( \mathbf{x}\right) \right\vert ^{2}\right) +\omega .
\end{equation*}%
According to (\ref{r10}) using monotonicity of the function $\ln _{+,\xi
}\left( s\right) $ with respect to $s$ \ we obtain for $\left\vert \mathbf{x}%
\right\vert \geq R\geq \bar{R}\geq 1,$ 
\begin{equation*}
V_{\kappa ,\xi }\left( \mathbf{x}\right) \leq \frac{1}{R}+\frac{\kappa ^{2}}{%
2}\left( \ln _{+,\xi }\left( \kappa ^{-3}R^{-2}C_{0}^{2}/C_{g}^{2}\right)
+3\right) +\omega .
\end{equation*}%
Note that 
\begin{equation*}
V_{\kappa ,\xi }\left( \mathbf{x}\right) \leq \frac{1}{R}+\frac{\kappa ^{2}}{%
2}\left( \xi +3\right) +\omega \text{ \ if \ }\ln \left( \kappa
^{-3}R^{-2}C_{0}^{2}/C_{g}^{2}\right) \leq \xi ,R\geq \bar{R}.
\end{equation*}%
Therefore, for any $0<\kappa \leq \kappa _{0},$ $\bar{\xi}\leq 0$ we can
find $R_{1}\geq \bar{R}$ such that 
\begin{equation*}
\ln \left( \kappa ^{-3}R_{1}^{-2}C_{0}^{2}/C_{g}^{2}\right) \leq \bar{\xi},
\end{equation*}%
and if $\left\vert \bar{\xi}\right\vert $ is so large that (\ref{xi0})
holds, we obtain for $\xi =$ $\bar{\xi}$ 
\begin{equation}
V_{\kappa ,\xi }\left( \mathbf{x}\right) \leq -\frac{\kappa ^{2}}{4}%
\left\vert \bar{\xi}\right\vert +\omega <0\text{ \ for \ }\left\vert \mathbf{%
x}\right\vert \geq R_{1}.  \label{Vle0}
\end{equation}%
According to definition (\ref{lnplus}) $V_{\kappa ,\xi }\left( \mathbf{x}%
\right) $ is a monotonely increasing function of $\xi ,$ therefore this
inequality holds for all $\xi \leq $ $\bar{\xi}.$

Note that $\Psi \left( \mathbf{x}\right) $ cannot change sign for $%
\left\vert \mathbf{x}\right\vert >R_{1}$. Indeed, if $\Psi \left( \mathbf{x}%
\right) =0$ at $\left\vert \mathbf{x}\right\vert =R_{1}^{\prime }>R_{1}$ we
could multiply (\ref{omv}) by $\Psi $ and integrate over $\left\{ \left\vert 
\mathbf{x}\right\vert >R_{1}^{\prime }\right\} $ which would yield 
\begin{equation*}
\int_{\left\{ \left\vert \mathbf{x}\right\vert >R_{1}^{\prime }\right\}
}\left( \frac{1}{2}\left\vert \nabla \Psi \right\vert ^{2}-V_{\kappa ,\xi
}\left( \mathbf{x}\right) \left\vert \Psi \right\vert ^{2}\right) d\mathbf{x}%
=0,
\end{equation*}%
and $\Psi \left( \mathbf{x}\right) =0$ for all $\left\vert \mathbf{x}%
\right\vert >R_{1}^{\prime }\ $which is impossible for a radial solution of (%
\ref{parbet}) with $\left\Vert \Psi \right\Vert =1.$ So we assume $\Psi
\left( \mathbf{x}\right) >0$ for $\left\vert \mathbf{x}\right\vert >R_{1}$
(the case of negative $\Psi $ is similar). We compare the solution $\Psi $
of this equation in the domain $\left\vert \mathbf{x}\right\vert >R$ with
the function 
\begin{equation*}
Y=C_{Y}\frac{1}{r}\mathrm{e}^{-pr},\quad p>0,C_{Y}>0,
\end{equation*}%
which is a solution of the equation 
\begin{equation*}
\nabla ^{2}Y-p^{2}Y=0.
\end{equation*}%
Subtracting half this equation from (\ref{omv}) we obtain%
\begin{equation}
\frac{1}{2}\nabla ^{2}\left( \Psi -Y\right) +V_{\kappa ,\xi }\left( \Psi
-Y\right) +\left( V_{\kappa ,\xi }+\frac{1}{2}p^{2}\right) Y=0.  \label{psmy}
\end{equation}%
We choose $p^{2}=\frac{\kappa ^{2}}{9}\left\vert \bar{\xi}\right\vert $ and (%
\ref{Vle0}) implies 
\begin{equation}
\left( V_{\kappa ,\xi }+\frac{1}{2}p^{2}\right) <\omega <0\text{\ \ for \ }%
\left\vert \mathbf{x}\right\vert \geq R_{1}.  \label{vp}
\end{equation}%
We choose $C_{Y}>0$ so that%
\begin{equation*}
C_{Y}\mathrm{e}^{-pR_{1}}>C_{0},
\end{equation*}%
where $C_{0}$ is the same as in (\ref{r10}). We assert that%
\begin{equation}
\Psi \left( \mathbf{x}\right) -Y\left( \mathbf{x}\right) \leq 0\text{ for\ }%
\left\vert \mathbf{x}\right\vert \geq R_{1}.  \label{Ypsi}
\end{equation}%
This is true at $\left\vert \mathbf{x}\right\vert =R_{1}$ according to the
choice of $C_{Y}$, the limit of $\Psi -Y\ $as $\left\vert \mathbf{x}%
\right\vert \rightarrow \infty $\ is zero. Assume now that (\ref{Ypsi}) does
not hold for some $\left\vert \mathbf{x}\right\vert >R_{1}$. Then $\Psi
\left( \mathbf{x}\right) -Y\left( \mathbf{x}\right) \ $must have a point of
local positive maximum at some $\left\vert \mathbf{x}\right\vert
=R_{1}^{\prime }>R_{1}$, at this point $\nabla ^{2}\left( \Psi -Y\right)
\leq 0$, by (\ref{vp}) $\left( V_{\kappa ,\xi }+p^{2}\right) Y<0$ and $%
V_{\kappa ,\xi }\left( \Psi -Y\right) \leq 0$ by (\ref{Vle0}). This
inequalities contradict (\ref{psmy}), hence (\ref{Ypsi}) holds. Inequality (%
\ref{psiexp}) follows from (\ref{Ypsi}).
\end{proof}

Now we prove the main result of this section for the nonlinear eigenvalue
problem with the original logarithmic inequality which corresponds to $\xi
=-\infty .$

\begin{theorem}
\label{Th:spectrum} Let $\xi =-\infty ,$ $n_{0}>0$ and let integer $2$ $\leq
n\leq n_{0}$ and $\kappa \leq 1$ be so small that (\ref{emine}) holds. Then
there exists a solution $\Psi _{n}\in \Xi _{\mathrm{rad}}\cap X$ of (\ref%
{parbet}) with eigenvalue $\omega =\omega _{\kappa ,n}$ \ with energy $%
E_{n}^{\kappa }=\omega _{\kappa ,n}+\frac{\kappa ^{2}}{2}$ which satisfies
the inequality 
\begin{equation}
\left\vert \omega _{\kappa ,n}-\omega _{0,n}\right\vert \leq C_{n}\kappa
^{2}\left( 1+\left\vert \ln \kappa \right\vert \right) .  \label{omomn}
\end{equation}%
The solution $\Psi _{n}$ decays superexponentially as $\left\vert \mathbf{x}%
\right\vert \rightarrow \infty ,$ namely (\ref{psiexp}), where $R_{1}$
depends on $p,$ holds with arbitrary large $p$.
\end{theorem}

\begin{proof}
We fix $n$ and consider a sequence $\xi _{m}\rightarrow -\infty .$ We have a
sequence of solutions $\Psi _{n,m}\in \Xi _{\mathrm{rad}}$ of (\ref{parbet})
with $\omega _{\kappa ,\xi _{m},n}$ satisfying (\ref{ommin}) which exist
according to Theorem \ref{Th:spectrum0} with $\xi =\xi _{m}$. Note that
according to (\ref{Ipm}) $\mathcal{E}_{\kappa }\left( \Psi _{n,m}\right) \in %
\left[ E^{\prime },E^{\prime \prime }\right] $ are bounded uniformly in $m$.
Therefore $\Psi _{n,m}$ are bounded in $H^{1}.$ Let us fix $\bar{\xi}$ which
satisfies (\ref{xi0}) and take such $m$ that $\xi _{m}\leq \bar{\xi}.$ All $%
\Psi _{n,m}$ satisfy (\ref{psiexp}). We can find a subsequence of $\Psi
_{n,m}$ which converges to $\Psi _{n,\infty }\in $ $\Xi _{\mathrm{rad}}$
weakly in $H^{1},$ strongly in $L^{5}\left( \left\{ \left\vert \mathbf{x}%
\right\vert \leq R\right\} \right) \cap L^{1}\left( \left\{ \left\vert 
\mathbf{x}\right\vert \leq R\right\} \right) $ for every $R>0$ and almost
everywhere in $\mathbb{R}^{3};$ we also choose the subsequence so that $%
\omega _{\kappa ,\xi _{m},n}\rightarrow \omega _{\kappa ,-\infty ,n}\in %
\left[ E^{\prime },E^{\prime \prime }\right] -\frac{\kappa ^{2}}{2}.$ We
obtain that $\nabla ^{2}\Psi _{n,m}$ converges weakly to $\nabla ^{2}\Psi
_{n,\infty }\in $ $\Xi _{\mathrm{rad}}$. Now we show that $\Psi _{n,\infty }$
is a solution of (\ref{parbet}) with $\omega =\omega _{\kappa ,n}=\omega
_{\kappa ,-\infty ,n}$ in the sense of distributions. The restriction of $%
g_{\kappa ,\xi _{m}}\left( \Psi _{n,m}\left( \mathbf{x}\right) \right) $ to
the ball $\left\{ \left\vert \mathbf{x}\right\vert \leq R\right\} $
converges to $g_{\kappa ,-\infty }\left( \Psi _{n,\infty }\left( \mathbf{x}%
\right) \right) $ in $L^{2}\left( \left\{ \left\vert \mathbf{x}\right\vert
\leq R\right\} \right) $. Multiplying (\ref{parbet}) with $\xi =\xi _{m}$ by
an infinitely smooth test function $f$ with compact support and integrating,
we observe that we can pass to the limit in every term of the equation,
obtaining that $\Psi _{n,\infty }$ is a solution of (\ref{parbet}). Now we
show that $\Psi _{n,\infty }\in $ $\Xi _{\mathrm{rad}}\cap X.$ Since
estimate (\ref{psiexp}) is uniform in $\xi _{m}\leq \bar{\xi}$ and $%
\left\vert G_{\kappa ,-\infty }\left( C^{\prime \prime }\mathrm{e}%
^{-pr}\right) \right\vert $ is integrable over $\mathbb{R}^{3}$ we easily
derive from dominated convergence theorem that 
\begin{equation}
\int_{\left\vert \mathbf{x}\right\vert \geq R_{1}}G_{\kappa ,\xi _{m}}\left(
\Psi _{n,m}\right) \,\mathrm{d}\mathbf{x}\rightarrow \int_{\left\vert 
\mathbf{x}\right\vert \geq R_{1}}G_{\kappa ,-\infty }\left( \Psi _{n,\infty
}\right) \,\mathrm{d}\mathbf{x.}  \label{glimi0}
\end{equation}

Note also that 
\begin{equation}
\int_{\left\vert \mathbf{x}\right\vert \leq R_{1}}G_{\kappa ,\xi _{m}}\left(
\Psi _{n,m}\right) \,\mathrm{d}\mathbf{x}\rightarrow \int_{\left\vert 
\mathbf{x}\right\vert \leq R_{1}}G_{\kappa ,-\infty }\left( \Psi _{n,\infty
}\right) \,\mathrm{d}\mathbf{x}  \label{glimi}
\end{equation}%
as $m\rightarrow \infty .$ Indeed, 
\begin{gather*}
G_{\kappa ,\xi _{m}}\left( \left\vert \Psi _{n,m}\right\vert ^{2}\right)
-G_{\kappa ,-\infty }\left( \left\vert \Psi _{n,\infty }\right\vert
^{2}\right) = \\
=G_{\kappa ,\xi _{m}}\left( \left\vert \Psi _{n,m}\right\vert ^{2}\right)
-G_{\kappa ,-\infty }\left( \left\vert \Psi _{n,m}\right\vert ^{2}\right)
+G_{\kappa ,-\infty }\left( \left\vert \Psi _{n,m}\right\vert ^{2}\right)
-G_{\kappa ,-\infty }\left( \left\vert \Psi _{n,\infty }\right\vert
^{2}\right)
\end{gather*}%
Note that $G_{\kappa ,-\infty }\left( \left\vert \Psi \right\vert
^{2}\right) $ defined by (\ref{g1gauss}) satisfies Lipschitz condition. 
\begin{equation*}
\left\vert G_{\kappa ,-\infty }\left( \left\vert \Psi _{n,m}\right\vert
^{2}\right) -G_{\kappa ,-\infty }\left( \left\vert \Psi _{n,\infty
}\right\vert ^{2}\right) \right\vert \leq C\left\vert \left\vert \Psi
_{n,m}\right\vert -\left\vert \Psi _{n,\infty }\right\vert \right\vert
\left( 1+\left\vert \Psi _{n,m}\right\vert ^{2}-\left\vert \Psi _{n,\infty
}\right\vert ^{2}\right) .
\end{equation*}%
Hence 
\begin{gather}
\int_{\left\vert \mathbf{x}\right\vert \leq R}\left\vert G_{\kappa ,-\infty
}\left( \left\vert \Psi _{n,m}\right\vert ^{2}\right) -G_{\kappa ,-\infty
}\left( \left\vert \Psi _{n,\infty }\right\vert ^{2}\right) \right\vert \,%
\mathrm{d}\mathbf{x}  \label{gmin} \\
\leq C^{\prime }\left\Vert \Psi _{n,m}-\Psi _{n,\infty }\right\Vert \left(
1+\left\Vert \Psi _{n,m}\right\Vert _{L^{4}}^{2}+\left\Vert \Psi _{n,\infty
}\right\Vert _{L^{4}}^{2}\right) +C^{\prime }\left\Vert \Psi _{n,m}-\Psi
_{n,\infty }\right\Vert _{L^{1}\left( \left\vert \mathbf{x}\right\vert \leq
R\right) }  \notag
\end{gather}%
>From (\ref{g11p}) we obtain 
\begin{equation}
\int_{\left\vert \mathbf{x}\right\vert \leq R}\left\vert G_{\kappa ,\xi
_{m}}\left( \left\vert \Psi _{n,m}\right\vert ^{2}\right) -G_{\kappa
,-\infty }\left( \left\vert \Psi _{n,m}\right\vert ^{2}\right) \right\vert \,%
\mathrm{d}\mathbf{x}\leq C\mathrm{e}^{\xi _{m}/4}\int_{\left\vert \mathbf{x}%
\right\vert \leq R}\left\vert \Psi _{n,m}\right\vert \,\mathrm{d}\mathbf{x}.
\label{gmin1}
\end{equation}%
Using (\ref{gmin}), (\ref{gmin1}) and convergence of $\Psi _{n,m}$ in $%
L^{1}\left( \left\{ \left\vert \mathbf{x}\right\vert \leq R\right\} \right) $
we obtain (\ref{glimi}). From (\ref{glimi0}) and (\ref{glimi}) we infer 
\begin{equation}
\int_{\mathbb{R}^{3}}G_{\kappa ,\xi _{m}}\left( \Psi _{n,m}\right) d\mathbf{x%
}\rightarrow \int_{\mathbb{R}^{3}}G_{\kappa ,-\infty }\left( \Psi _{n,\infty
}\right) d\mathbf{x}.  \label{glimi1}
\end{equation}%
This implies that $\Psi _{n,\infty }\in X$ with $X$ defined by (\ref{Xspace}%
). From (\ref{glimi1}) and weak convergence in $H^{1}$ we infer that 
\begin{equation*}
\mathcal{E}_{\kappa }\left( \Psi _{n,\infty }\right) \leq \lim_{m\rightarrow
\infty }\sup \mathcal{E}_{\kappa }\left( \Psi _{n,m}\right) .
\end{equation*}%
Since $\mathcal{E}_{\kappa }\left( \Psi _{n,m}\right) $ are bounded
uniformly, $\mathcal{E}_{\kappa }\left( \Psi _{n,\infty }\right) $ is
bounded, hence $\Psi _{n,\infty }\in W$. Therefore (\ref{parbet}) holds in $%
H^{-1}+X^{\prime }$. Similarly\ to (\ref{glimi1}) we obtain that $\left\Vert
\Psi _{n,m}\right\Vert ^{2}\rightarrow \left\Vert \Psi _{n,\infty
}\right\Vert ^{2}=1$, hence $\Psi _{n,\infty }\in $ $\Xi _{\mathrm{rad}}.$
Since we can choose $\bar{\xi}$ arbitrary large, $p$ in (\ref{psiexp}) can
also be taken arbitrary large, and from convergence almost everywhere we
derive that $\Psi _{n,\infty }\left( r\right) $ decays superexponentially.
\end{proof}

The following theorem shows, roughly speaking, that there is no eigenvalues $%
\omega $ of the nonlinear problem in the gaps between small neighborhoods of
eigenvalues of the linear problem.$.$

\begin{theorem}
\label{Th:spec1} Let $\omega $ satisfy the inequalities 
\begin{equation}
-\frac{1}{2}\leq \omega _{0,n}+\delta <\omega <\omega _{0,n+1}-\delta <0,
\label{omom}
\end{equation}%
where $\delta >0.$ There exists $C_{4}\left( n\right) >0$ such that if 
\begin{equation}
\delta \geq C_{4}\left( n\right) \kappa ^{2}\left( 1+\left\vert \ln \kappa
\right\vert \right) ,  \label{delc4}
\end{equation}%
then the set of (radial or non-radial) solutions $\Psi \in \Xi \cap X$ of
equation (\ref{parbet}) with such $\omega $ is empty.
\end{theorem}

\begin{proof}
Consider in $L^{2}\left( \mathbb{R}^{3}\right) $ a linear operator $\mathcal{%
O}_{\omega }$ (the classical Schr\"{o}dinger operator with the Coulomb
potential shifted by $\omega $ ) 
\begin{equation*}
\mathcal{O}_{\omega }\Psi =\mathcal{O}\Psi -\omega \Psi =-\omega \Psi -\frac{%
1}{2}\nabla ^{2}\Psi -\frac{1}{\left\vert \mathbf{x}\right\vert }\Psi .
\end{equation*}%
Consider solution of (\ref{parbet}) written in the form 
\begin{equation}
\mathcal{O}_{\omega }\Psi =-\frac{1}{2}g_{\kappa }\left( \Psi \right)
\label{parbet1}
\end{equation}%
with$\ \omega $ in the gap between two consecutive points of spectrum of $%
\mathcal{O}$. Multiplying (\ref{parbet}) (where $\xi =-\infty )$\ in $%
H=L^{2}\left( \mathbb{R}^{3}\right) $ by $\Psi \in \Xi \cap X$ we obtain%
\begin{equation}
\left\langle \mathcal{O}_{\omega }\Psi ,\Psi \right\rangle =-\frac{1}{2}%
\left\langle g_{\kappa }\left( \Psi \right) ,\Psi \right\rangle ,
\end{equation}%
where $g_{\kappa }\left( \Psi \right) $ is defined in (\ref{gkapp}), $%
\left\langle u,v\right\rangle \ $coincides with scalar product in $%
H=L^{2}\left( \mathbb{R}^{3}\right) $ and determines duality between $H^{1}$
and $H^{-1}$ and $X$ and $X^{\prime }$ in (\ref{Xspace}). Note that $%
g_{\kappa }\left( \Psi \right) \in X^{\prime }$ according to Lemma 9.3.3 in 
\cite{Cazenave03}, hence the expressions are well-defined. Since 
\begin{equation*}
\int \frac{1}{\left\vert \mathbf{x}\right\vert }\left\vert \Psi \right\vert
^{2}\,\mathrm{d}\mathbf{x}\leq C_{\epsilon }||\Psi ||^{2}+\epsilon ||\Psi
||_{H^{1}}||\Psi ||\text{ for }\Psi \in H^{1}
\end{equation*}%
with arbitrary small $\epsilon >0,$ we have 
\begin{equation*}
\left\langle \mathcal{O}_{\omega }\Psi ,\Psi \right\rangle \geq \frac{1}{4}%
||\Psi ||_{H^{1}}^{2}-C_{\epsilon }^{\prime }||\Psi ||^{2}.
\end{equation*}%
Using (\ref{Weis}) we estimate $\left\langle g_{\kappa }\left( \Psi \right)
,\Psi \right\rangle $ and we obtain for solutions of (\ref{parbet}) the
following inequality: 
\begin{equation*}
\frac{1}{4}||\Psi ||_{H^{1}}^{2}-C^{\prime \prime }||\Psi ||^{2}\leq \frac{3%
}{2}\kappa ^{2}\left\vert \ln \int \left\vert \nabla \Psi \right\vert ^{2}d%
\mathbf{x}\right\vert .
\end{equation*}%
Hence$\left\Vert \Psi \right\Vert _{H^{1}}\leq M$ for solutions $\Psi \in
\Xi \cap X.$

Let $\breve{E}_{n}^{0}\subset H^{1}\left( \mathbb{R}^{3}\right) $ be the
finite-dimensional space with the orthonormal basis of eigenfunctions $%
\check{\Psi}_{j}$ of the operator $\mathcal{O}_{0}=\mathcal{O}$ with
negative eigenvalues $\check{\omega}_{j}\leq \omega _{0,n}$ (since we are
not restricted to radial functions, the eigenvalues $\check{\omega}_{j}\leq 
\check{\omega}_{j+1}$ may have high multiplicity and $\check{\omega}%
_{n}<\omega _{0,n}$ for $n>2$). Let $\Pi _{n}$ be orthoprojection in $%
L^{2}\left( \mathbb{R}^{3}\right) $ onto $\breve{E}_{+}^{0}.$ We have by (%
\ref{omom}) 
\begin{equation*}
\left\langle \left( \mathcal{O}_{\omega }-\mathcal{O}_{\omega }\Pi
_{n}\right) \Psi ,\Psi \right\rangle \geq \left( \omega _{0,n+1}-\omega
\right) ||\Psi ||^{2}.
\end{equation*}%
Since the eigenfunctions $\check{\Psi}_{j}$ of $\mathcal{O}$ are bounded,
continuous and decay exponentially, we see that operators 
\begin{equation*}
\Pi _{n}\Psi =\sum_{j=1}^{n}\left\langle \Psi ,\check{\Psi}_{j}\right\rangle 
\check{\Psi}_{j},\qquad \mathcal{O}_{\omega }\Pi _{n}\Psi
=\sum_{j=1}^{n}\left( \check{\omega}_{j}-\omega \right) \left\langle \Psi ,%
\check{\Psi}_{j}\right\rangle \check{\Psi}_{j}
\end{equation*}%
are bounded from $L^{p}\left( \mathbb{R}^{3}\right) $ to $L^{p^{\prime
}}\left( \mathbb{R}^{3}\right) $ for any given $p,p^{\prime },$ $\infty \geq
p\geq 1$, $\infty \geq p^{\prime }\geq 1.$ Solutions $\Psi \in \Xi \cap
X^{\prime }$ are bounded in $H^{1},$ and by Sobolev imbedding in $%
L^{6}\left( \mathbb{R}^{3}\right) .$ Since $\left\vert g_{\kappa }\left(
\Psi \right) \right\vert \leq C\kappa ^{2}\left( 1+\left\vert \ln \kappa
\right\vert \right) \left( \left\vert \Psi \right\vert ^{3/2}+\left\vert
\Psi \right\vert ^{1/2}\right) $ we obtain 
\begin{equation}
\left\Vert \Pi _{n}g_{\kappa }\left( \Psi \right) \right\Vert \leq C^{\prime
\prime }\kappa ^{2}\left( 1+\left\vert \ln \kappa \right\vert \right)
\label{ping}
\end{equation}%
Multiplying (\ref{parbet}) by $\Psi -\Pi _{n}\Psi $ we obtain 
\begin{equation*}
\left\langle \mathcal{O}_{\omega }\Psi ,\left( \Psi -\Pi _{n}\Psi \right)
\right\rangle =-\frac{1}{2}\left\langle g_{\kappa }\left( \Psi \right)
,\left( \Psi -\Pi _{n}\Psi \right) \right\rangle
\end{equation*}%
Therefore%
\begin{gather}
\left( \omega _{0,n+1}-\omega \right) ||\left( 1-\Pi _{n}\right) \Psi
||^{2}\leq -\frac{1}{2}\left\langle g_{\kappa }\left( \Psi \right) ,\left(
\Psi -\Pi _{n}\Psi \right) \right\rangle \\
\leq -\frac{1}{2}\int g_{\kappa }\left( \Psi \right) \Psi d\mathbf{x+}\frac{1%
}{2}\left\Vert \Pi _{n}g_{\kappa }\left( \Psi \right) \right\Vert \left\Vert
\Psi \right\Vert  \notag
\end{gather}%
Using (\ref{Weis}) and (\ref{ping}) to estimate the right-hand side we obtain%
\begin{equation}
\left( \omega _{0,n+1}-\omega \right) ||\left( 1-\Pi _{n}\right) \Psi
||^{2}\leq C_{1}\kappa ^{2}\left( 1+\left\vert \ln \kappa \right\vert
\right) .  \label{omnp1}
\end{equation}%
Multiplying (\ref{parbet1}) by $\Pi _{n}\Psi $ we obtain 
\begin{gather}
\left\vert \omega _{0,n}-\omega \right\vert ||\Pi _{n}\Psi ||^{2}\leq
\left\vert \left\langle \mathcal{O}_{\omega }\Psi ,\Pi _{n}\Psi
\right\rangle \right\vert =\frac{1}{2}\left\vert \left\langle \Pi
_{n}g_{\kappa }\left( \Psi \right) ,\Pi _{n}\Psi \right\rangle \right\vert
\label{omnp2} \\
\leq \frac{1}{2}\left\Vert \Pi _{n}g_{\kappa }\left( \Psi \right)
\right\Vert \leq C_{2}\kappa ^{2}\left( 1+\left\vert \ln \kappa \right\vert
\right) .  \notag
\end{gather}%
>From (\ref{omnp1}) and (\ref{omnp2}) using (\ref{omom}) we obtain 
\begin{equation*}
\delta ||\Psi ||^{2}\leq C_{3}\kappa ^{2}\left( 1+\left\vert \ln \kappa
\right\vert \right)
\end{equation*}%
If we take in (\ref{delc4}) $C_{4}\left( n\right) =2C_{3}$ we obtain a
contradiction with $\left\Vert \Psi \right\Vert =1$. Hence, a solution of (%
\ref{parbet}) with such $\omega $ does not exist.
\end{proof}

\begin{remark}
Theorem \ref{Th:spectrum} states that for every energy level of the linear
Schr\"{o}dinger hydrogen operator there exists an energy level of nonlinear
hydrogen equation with the Coulomb potential with an estimate of difference
of order $\sim $ $\kappa ^{2}\ln \kappa $. Based on (\ref{omom1}) we see
that if $\chi =\hbar $ where $\hbar $ is the Planck constant and $\kappa =0$
then $\omega ^{1}=\omega _{n}^{1}$ coincide with standard expression for
hydrogen frequencies, and a natural choice for parameter $\chi $ is $\chi
=\hbar $.
\end{remark}

\begin{remark}
A theorem similar to Theorem \ref{Th:spectrum} (without superexponential
decay of solutions) can be proven not only for the logarithmic nonlinearity,
but for more general subcritical nonlinearities which involve a small
parameter $\kappa $. Moreover, if corresponding functions $g_{\kappa }$ are
of class $C^{1}$ and their derivatives satisfy natural growth conditions,
then the standard bifurcation analysis shows that every $E_{n}^{0}$
generates a branch $E_{\kappa ,n}$ of solutions $\Psi _{n}\in \Xi _{\mathrm{%
rad}}$ of (\ref{parbet}) for small $\kappa .$
\end{remark}

\begin{remark}
In Theorem \ref{Thepsel} we assumed in the macroscopic case electron size $%
a\ $to be very small and in this section we assume $\kappa =a_{1}/a$ to be
very small, which seem to contradict one another. But a closer look shows
that in fact the both cases are consistent one with another, if we assume
the possibility $a_{1}\rightarrow 0$. To show that this assumption is quite
reasonable, we look at the values of physical constants. Note that a typical
macroscopic scale is $a_{\mathrm{macr}}\sim 10^{-3}m$ and the Bohr radius $%
a_{1}\sim 5.3\times 10^{-11}$ $m$. Note also that the error in (\ref{eps4lim}%
) is of order $a^{2}/a_{\mathrm{macr}}^{2}\ll 1\ $and here $\kappa
^{2}=a_{1}^{2}/a^{2}\ll 1$ which is possible if $a_{1}^{2}\ll a^{2}\ll a_{%
\mathrm{macr}}^{2}$. Since $a_{1}^{2}/a_{\mathrm{macr}}^{2}\sim 2.8\times
10^{-15}$ we consider $a_{1}^{2}/a_{\mathrm{macr}}^{2}\rightarrow 0$ as a
reasonable approximation.
\end{remark}

\begin{remark}
In the case when the energy functional with the Coulomb potential $\mathcal{E%
}_{\mathrm{Cb}}\left( \Psi \right) $ is considered not only on radial
functions, the eigenvalues of the linear problem are not simple and have
very high degeneracy. Note that if we start from the relativistic equation (%
\ref{KG}) and look for $\omega $-harmonic solutions we arrive to a nonlinear
perturbation of so-called \emph{relativistic Schr\"{o}dinger equation}, see 
\cite{Schiff}. The spectrum of this linear problem has a fine structure and
the degeneracy with respect to the total angular momentum disappears. If we
fix the value of $z$-component $L_{z}$ of the angular momentum $L$, for
example by looking for solutions of the form $\Psi _{1}=$\textrm{$e$}$^{%
\mathrm{i}m_{z}\theta }\Psi _{10}\left( r,z\right) $ with a fixed integer $%
m_{z}$, one can eliminate the remaining degeneracy in the linear problem.
This provides a ground to expect that the lower energy levels of the
nonlinear relativistic Schr\"{o}dinger equation can be approximated by the
lower energy levels of the linear relativistic Schr\"{o}dinger equation if $%
\kappa $ is small.
\end{remark}

\begin{remark}
\label{R:prel}Note that the decoupled equation (\ref{eqh2}) with $b=0$ for
proton has a unique ground state solution (gausson) which corresponds to the
minimum $E_{1}^{p}$ of energy (see Theorem \ref{Thglmin} ) and has a
sequence of solutions with higher critical energy levels $E_{j}^{p}$, $j\geq
2$ (see \cite{Cazenave83}). Since the ground state is unique, $%
E_{j}^{p}>E_{1}^{p}$ if $j>1$. The logarithmic nonlinearity $G_{2}^{\prime
}\left( \left\vert \psi \right\vert ^{2}\right) =-\kappa _{p}^{2}\ln \left(
\kappa _{p}^{-3}\left\vert \psi \right\vert ^{2}/C_{g}^{2}\right) -3$ for
proton involves the parameter $\kappa _{p}$ which is different from the
parameter $\kappa $ for electron. A simple rescaling and use of (\ref{ommine}%
) shows that dependence of $E_{n}^{p}\left( \kappa _{p}\right) $ on $\kappa
_{p}$ is very simple: $E_{n}^{p}\left( \kappa _{p}\right) =$ $\kappa
_{p}^{2}E_{n}^{p}\left( 1\right) $. There are physical motivations based on
experimental estimates of the size of the proton to assume that $\kappa _{p}$
is large, $\kappa _{p}\gg 1$. So we can assume that the gap $E_{j}^{p}\left(
\kappa _{p}\right) -E_{1}^{p}$ $\left( \kappa _{p}\right) \geq $\ $\left(
E_{j}^{p}\left( 1\right) -E_{1}^{p}\left( 1\right) \right) \kappa _{p}^{2}$
is larger than $1$ for $j\geq 2$. (Numerically calculated values of $%
E_{2}^{p}\left( 1\right) -E_{1}^{p}\left( 1\right) $ are given in \cite%
{Bialynicki1}, they are of order $1$). At the same time $\omega _{0,n}$
given by (\ref{om0n}) satisfy inequality $-\frac{1}{2}\leq \omega _{0,n}<0$.
Consequently, since the total energy $\mathcal{E}$ defined by (\ref{Epe}) of
the system (\ref{H1})-(\ref{v1v2}) equals the sum of energies of electron
and proton one may expect for small $b$ and $\kappa $\ and large $\kappa
_{p} $ that the lower negative energy levels of the functional $\mathcal{E}$
cannot branch as $b$ varies from $E_{j}^{p}+$ $\omega _{0,n}$ with $j\geq 2$
and must branch from $E_{1}^{p}+\omega _{0,n}$. Hence we expect that if $b$
and $\kappa $\ are small and $\kappa _{p}$ is large, the lower negative
energy levels of the functional $\mathcal{E}$ which corresponds to the
system (\ref{H1})-(\ref{v1v2}) are close to lower discrete energy levels of $%
\mathcal{E}_{\mathrm{Cb}}$ which are described in Theorem \ref{Th:spectrum}.
\end{remark}

\textbf{Acknowledgment.} The research was supported through Dr. A. Nachman
of the U.S. Air Force Office of Scientific Research (AFOSR), under grant
number FA9550-04-1-0359.

\end{document}